\documentclass[a4paper,11pt]{article}
\usepackage[title,titletoc,toc]{appendix}
\usepackage[utf8]{inputenc}
\usepackage[affil-it]{authblk}
\usepackage{graphicx} 
\usepackage{amsmath}
\usepackage{amsfonts}
\usepackage{amsthm}
\usepackage{amssymb}
\usepackage{blindtext}
\usepackage{titlefoot}
\usepackage{bbm}
\usepackage{xcolor}
\usepackage{tikz}
\usepackage{mathtools}
\usepackage[square, numbers]{natbib}
\usepackage{hyperref}
\usepackage{mathrsfs}
\usepackage{enumitem}
\usepackage{csvsimple}
\usepackage{alphabeta}
\usepackage{pdflscape}
\usepackage{pdfpages}
\usepackage{verbatim}
\usepackage{float}
\usepackage{tcolorbox}

\makeatletter 

\newcommand*\bigcdot{\mathpalette\bigcdot@{.5}}
\newcommand*\bigcdot@[2]{\mathbin{\vcenter{\hbox{\scalebox{#2}{$\m@th#1\bullet$}}}}}

\makeatother 

\makeatletter
\def\set@csv@head{\relax}
\makeatother

\newcommand{\is}{\bigcdot}

\newtheorem{theorem}{Theorem}[section]

\newtheorem{proposition}[theorem]{Proposition}
\newtheorem{remark}[theorem]{Remark}

\newtheorem{lemma}[theorem]{Lemma}

\newtheorem{corollary}[theorem]{Corollary}
\newtheorem{definition}[theorem]{Definition}
\hypersetup{
    colorlinks=true,
    linkcolor=blue,
    filecolor=magenta,
    urlcolor=cyan,
    citecolor=purple 
}
\title{Second-order Esscher pricing for L\'evy models with applications: Risk \\management and fear quantification 
}
\author{Tahir Choulli and  Ella Elazkany \\
Mathematical and Statistical Sciences Dept.\\
University of Alberta, Edmonton, Canada
\and 
Mich\`ele Vanmaele\\ 
Department of Mathematics, Computer Science and Statistics,\\
Ghent University, Ghent, Belgium}
\begin{document}

\maketitle

\begin{abstract}
 This paper proposes the second-order Esscher transform as a tractable extension of the classical Esscher framework for option pricing and risk management in L\'evy-driven markets. For a general L\'evy process, we derive the associated densities and equivalent pricing measures, characterize the martingale condition in closed form, and obtain FFT-based valuation formulas for European call options. For jump-diffusion models, we establish explicit pricing formulas under the second-order Esscher measure and show that the resulting option prices lie in an interval bounded below by the Black--Scholes price and above by the underlying asset value. For the constant jump-diffusion model, we further prove monotonicity of option prices with respect to the second-order Esscher parameter.
		 
		 An empirical analysis based on market data shows that this additional parameter provides a tractable tool for stress testing, delta-hedging evaluation, and the construction of interval-valued risk measures in incomplete markets. We further document a strong association between the estimated second-order Esscher parameter and standard indicators of market stress, including the VIX, news sentiment, and crisis regimes. The proposed framework preserves analytical tractability while enlarging the class of admissible pricing measures, thereby supporting pricing, hedging, and stress-based risk assessment in incomplete markets with jump and general L\'evy dynamics.
\end{abstract}
\section{Introduction}

The Esscher transform is one of the most widely used tools for constructing equivalent martingale measures in incomplete markets, particularly for models driven by Lévy processes. Its appeal lies in its tractability, its compatibility with exponential Lévy dynamics, and its long history in actuarial science and mathematical finance. However, the classical Esscher transform is inherently one-dimensional: it adjusts the drift of the underlying process through a single parameter, thereby capturing only mean-based sources of risk. Recent work by \cite{choulli2025second} introduced the \emph{second-order Esscher transform} for continuous-time models, a two-parameter extension that incorporates both linear and quadratic terms in the underlying semimartingale. This additional degree of freedom produces a family of equivalent martingale measures and, consequently, a \emph{pricing interval} rather than a single arbitrage-free price.\\
	
	The present paper develops the practical and theoretical implications of this second-order framework.  
	The present paper develops the theoretical foundations and practical implications of a second-order Esscher framework for exponential L\'evy models.
				Our \textit{first contribution} is to characterise the second-order Esscher density and the associated family of pricing measures for general L\'evy processes. We derive the martingale condition in closed form that relates  the first-order Esscher parameter and the second-order parameter. Furthermore, we analyse the impact of the transformation on the L\'evy property, identifying conditions under which it is preserved. In the case of constant second-order parameter, we provide a complete characterisation of the class of L\'evy models admitting second-order Esscher densities and derive Fourier-based valuation formulas for European options.\\
	Our \textit{second contribution} concerns explicit valuation results for European contingent claims in jump-diffusion settings. For the constant jump-diffusion (CJD) model, we show that the second-order parameter induces a continuum of arbitrage-free option prices, bounded below by the Black--Scholes price and above by the underlying asset value. We prove monotonicity of option prices with respect to the second-order parameter and characterize the corresponding  Esscher-adjusted jump intensities. These results provide a transparent structural interpretation of the second-order parameter as a lever that encodes additional information, preferences, or risk considerations not specified by the model dynamics.\\
		The \textit{third contribution} addresses implications for risk management. Exploiting the fact that the second-order Esscher transform generates a parametric family of equivalent martingale measures, we construct by means of real market data interval-valued risk measures, including value-at-risk (VaR) and expected shortfall (ES). 
		For a fixed model, variation in the second-order parameter induces a family of profit-and-loss distributions, thereby providing a coherent framework for scenario analysis and stress testing. This approach is particularly relevant in settings where reliable calibration is challenging, such as illiquid or incomplete markets, e.g., illiquid commodity, insurance-linked, or cryptocurrency markets.\\
		\textit{Finally}, we demonstrate that the second-order parameter~$\psi$ provides a parsimonious and theoretically consistent mechanism for incorporating external information into option valuation. Empirical evidence shows that models calibrated exclusively under the historical measure via maximum likelihood display systematic pricing discrepancies compared to market option data. Allowing for variation in~$\psi$ significantly improves pricing accuracy, indicating that the second-order version captures features of the risk-neutral distribution not identifiable from return data alone. Importantly, $\psi$ can be interpreted as a latent state variable governing deviations between physical and risk-neutral dynamics, thereby encoding effects such as regime shifts, behavioral distortions, or market sentiment, while preserving martingale consistency.\\
			Taken together, these results position the second-order Esscher transform as a flexible and mathematically coherent extension of the classical first-order Esscher transform.
	 The framework combines tractability with enhanced expressive power, making it well suited for pricing, hedging, and risk quantification in L\'evy-driven markets.
	
	The remainder of the paper is organized as follows. Section~\ref{section: pricing} introduces the general pricing framework and specifies the class of L\'evy models under consideration. Section~\ref{section:Levy} derives the second-order Esscher densities for L\'evy processes and analyses their structural properties. Section~\ref{section:jumpdiffusion} specializes the framework to jump-diffusion models and develops the corresponding second-order Esscher representation. 	
	Section~\ref{section: risk management} provides a detailed analysis of the constant jump-diffusion model under the second-order Esscher measure. In particular, we show that the Esscher parameter becomes functionally dependent on the jump structure, and we characterise the resulting pricing interval. This interval is subsequently used to assess hedging performance and residual errors within a risk management setting.
		Finally, Section~\ref{section: fear} investigates the economic interpretation of the second-order Esscher parameter. Using empirical evidence across different market regimes, we illustrate how this parameter captures latent market information and enhances the ability of the proposed models to reproduce observed option prices.
\section{Notation and the mathematical and financial model}\label{section: pricing}
Our mathematical model starts by a given filtered probability space ${\cal{B}}:=(\Omega, \mathcal{F}, (\mathcal{F}_t)_{t\geq 0}, \mathbb P)$ (called hereafter a stochastic basis), which is supposed to satisfy the usual conditions. This section has two subsections. The first subsection recalls some general notation that will be used throughout the paper, while the second subsection focuses on introducing the model and outlining its crucial ingredients and important properties.
 \subsection{General notations} 
 Throughout the paper, for any probability $Q$ on $(\Omega,{\cal{F}})$, we denote ${\cal A}(Q)$ (respectively ${\cal M}(Q)$) the set
of right-continuous with left limits (RCLL hereafter) and $\mathbb{F}$-adapted processes with $Q$-integrable variation (are uniformly $Q$-integrable martingales). When the probability measure is not mentioned, then by default we are using the probability measure $P$. The set ${\cal{V}}$ (respectively ${\cal{V}}^+$) denotes the set of RCLL and $\mathbb{F}$-adapted processes with finite variation (respectively nondecreasing).  
For any process $Y$, we denote by $^{o,\mathbb H}(Y)$  (respectively $^{p,\mathbb H}(Y)$)  the
$\mathbb{F}$-optional (respectively $\mathbb{F}$-predictable) projection of $Y$. For an increasing process $V$, we denote $V^{o,\mathbb{F}}$ (respectively $V^{p,\mathbb{F}}$) its dual $\mathbb{F}$-optional (respectively $\mathbb{F}$-predictable) projection. ${\cal O}$, ${\cal P}$ and  $\mbox{\bf{Prog}}$ represent the $\mathbb{F}$-optional, the $\mathbb{F}$-predictable and the $\mathbb{F}$-progressive $\sigma$-fields  respectively on $\Omega\times[0,+\infty[$. For a semimartingale $Y$, we denote by $L(Y)$ the set of all $Y$-integrable processes in the It\^o sense, and for $H\in L(Y)$, the resulting integral is a one-dimensional semimartingale denoted by $H\is Y:=\int_0^{\cdot} H_udY_u$. If $\mathcal{C}$ is a set of $\mathbb{F}$-adapted processes,
then $\mathcal{C}_{loc}$ --except when it is stated otherwise-- is the set of processes, $Y$,
for which there exists a sequence of stopping times,
$(T_n)_{n\geq 1}$, that increases to infinity and $Y^{T_n}$ belongs to $\mathcal{C}$, for each $n\geq 1$. 
\subsection{Mathematical model and preliminaries}
 On the stochastic basis ${\cal{B}}$, we consider given a one-dimensional  L\'evy process $X=(X_t)_{t\geq 0}$ which is a semimartingale. We refer the reader to \cite{ken1999levy, jacod2013limit,eberlein2019mathematical,schoutens2003levy}  and the references therein for the definition of L\'evy processes, their properties and their analysis. Throughout the paper, we consider the truncation function $h(x) = x 1_{\{|x|\leq 1\}}$. Then with the pair $(X,h)$, we associate $(b(h), \sigma^2, \nu_L)$, which are are called the predictable characteristics of $X$ and also known as the L\'evy-Khintchine triplet of $X$. $\sigma\geq0$ represents the diffusion coefficient, $b(h)\in\mathbb{R}$, and $\nu_L(dx)$ is called the L\'evy measure of $X$  and satisfies
\begin{equation}\label{Levy_measure}
    \int \min(1,x^2) \nu_L(dx) < + \infty \quad \text{and} \quad \nu_L(\{0\}) = 0.
\end{equation}
Furthermore, we associate with $X$ the random measure of its jumps denoted by $\mu_X$ given by
\begin{equation*}
    \mu_{X}(dt,dx)
    := \sum_{0 \le s} \delta_{(s,\Delta X_{s})}(dt,dx)\, \mathbf{1}_{\{\Delta X_{s} \neq 0\}},\quad\mbox{and}\quad     \nu(dt,dx) = \nu_L(dx)dt,
\end{equation*}
is its dual predictable projection measure (its compensator random measure). Thus, $X$ is fully characterized by its L\'evy-Khintchine triplet  via {\it L\'evy-Khintchine formula}
\begin{equation}\label{Khintchine_formula}
\begin{split}
   \Phi_X(t,u) &:= \mathbb E[e^{iuX_t}] = \exp \left( \varphi_X(u)t \right), \quad u\in \mathbb R,\\
    \varphi_X(u) &:= iu b(h) - \frac{\sigma^2 u^2}{2} + \int \left(e^{iu x} -1 - iu h(x) \right)\nu_L(dx).\end{split}
\end{equation}
Or equivalently the canonical representation of $X$ is given  by 
\begin{equation}\label{Canonical_representation4_X}
    X
    = b(h) t
    + \sigma{W}
    + h\star (\mu_X - \nu) 
    +(x-h) \star \mu_X.
\end{equation}
The last two stochastic integrals above are defined as in \cite{jacod2013limit}.\\

Our financial model consists of two assets, one bank account and one stock, having the price processes  $B$ and $S$ respectively given by 
 \begin{equation}\label{Stock_model}
    S_t = S_0 \exp(X_t),\quad B_t:=\exp(rt),\quad \mbox{where}\ r\in\mathbb{R},\quad\mbox{and}\ S_0>0.
\end{equation}

\begin{lemma}\label{Stock_properties} The following assertions hold.\\
{\rm{(a)}} The stock's price can be written as 
            \begin{equation}\label{S2Xtile}
            \begin{split}
    &S= S_0 \exp(X) = S_0 {\mathcal{E}}(\widetilde X),\\
    & \widetilde X:= \widetilde b(h)\ t + \sigma W + h \star (\mu_X - \nu)+(e^x-1-h)\star\mu_X,\quad  \widetilde b(h) : = b(h) + \frac{\sigma^2}{2}.\end{split}
\end{equation}
{\rm{(b)}} $X\in{\cal{V}}$ (i.e. it has a finite variation) if and only if $\vert{h}\vert\star \mu_X\in {\cal{V}}^+$. In this case, the L\'evy-Khintchine triplet of $X$ is the triplet $(b_0, \sigma^2, \nu)$ and 
        \begin{equation}\label{model: general Levy FV}
        \begin{split}
            &X = b_0 t +\sigma W+ x \star \mu_X,\quad b_0 := b(h) - \int h(x) \nu_L(dx)\\
   &\varphi_X(u)= iu b_0 - \frac{\sigma^2 u^2}{2} + \int_\mathbb R (e^{iu x} -1) \nu_L(dx).\end{split}
        \end{equation}
{\rm{(c)}} $X$ is integrable (i.e. $E[\vert X_t\vert]<+\infty$ for any $t\geq 0$) if and only if $\int_{\vert{x}\vert>1} \vert{x}\vert\nu_L(dx) <+\infty$. In this case, the  L\'evy-Khintchine triplet of $X$ is $(b_1, \sigma^2, \nu)$, and 
               \begin{equation*}\begin{split}
            &X = b_1 t + \sigma W +  x \star (\mu_X - \nu),\quad b_1 := b(h) + \int_{|x| >1} x \nu_L(dx),\\
            &\varphi_X (u) = iu b_1 - \frac{\sigma^2 u^2}{2} + \int_\mathbb R (e^{iu x} -1 - iu x) \nu_L(dx).\end{split}
        \end{equation*}
        
{\rm{(d)}}  If $X$ is integrable and has a finite variation, then 
\begin{equation*}
            X= b_1 t + \sigma W +  x \star \mu_X - x\star \nu,\quad b_1 := b(h) + \int_{|x| >1} x \nu_L(dx).
        \end{equation*}
\end{lemma}
The proof of the lemma is obvious and hence it will be omitted herein. Examples of L\'evy processes with finite variation are the constant jump-diffusion (CJD), the log-normal jump-diffusion (LJD) and the variance gamma models (VG) (see \cite{Honore1998Pitfalls, merton1976option, madanVG1990, madan1998variance} for these models). The normal inverse gaussian (NIG) model has infinite variation but it is an integrable process (see \cite{barndo1995normal, Barndorff1997NIGtype}).\\
We end this section by recalling the definition of the second-order Esscher densities/measures introduced in \cite{choulli2025second}. To this end, we recall the following two spaces of processes
\begin{equation}\label{Theta_Z(loc)}
\begin{split}
&\Theta(X):=\left\{(\theta,\psi)\ \Big|\quad \begin{array}{ll}
&\theta\in L(X),\\
&\psi\ \text{is predictable  and locally bounded}
\end{array}
\right\}.\\
&\mathcal Z_{loc}(S):=\left\{Z\ \mbox{positive process}\Big|\ Z\in{\cal{M}}_{loc}\quad \mbox{and}\quad (SZ)/B\in {\cal{M}}_{loc}\right\}.
\end{split}
\end{equation}
\begin{definition}
    Let $Z$ be a positive, RCLL, and adapted process.\\
    {\rm{(a)}} ${Z}$ is called a second-order exponential Esscher density for $(S,\mathbb F)$ if there exist $(\theta, \psi) \in \Theta(X)$ and a predictable process $K\in{\cal{V}}$ such that
        \begin{equation*}
            Z = Z^{\psi}:= \exp \bigg(\theta \bigcdot X + \psi{x}^2\star \mu_X - K \bigg) \in \mathcal Z_{loc}(S).
        \end{equation*}
     $\psi$ will be called the Esscher parameter of the density $Z^{\psi}$. If $Z^{\psi}$ is uniformly integrable, then $\mathbb{Q}^{\psi}:= Z_\infty^{\psi} \bigcdot \mathbb P$ is called a second-order exponential Esscher pricing measure for $S$. The set of these pricing densities (respectively measures) is denoted by $\mathcal Z^{EE}(S)$ (respectively $\mathbb Q^{EE}(S)$).\\
{\rm{(b)}} $Z$ is called a second-order linear Esscher density for $S$ if there exist $(\theta,\psi) \in \Theta(S)$ and a predictable process $\widetilde K\in{\cal{V}}$ such that
        \begin{equation*}
            Z = Z^{\psi}:= \exp \bigg(\theta \bigcdot S + \sum_{0<s \leq \cdot} \psi_s (\Delta  S_s)^2 - \widetilde K \bigg) \in \mathcal Z_{loc}(S).
        \end{equation*}
        The set of second-order linear Esscher pricing densities (respectively measures) for $S$ is denoted by $\mathcal{Z}^{LE}(S)$ (respectively $\mathbb Q^{LE}(S)$).
\end{definition}

\begin{remark} It is clear from the definition above that, the exponential Esscher measure is defined using the process $X$ in (\ref{Canonical_representation4_X}), whereas the linear Esscher measure corresponds to the process $S$ itself or equivalently $\widetilde X\ (:=S_{-}^{-1}\is{S})$ specified in (\ref{S2Xtile}). Throughout the rest of the paper, we will mainly focus on the \emph{exponential Esscher} due to its relevance in the empirical experiments on the one hand. On the other hand, in the case of one-dimensional setting as is the case of the current paper, the linear and exponential Esscher are intimately related to each via an equivalent local change of probability. For more details about this latter fact, we refer the reader to \cite{choulli2025second}. \\
\end{remark}
\section{Second-order Esscher densities for L\'evy models}\label{section:Levy} 
In this section, we derive the second-order Esscher densities/measures and their properties for $X$. To this end, for any pair of processes $(\theta,\psi)$, we put 
\begin{equation}\label{I(theta,psi)}
 I_s(\theta,\psi):=\int \big| (e^x -1)e^{\theta_s x +\psi_s x^2} - h(x) \big|\nu_L(dx) + \int\big|e^{\theta_s x+\psi_s x^2} - 1 -h(\theta_s x) \big| \nu_L(dx),\ s\geq 0.
 \end{equation}

\begin{theorem}\label{General_Levy_Existence}
    Let $Z$ be a positive process. Then the following assertions are equivalent.\\
    {\rm{(a)}} $Z$ is an exponential Esscher density of order-two for $S$ (i.e. $Z \in \mathcal Z^{EE} (S)$),\\
     {\rm{(b)}}  There exists a pair $(\theta, \psi) \in \Theta (X)$ such that
        \begin{eqnarray}   
           && \int_0^{\cdot} I_s(\theta,\psi)ds\in{\cal{V}}^+,\label{Integrability_Condition1}\\
           && b(h) + \frac{\sigma^2}{2} + \theta \sigma^2 + \int \bigg[(e^x - 1) \exp(\theta x +\psi x^2) - h(x) \bigg] \nu_L(dx) = r \quad \mathbb P \otimes dt \text{-a.e.}      \label{Martingale_Equation}\\
           &&Z = \mathcal E(N) \quad \text{with} \quad N = \theta \sigma \bigcdot W + \left(\exp(\theta x + \psi x^2) -1\right) \star (\mu_X - \nu).     \label{Esscher_Density_Form}
        \end{eqnarray}
         {\rm{(c)}}  There exists a pair $(\theta, \psi) \in \Theta (X)$ satisfying (\ref{Integrability_Condition1}), (\ref{Martingale_Equation}) and
        \begin{equation}\label{eq: Esscher definition for Levy and psi a process}
            Z= \exp \bigg(\theta \bigcdot X + \sum_{0<s\leq\cdot} \psi_s (\Delta X_s)^2 - \int_0^{\cdot}\kappa_s(\theta, \psi)ds\bigg)
        \end{equation}
        where $\kappa(\theta, \psi)$ is a process given by 
        \begin{equation}\label{kappa(theta,psi)}
            \kappa_s(\theta, \psi):= \theta_s{ b}(h) + \frac{\theta^2_s\sigma^2}{2} + \int \bigg(\exp(\theta_s{x} +\psi_s{x}^2) - 1 - \theta_s{h}(x) \bigg) \nu_L(dx).
        \end{equation}
    Furthermore, for any predictable and locally bounded $\psi$, (\ref{Martingale_Equation}) has at most one solution.
\end{theorem}
\begin{proof} The proof has three parts. The first and the second parts prove some general claims that are useful in themselves, while the third part  proves the theorem.\\
    {\bf Part 1.} This part proves that, for any triplet of predictable processes $(\theta,\psi,K)$ which belongs to $\Theta(X)\times {\cal{V}}$, we have 
    \begin{equation}\label{Conditions}
    \begin{split}
    Z:=\exp(\theta\is{X}+\sum\psi(\Delta{X})^2-K)\in{\cal{M}}_{loc}\ \mbox{iff}\ \begin{cases}\displaystyle\int_0^{\cdot} \int\big|e^{\theta_s x+\psi_s x^2} - 1 -h(\theta_s{x}) \big| \nu_L(dx)ds\in{\cal{V}}^+,\\
    \\
    K=\int_0^{\cdot}\kappa_s(\theta,\psi)ds,\quad\mbox{and (\ref{Esscher_Density_Form}) holds},\end{cases}
    \end{split}
    \end{equation}
 where $\kappa(\theta,\psi)$ is the process defined in (\ref{kappa(theta,psi)}). To this end, we apply It\^o's formula to the process $Z:=\exp(\theta\is{X}+\sum\psi(\Delta{X})^2-K)$ and derive
    \begin{equation}\label{Ito4Z}
        \begin{split}
           (Z_- )^{-1} \is{Z} &= \theta\is{X}-K+\frac{1}{2}\theta^2 \sigma^2\is{t}+\sum\left(\exp(\theta\Delta{X}+\psi(\Delta{X})^2-\Delta{K})-1-\theta\Delta{X}+\Delta{K}\right),\\
           &=\theta\is{X}+ (e^{\theta x + \psi x^2} -1- \theta x)\star\mu_X+\frac{1}{2}\theta^2 \sigma^2\is{t} - K + \sum (e^{-\Delta K} -1 + \Delta K).
        \end{split}
    \end{equation}
The last equality is due to the fact that $\Delta X\Delta K =0$, which leads to $\exp(\theta\Delta{X}+\psi(\Delta{X})^2-\Delta{K})-1-\theta\Delta{X}+\Delta{K}=\left(\exp(\theta\Delta{X}+\psi(\Delta{X})^2)-1-\theta\Delta{X}\right)+\left(\exp(-\Delta{K})-1+\Delta{K}\right).$ Thanks to \cite[Lemma 3.19-(c)]{choulli2025second}, we have 
\begin{equation*}
\begin{split}
\theta\is{X}&=\theta\sigma\is{W}+h(x\theta)\star(\mu_X-\nu)+\theta{x}I_{\{\vert\theta{x}\vert>1\}}\star\mu_X+ B^{\theta},\\
B^{\theta}&:= \int_0^{\cdot}\left\{\theta_s{b}(h)+\int (-I_{\{\vert \theta_s{x}\vert>1\}} \theta_s{h}(x)+I_{\{\vert{x}\vert>1\}}h(x\theta_s))\nu_L(dx)\right\}ds.\end{split}\end{equation*}
 Thus, by inserting this in (\ref{Ito4Z}) and arranging terms together afterwards, we obtain
\begin{equation*}
        \begin{split}
       \frac{1}{Z_-} \is{Z} &=\mbox{local martingale}+  (e^{\theta x + \psi x^2} -1- h(\theta x))\star\mu_X +B^{\theta}+\frac{1}{2}\theta^2 \sigma^2\is{t}- K + \sum (e^{-\Delta K} -1 + \Delta K).
            \end{split}
    \end{equation*}
    Therefore, $Z\in{\cal{M}}_{loc}$ if and only if  $(Z_- )^{-1} \is{Z} \in{\cal{M}}_{loc}$ if and only if 
    \begin{eqnarray}
    &&(\exp(\theta x + \psi x^2) -1- h(\theta x))\star\mu_X \in{\cal{A}}_{loc},\label{Condition1}\\
    &&K-\sum (e^{-\Delta K} -1 + \Delta K)=B^{\theta}+\frac{1}{2}\theta^2 \sigma^2\is{t}+(e^{\theta x + \psi x^2} -1- h(\theta x))\star\nu,\label{Condition2}\\
    &&(Z_- )^{-1} \is{Z} =\theta\sigma\is{W}+( e^{\theta x + \psi x^2} -1)\star(\mu_X-\nu).\label{Condition3}
    \end{eqnarray}
   Hence, it is clear that (\ref{Condition3}) is equivalent to (\ref{Esscher_Density_Form}), (\ref{Condition1}) is equivalent to the first condition in the right-hand-side of (\ref{Conditions}), and direct calculations show that (\ref{Condition2}) holds if and only if $K$ is continuous and the second condition in the right-hand-side of (\ref{Conditions}) holds. This proves (\ref{Conditions}) and completes the first part.\\
  {\bf Part 2.} Herein, we prove that for any pair $(\theta,\psi)\in\Theta(X)$ such that the process $N:= \theta \sigma \is W + \left(\exp(\theta x + \psi x^2) -1\right) \star (\mu_X - \nu)$ is a well defined local martingale, we have 
  \begin{equation}\label{Claim4Z}
  \begin{split}
  Z:={\cal{E}}(N)\in{\cal{Z}}_{loc}(S)\ \mbox{iff}\  \begin{cases}\int_0^{\cdot} \int \big| (e^x -1)\exp(\theta_s x +\psi_s x^2) - h(x) \big|\nu_L(dx)ds\in{\cal{V}}^+\\
  \\
  \mbox{ and equation (\ref{Martingale_Equation}) holds}.\end{cases}
  \end{split}
    \end{equation}
    Thanks to It\^o's formula and  (\ref{S2Xtile}), we deduce that  $Z S/B \in \mathcal M_{loc}$ if and only if $\widetilde X + [N, \widetilde X]-rt $ is a local martingale, or equivalently 
    $$-rt+\sigma {W}+h(x)\star(\mu_X-\nu)+\widetilde{b}(h) t+\theta\sigma^2\is t+\left((e^x-1)\exp(\theta{x}+\psi{x}^2)-h(x)\right)\star\mu_X  \in \mathcal M_{loc}.$$ 
    This is equivalent to 
\begin{equation}\label{Equivalences}
  \begin{split}
 & \left((e^x-1)\exp(\theta{x}+\psi{x}^2)-h(x)\right)\star\mu_X\in{\cal{A}}_{loc},\\
 \mbox{and}\ &r={b}(h)+{\frac{\sigma^2}{2}}+\theta\sigma^2+\int \left((e^x-1)\exp(\theta{x}+\psi{x}^2)-h(x)\right)\nu_L(dx).
\end{split}
    \end{equation}
Then, it is clear that the first condition in the above is equivalent to the first condition in the right-hand side of (\ref{Claim4Z}). This ends the proof of (\ref{Claim4Z}) and the second part is complete.\\
{\bf Part 3.} This part proves the theorem. In fact, both the equivalence (b) $\Longleftrightarrow$ (c) and the implication (a) $\Longrightarrow$ (b) follow immediately from combining part 1 and part 2 above, while the implication (b) $\Longrightarrow$ (a) is due to direct calculations as in  part 2. Thus the rest of this part will prove the last statement of the theorem. To this end, we suppose that, for a fixed $\psi$, there exists two solutions $\theta$ and $\theta'$ to equation (\ref{Martingale_Equation}), and this yields     \begin{equation*}
        (\theta - \theta')\sigma^2 + \int (e^x-1)e^{\theta' x +\psi x^2} (e^{(\theta - \theta')x} -1) \nu_L(dx) = 0,\quad P\otimes dt\mbox{-a.e.}.
    \end{equation*}
As, $(e^u-1)/u > 0$ and $u(e^u-1)> 0$ for any $u\not=0$, then the above equation holds iff $\theta = \theta'$ $P\otimes dt$-a.e., and the uniqueness of the solution to (\ref{Martingale_Equation}) follows immediately. This ends the proof of theorem.
    \end{proof}


Throughout the rest of the paper, for any predictable locally bounded process $\psi$, we put
\begin{equation}\label{ZEE-QEE}
\begin{split}
\mathcal Z^{EE}(S,\psi)&:=\left\{Z\in \mathcal Z^{EE}(S)\ \big|\ \  \psi\ \mbox{is the Esscher parameter of}\ Z\right\},\\
\mathcal Q^{EE}(S,\psi)&:= \left\{Z^\psi_T \bigcdot \mathbb P\ \big|\quad Z^\psi \in \mathcal Z^{EE}(S,\psi) \cap \mathcal M \right\}.
\end{split}
\end{equation}
Our next theorem investigates how the L\'evy property is affected by the second-order Esscher density with Esscher-parameter $\psi$. In fact, when this parameter is a constant (does not depend on $\omega\in\Omega$ nor on time $t$), we prove that the L\'evy property is preserved under the second-order Esscher measure and the characterization of the density of this latter measure becomes simpler.
\begin{theorem}\label{Levy_Psi_Constant_Existence}
    Let $\psi \in \mathbb R$ and $Z$ be a positive process. Then the following assertions hold.\\
    {\rm{(a)}} $Z\in \mathcal Z^{EE}(S,\psi)$ if and only if there exists $\theta \in \mathbb R$ satisfying
    \begin{eqnarray}
       && \int_{-\infty}^{-1} \exp( \theta{x}+\psi{x}^2 )\nu_L(dx) + \int_{1}^{\infty} \exp\left( (1+\theta){x}+\psi{x}^2 \right)\nu_L(dx) <+\infty, \label{Integrability_Condition_Psi_Constant}\\
       && b(h) + \frac{\sigma^2}{2} + \theta \sigma^2 + \int \bigg[(e^x-1)\exp(\theta x + \psi x^2) - h(x) \bigg]\nu_L(dx) = r, \label{Martingale_Equation_Psi_Constant}
\\&&\mbox{and}\quad Z = {\mathcal{E}} \bigg(\theta \sigma W + \left(\exp(\theta x+ \psi x^2) -1\right)\star (\mu_X - \nu) \bigg).  \label{eq: Esscher density general levy psi constant}
    \end{eqnarray}
   {\rm{(b)}} Suppose that $\mathcal Q^{EE}(S,\psi) \neq \emptyset$, and let $\mathbb Q^\psi \in \mathcal Q^{EE}(S,\psi)$. Then $X$ remains a L\'evy process under $\mathbb Q^\psi$ with the L\'evy-Khintchine triplet $(b^\psi(h), \sigma^\psi, \nu_L^\psi)$ given by
    \begin{equation}\label{Levy_Characteristics_Psi}
        \begin{split}
            b^\psi(h) &:= b(h) + \theta \sigma^2 + \int\left(\exp(\theta x +\psi x^2) -1\right) h(x) \nu_L(dx),\\
            \sigma^\psi &:= \sigma,\quad\quad \nu_L^\psi(dx) := e^{\theta x +\psi x^2} \nu_L(dx).
        \end{split}
    \end{equation}
\end{theorem}
\begin{proof} This proof has two parts, where we prove assertions (a) and (b) respectively.\\
{\bf Part 1.} Herein we prove assertion (a). To this end, in virtue of Theorem \ref{General_Levy_Existence}, we remark that the proof of this assertion follows immediately as soon as we prove that in the current case of $\psi\in\mathbb{R}$ (i.e. $\psi$ being a constant in $(\omega,t)\in\Omega\times[0,T]$) the root of (\ref{Martingale_Equation_Psi_Constant}) when it exists is also a constant  and the conditions (\ref{Integrability_Condition1}) and (\ref{Integrability_Condition_Psi_Constant}) are equivalent. \\
As both $\theta$ and $\psi$ are constant, then it is clear that (\ref{Integrability_Condition1}) holds if and only if 
 \begin{equation}
        \int \bigg| (e^x -1)e^{\theta x +\psi x^2} - h(x) \bigg|\nu_L(dx)  + \int\bigg|e^{\theta x+\psi x^2} - 1 - h(\theta x) \bigg|\nu_L(dx) < +\infty 
    \end{equation}
On the one hand, thanks to the first property in (\ref{Levy_measure}) and the two facts that $(e^x -1)e^{\theta x +\psi x^2} - h(x)=((e^x -1)e^{\theta x +\psi x^2} -x)I_{\{\vert{x}\vert\leq 1\}}+(e^x -1)e^{\theta x +\psi x^2}I_{\{\vert{x}\vert>1\}}$ and the fact that there exists a positive constant $\alpha(\theta,\psi)$ such that 
$\vert((e^x -1)e^{\theta x +\psi x^2} -x)I_{\{\vert{x}\vert\leq 1\}}\vert\leq \alpha(\theta,\psi)x^2 I_{\{\vert{x}\vert\leq 1\}}$, we deduce that 
 \begin{equation}\label{Integrability_Equivalence1}
        \int \bigg| (e^x -1)e^{\theta x +\psi x^2} - h(x) \bigg|\nu_L(dx)< +\infty \ \mbox{iff}\        \int_{\vert{x}\vert>1}\vert{e}^x -1\vert\exp(\theta x +\psi x^2)\nu_L(dx) < +\infty.    \end{equation}
On the other hand, it is easy to check that $ \int_{-\infty}^{-1} \vert{e}^x -1\vert\exp(\theta x +\psi x^2)\nu_L(dx) < +\infty$ if and only if $ \int_{-\infty}^{-1}\exp(\theta x +\psi x^2)\nu_L(dx) < +\infty$ and $ \int_{1}^{\infty} \vert{e}^x -1\vert\exp(\theta x +\psi x^2)\nu_L(dx) < +\infty$ if and only if $ \int_{1}^{\infty}\exp((\theta+1) x +\psi x^2)\nu_L(dx) < +\infty$. Then, by combining these latter remarks and (\ref{Integrability_Equivalence1}), direct calculations afterwards show that 
\begin{equation}\label{Integrability_Equivalence2}
 \int \bigg| (e^x -1)e^{\theta x +\psi x^2} - h(x) \bigg|\nu_L(dx)  <\infty\  \mbox{if and only if (\ref{Integrability_Condition_Psi_Constant}) holds}.\end{equation}
Similarly, we will prove that 
\begin{equation}\label{Integrability_Equivalence3}
        \int \bigg| \exp(\theta x +\psi x^2)-1-h(\theta{x}) \bigg|\nu_L(dx)< +\infty \ \mbox{iff}\        \int_{\vert{x}\vert>1}\exp(\theta x +\psi x^2)\nu_L(dx) < +\infty.    \end{equation}
For the case of $\theta\not=0$, we have $e^{\theta x+\psi x^2} - 1 -h(\theta {x})=(e^{\theta x+\psi x^2} - 1 -\theta{x})I_{\{\vert{x}\vert\leq 1/\vert\theta\vert\}}+ (e^{\theta x+\psi x^2} - 1)I_{\{\vert{x}\vert>1/\vert\theta\vert\}}$, and there exists a positive constant $\beta(\theta, \psi)$ such that $\vert{e}^{\theta x+\psi x^2} - 1 -\theta{x}\vert{I}_{\{\vert{x}\vert\leq 1/\vert\theta\vert\}}\leq \beta(\theta,\psi)x^2I_{\{\vert{x}\vert\leq 1/\vert\theta\vert\}}$. Then, by combining these two remarks and the first property in (\ref{Levy_measure}) ,  we deduce that  (\ref{Integrability_Equivalence3}) holds when $\theta\not=0$. For the case when $\theta=0$, we get 
 \begin{equation*}
        \int \bigg| \exp(\theta x +\psi x^2)-1-h(\theta{x}) \bigg|\nu_L(dx)=  \int_{\vert{x}\vert>1} \vert \exp(\psi x^2)-1\vert\nu_L(dx)+\int_{\vert{x}\vert\leq1} \vert \exp(\psi x^2)-1\vert\nu_L(dx).    \end{equation*}
Thus, there exists a positive constant $C_\psi$ such that $\int_{\vert{x}\vert\leq1} \vert \exp(\psi x^2)-1\vert\nu_L(dx)\leq C_{\psi} \int x^2{I}_{\{\vert{x}\vert\leq1\}}\nu_L(dx)<+\infty$ due to (\ref{Levy_measure}). Combining this with $ \int_{\vert{x}\vert>1} \vert \exp(\psi x^2)-1\vert\nu_L(dx)<\infty$ iff $\int_{\vert{x}\vert>1} \exp(\psi x^2)\nu_L(dx)<\infty$ as $ \nu_L(\vert{x}\vert>1)<\infty$ due to (\ref{Levy_measure}), proves that (\ref{Integrability_Equivalence3}) holds for the case of $\theta=0$ also.\\
 Therefore, it is clear that the right-hand-side condition in (\ref{Integrability_Equivalence3}) is implied by  (\ref{Integrability_Condition_Psi_Constant}). Therefore, by combining this with (\ref{Integrability_Equivalence3}) and (\ref{Integrability_Equivalence1}), the proof of assertion (a) is complete.\\
{\bf Part 2.} Herein we prove assertion (b). To this end, suppose that there exists $\mathbb{Q}^{\psi}\in\mathcal Q^{EE}(S,\psi)$. Then, thanks to \cite[Chapter II, Theorem 2.42 and Corollary 4.19]{jacod2013limit}, it is enough to prove that, for any $u\in \mathbb{R}$, the process  
\begin {equation}\label{ToProve}
\begin{split}
\exp(iuX)- \varphi^{\psi}_X(u)\int_0^{\cdot} e^{iuX_{s-}}ds\quad\mbox{is a complex-valued local maringale under}\ \mathbb{Q}^{\psi},\\
\mbox{where}\quad \varphi^{\psi}_X(u):= iu b^\psi(h)- \frac{u^2}{2}\sigma^2 + \int \bigg(e^{iu x} -1 - iu h(x) \bigg)\nu^\psi_L(dx).
\end{split}
\end{equation}
To prove this, we first remark that due to Girsanov's theorem we have 
\begin{equation*}\label{definition: W^psi and (mu - nu^psi) martingales}
    W^\psi_t := W_t - \theta \sigma{t}\in{\cal{M}}_{loc}(\mathbb{Q}^{\psi}) \quad \text{and} \quad M^{\psi}:= h(x)\star (\mu_X - \nu) - h(x)\big(e^{\theta x + \psi x^2} -1 \big) \star \nu\in{\cal{M}}_{loc}(\mathbb{Q}^{\psi}),
\end{equation*} 
and direct calculations show that the dual predictable projection of the random measure $\mu_X$ under $\mathbb{Q}^{\psi}$ is $\nu^{\psi}(dx,dt):=\exp(\theta{x}+\psi{x}^2)\nu_L(dx)dt$.
Thus, by combining these with It\^o's formula applied to $e^{iuX}$ and (\ref{Canonical_representation4_X}), we obtain
    \begin{equation*}
        \begin{split}
            &e^{-iuX_{t-}}d\big(e^{iuX_t} \big)\\
             =&iudX_t-{\frac{u^2\sigma^2}{2}}dt+d\sum_{0<s\leq t}(e^{iu\Delta{X}_s}-1-iu\Delta{X}_s)\\
            =& \big(iu b(h) - \frac{u^2}{2} \sigma^2 \big) dt + iu{\sigma}dW_t +iud(h(x)\star(\mu_X-\nu))_t+d((e^{iux}-1-iuh(x))\star\mu_X)_t\\
        =& \left\{iu b(h) +iu \theta \sigma^2 - \frac{u^2}{2} \sigma^2 +iu\int h(x)\big(e^{\theta x + \psi x^2} -1 \big)\nu_L(dx) \right\} dt + iu{\sigma}dW_t^\psi + iudM_t^\psi \\
     &+ \bigg(e^{iu x} -1- iu h(x) \bigg)e^{\theta x + \psi x^2} \nu_L(dx) dt + d\left(( e^{iux} - 1-iuh(x))\star(\mu_X - \nu^\psi)\right)_t\\
     =&d\left(\mathbb{Q}^{\psi}\mbox{-local martingale}\right)+\left\{iu b^{\psi}(h) - \frac{u^2}{2} \sigma^2 +\int (e^{iux}-1-iuh(x))e^{\theta{x}+\psi{x}^2}\nu_L(dx)\right\} dt
        \end{split}
    \end{equation*}
    This proves (\ref{ToProve}), and the proof of the theorem is complete. 
\end{proof}

Under some integrability assumption on the L\'evy measure, the next theorem completely outlines the L\'evy models which admit second-order Esscher density with constant parameter $\psi$.
\begin{theorem}\label{Theorem4ExistenceZ(psi)}
    Let $\psi \in \mathbb R$  such that 
    \begin{equation}\label{Integrability_Condition2}
        \int_{\vert{x}\vert>1}\exp\left(\theta{x}+ \psi x^2\right) \nu_L(dx)  < +\infty,\quad \mbox{for any}\ \theta \in \mathbb{R}.
    \end{equation}
      Then the following assertions hold.\\
    {\rm{(a)}} If the condition
    \begin{equation}\label{Sufficient_Condition1}
        \sigma+\min \bigg(\nu_L(0,\infty), \nu_L(-\infty, 0) \bigg) >0
    \end{equation}
    holds, then there exists unique $\theta(\psi)=F_{\psi}^{-1}(0) \in \mathbb R$ such that $\emptyset\not={\cal{Z}}^{EE}(S,\psi)\subset{\cal{M}}^p$, $p\in(1,\infty)$.\\
    {\rm{(b)}} If the condition
    \begin{equation}\label{Sufficient_Condition2}
    0= \sigma+\min \bigg(\nu_L(0,\infty), \nu_L(-\infty, 0) \bigg)<\max\bigg(\nu_L(0,\infty), \nu_L(-\infty, 0) \bigg)
    \end{equation}
     holds, then ${\cal{Z}}^{EE}(S,\psi)\not=\emptyset$ if and only if
    \begin{equation}\label{NS_Conditions}
        \left(b(h) -r- \int_{-1}^1 x \nu_L(dx)\right)\int_{-1}^1 x \nu_L(dx)<0.
    \end{equation}
    Furthermore, in this case we have ${\cal{Z}}^{EE}(S,\psi)\subset{\cal{M}}^p$, for any $p\in(1,\infty)$.
\end{theorem}
\begin{proof} Let $\psi\in\mathbb{R}$ such that (\ref{Integrability_Condition2}) holds. The proof of the theorem is achieved in two parts.\\
{\bf Part 1.} Here we prove that when there exists $Z^{\psi}\in{\cal{Z}}^{EE}(S,\psi)$, then necessarily $Z^{\psi}$ is a martingale that is $p$-integrable for any $p\in(1,\infty)$. To this end, thanks to Theorem \ref{Levy_Psi_Constant_Existence} we conclude that there exists $\theta\in\mathbb{R}$ such that 
 \begin{equation*}
 Z^{\psi}_t=
\exp\!\bigg(\theta X_t+\psi \sum_{0<s\le t} (\Delta X_s)^2-\kappa(\theta,\psi)\,t\bigg),\qquad t\in[0,T],
\end{equation*}
Thus, for $p\in(1,\infty)$, we get
 \begin{equation}\label{Z(p)}
 \begin{split}
( Z^{\psi}_t)^p&=\exp\!\bigg(p\theta X_t+p\psi \sum_{0<s\le t} (\Delta X_s)^2-p\kappa(\theta,\psi)\,t\bigg)\\
&=\exp\!\bigg(p\theta X_t+p\psi \sum_{0<s\le t} (\Delta X_s)^2-\kappa(p\theta,p\psi)\,t\bigg)\,\exp\bigl(\kappa(p\theta,p\psi)t - p\kappa(\theta,\psi)t\bigr).\end{split}
\end{equation}
By definition of $\kappa(\theta,\psi)$ and part 1 in the proof of Theorem \ref{General_Levy_Existence} (see (\ref{Conditions}) more precisely), we have
 \begin{equation*}
\mathbb E\Big[\exp\!\big(p\theta X_t+p\psi \sum_{0<s\le t} (\Delta X_s)^2-\kappa(p\theta,p\psi)\,t\big)\Big]= 1.
\end{equation*}
Then by combining this with (\ref{Z(p)}), we obtain 
\begin{equation*}
\mathbb E[( Z^{\psi}_t)^p]=
\exp\!\big( t\,[\kappa(p\theta,p\psi) - p\kappa(\theta,\psi)]\big)< \infty,\qquad t\in[0,T].\end{equation*}
In particular,
$$
\mathbb E[Z_T^p]
=
\exp\!\big( T\,[\kappa(p\theta,p\psi) - p\kappa(\theta,\psi)]\big)
< \infty.
$$
Since $Z$ is a positive local martingale and hence is a supermartingale, thanks to Doob's maximal inequality we derive
\begin{equation*}
\mathbb E\Big[\sup_{0\le t\le T} \vert{Z}^{\psi}_t\vert^p\Big]\le\Big(\frac{p}{p-1}\Big)^p \mathbb E[|Z_T|^p]=\Big(\frac{p}{p-1}\Big)^p\exp\!\big( T\,[\kappa(p\theta,p\psi) - p\kappa(\theta,\psi)]\big)< \infty.
\end{equation*}
This proves that  $Z\in\mathcal M^p$ for every $p\in(1,\infty)$, and the inclusion
$\mathcal Z^{EE}(S,\psi)\subseteq\mathcal M^p$ follows. This ends the first part.\\
{\bf Part 2.} This part proves assertion (a). To this end, we suppose (\ref{Sufficient_Condition1}) holds and consider 
    \begin{equation*}
        F_{\psi}(\theta) := b(h)-r +\frac{\sigma^2}{2} + \theta \sigma^2 + \int\bigg[(e^x- 1)\exp(\theta x +\psi x^2) - h(x) \bigg] \nu_L(dx),\quad \theta\in\mathbb{R}.
    \end{equation*}
Then, thanks to the assumption (\ref{Integrability_Condition2}),  this function is well defined on $\mathbb{R}$ and is continuously differentiable with $F_{\psi}'(\theta)=\sigma^2+ \int{x}(e^x- 1)\exp(\theta x +\psi x^2) \nu_L(dx)>0$ for any $\theta\in\mathbb{R}$. To ease notation, we introduce the following auxiliary functions
    \begin{equation*}
        \begin{split}
            F_\psi^{(1)}(\theta) &:= \int f_\theta (x) \nu_L(dx),\quad  f_\theta (x) := \bigg[(e^x-1)e^{\theta x +\psi x^2} - x\bigg] 1_{\{|x| \leq 1\}} \\
            F_\psi^{(2)}(\theta) &:= \int g_\theta (x) \nu_L(dx),\quad g_\theta (x) :=(e^x-1)e^{\theta x +\psi x^2} 1_{\{|x| > 1\}}\\
            F^{(3)}(\theta) &:= b(h) -r+\frac{\sigma^2}{2} + \theta \sigma^2,
        \end{split}
    \end{equation*}
   and we have $F_{\psi}(\theta) = F^{(1)}_\psi(\theta) + F^{(2)}_\psi(\theta) +F^{(3)}(\theta)$.\\
   Notice that for any $x \in \mathbb R$, the maps $\theta \mapsto f_\theta(x)$ and $\theta \mapsto g_\theta(x)$ are increasing and 
   \begin{equation*}
       \int\vert{f}_\theta (x)\vert\nu_L(dx) +       \int\vert{g}_\theta (x)\vert\nu_L(dx) < +\infty,\quad \mbox{for any}\ \theta\in\mathbb{R}.    \end{equation*}
Thus, on the one hand, for any $\theta>0$, we have $f_{\theta}(x)\geq f_0(x)$ and $g_{\theta}(x)\geq g_0(x)$ and 
 \begin{equation*}
 \begin{split}
       &\int\vert{f}_0 (x)\vert\nu_L(dx) +       \int\vert{g}_0 (x)\vert\nu_L(dx) < +\infty,\\
       &\lim_{\theta \uparrow \infty} f_{\theta}(x)=\begin{cases}\infty\ \ \mbox{for}\ x\in(0,1)\cr\\
       -x\ \ \mbox{for}\ x\in(-1,0)\end{cases},\quad \lim_{\theta \uparrow \infty} g_{\theta}(x)=\begin{cases}\infty\ \ \mbox{for}\ x\in(1,\infty)\cr\\
       0\ \ \mbox{for}\ x\in(-\infty,-1)\end{cases}
     \end{split} \end{equation*}
Then, by combining these with the monotone convergence theorem, we get 
 \begin{equation*}
 \begin{split}
\lim_{\theta \uparrow \infty}F_\psi^{(1)}(\theta)= \begin{cases}
            +\infty &\text{if} \ \nu_L(0,1) >0\\
            -\int_{-1}^0 x \nu_L(dx) &\text{if} \ \nu_L(0,1) = 0
        \end{cases},\quad \lim_{\theta \uparrow \infty} F^{(2)}_\psi(\theta)=\begin{cases}
            +\infty &\text{if} \ \nu_L(1,\infty) >0\\
            0 &\text{if} \ \nu_L(1,\infty) = 0.
        \end{cases} \end{split} \end{equation*}
Therefore, by combining these with the facts that $\int_{-1}^0-x\nu_L(dx)\in [0,\infty]$ and $\displaystyle\lim_{\theta \uparrow \infty}F^{(3)}(\theta)=\infty$ if $\sigma>0$ and $ F^{(3)}(\theta)= b(h) -r$ if $\sigma=0$, we deduce that
 \begin{equation}\label{Limit_F(psi)}
\lim_{\theta \uparrow \infty}F_\psi(\theta)=\begin{cases}
        +\infty \quad &\text{if} \quad \sigma >0\ \mbox{or}\  \nu(0,\infty)>0\\
        b(h)-r-\int_{-1}^0x\nu_L(dx)&\text{otherwise}.
    \end{cases}
\end{equation}
On the other hand, for $\theta<0$, we have $f_{\theta}(x)\leq f_0(x)$ and $g_{\theta}(x)\leq g_0(x)$ and 
 \begin{equation*}
 \begin{split}
      \lim_{\theta \downarrow -\infty} f_{\theta}(x)=\begin{cases}\infty\ \ \mbox{for}\ x\in(-1,0)\cr\\
       -x\ \ \mbox{for}\ x\in(0,1)\end{cases},\quad \lim_{\theta \downarrow -\infty} g_{\theta}(x)=\begin{cases}\infty\ \ \mbox{for}\ x\in (-\infty,-1)\cr\\
       0\ \ \mbox{for}\ x\in(1,\infty).\end{cases}
     \end{split} \end{equation*}
Then, similar arguments and analyses yield 
 \begin{equation}\label{Limit_F(psiNegative)}
\lim_{\theta \downarrow -\infty}F_\psi(\theta)=\begin{cases}
        -\infty \quad &\text{if} \quad \sigma +\nu(-\infty,0)>0\\
        b(h)-r-\int_{0}^1 x\nu_L(dx)&\text{otherwise}.
    \end{cases}
\end{equation}
Then both assertions (a) and (b) follow immediately from combining (\ref{Limit_F(psiNegative)}), (\ref{Limit_F(psi)}) and the properties of the function $F^{\psi}$ discussed above. This ends the proof of the theorem.\end{proof}
The condition (\ref{Integrability_Condition2}) in the previous theorem is not very restrictive, and indeed it is fulfilled for L\'evy models when the Esscher parameter is negative (i.e. for $\psi<0$). Our next result shows that for a negative Esscher-parameter $\psi$, the second-order density for any L\'evy model --when it exists-- is a $p$-integrable martingale for any $p\in(1,\infty)$. 
\begin{theorem}\label{Theorem4PsiNegative}
    If $\psi \in (-\infty, 0)$, then $\mathcal Z^{EE}(S, \psi)\not=\emptyset$ if and only if either (\ref{Sufficient_Condition1}) holds or both (\ref{Sufficient_Condition2}) and (\ref{NS_Conditions}) hold. Furthermore, in both cases,  $\mathcal Z^{EE}(S, \psi)$ has a unique element $Z^\psi \in \mathcal M^p$ (i.e. it is a  $p$-integrable martingale) for any $p\in (1,\infty)$.
\end{theorem}

\begin{proof} Thanks to Theorem \ref{Theorem4ExistenceZ(psi)}, the proof of this theorem follows immediately as soon as we prove that when $\psi \in (-\infty,0)$ the assumption (\ref{Integrability_Condition2}) holds. This is the aim of the remaining part of the proof. To this end, we remark that for any $\theta\in\mathbb{R}$, we have 
\begin{equation*}
\begin{split}
  \int_{\vert{x}\vert>1}\exp\left(\theta{x}+ \psi x^2\right) \nu_L(dx)&=\exp(-\theta^2(4\psi)^{-1}) \int_{\vert{x}\vert>1}\exp\left( \psi (x+\theta(2\psi)^{-1})^2\right) \nu_L(dx)\\
  &\leq \exp(-\theta^2(4\psi)^{-1})\left(\nu_L(1,\infty)+\nu_L(-\infty,-1)\right)<+\infty.
\end{split}
\end{equation*}
The last inequality follows from the first property in (\ref{Levy_measure}). Since $\nu_L$ is a L\'evy measure, both $\nu_L(1,\infty)$ and $\nu_L(-\infty,-1)$ are finite, and the exponential term is finite as well. Hence the right-hand side is finite, which proves the theorem.
\end{proof}
We end this section by a direct application of the analysis of this section. In fact, we address the pricing of European call options under the second-order Esscher measure for the general L\'evy model represented by $X$. As it is not always possible to obtain an explicit formula for the option price, the use of the Fast Fourier transform algorithm (see \cite{carr1999option}) is a convenient way to calculate the price. For a brief review of the Fourier transform method see \cite{schmelzle2010option}.
\begin{proposition} Let $\psi\in\mathbb{R}$ such that there exists $\mathbb{Q}^{\psi}\in\mathcal Q^{EE}(S,\psi)$. For $t\in[0,T]$ and $u\in\mathbb{R}$, put 
\begin{equation*}
\begin{split}
 &\Phi^{\psi}_X\big(t, u):=\mathbb E^{\mathbb{Q}^{\psi}}\left[\exp(iuX_t)\right]=\exp (t\varphi^{\psi}_X (u)),\\
 & \varphi^{\psi}_X (u) := iu b^{\psi}(h) - \frac{u^2}{2} \sigma^2 + \int e^{\theta x +\psi x^2} \big[ e^{iux}-1 - iu h(x) \big] \nu_L(dx)
\end{split}
\end{equation*}
Then the following assertions hold.\\
 {\rm{(a)}} The European call option price at time $0$ with maturity $T$ and strike price $K$ ($:=e^k$), is given under $\mathbb{Q}^{\psi}$ by
    \begin{align*}
    C_0^{EE}(\psi,T) = \frac{e^{-\alpha k}e^{-rT}}{2\pi} \int_{-\infty}^{\infty} e^{-iu k} \frac{ \Phi^{\psi}_X\big(T, u-(1+\alpha)i\big)}{\alpha^2 + \alpha -u^2 +i(1+2\alpha )u}  du, \quad \forall u \in \mathbb{R},
\end{align*}
where $\alpha$ is a damping factor.\\
 {\rm{(b)}} The European put option price at time $0$ with maturity  $T$ and strike price $K$ ($:=e^k$), is given under $\mathbb{Q}^{\psi}$ by
\begin{equation*}
    P_0^{EE}(\psi,T) = \frac{e^{\alpha k} e^{-rT}}{2\pi} \int_{-\infty}^\infty e^{-iuk} \frac{\Phi^{\psi}_X \big(T, u-(1-\alpha)i\big)}{\alpha^2 - \alpha -u^2 +i(1 - 2\alpha)u} du,  \quad \forall u \in \mathbb{R},
\end{equation*}
where $\alpha$ is a damping factor.
\end{proposition}

\section{Esscher pricing for jump-diffusion models}\label{section:jumpdiffusion}
Throughout this section, we concentrate on the jump-diffusion models, and suppose that our L\'evy process $X$ takes the form of 
\begin{equation}\label{Model_GeneralJumpDiffusion}
    X_t = bt + \sigma W_t + \sum_{i=1}^{N_t} J_i.
\end{equation}
Here $N$ is  the Poisson process with intensity $\lambda\in(0,\infty)$, $W$ is a one-dimensional Brownian motion, and $(J_i)_{i\geq 1}$ is a sequence of independent and identically distributed random variables, with cumulative distribution function $F_J(dx)$, and the three components $(W,N, (J_i)_{i\geq 1})$ are independent.  In this section we derive, as explicit as possible, the second-order Esscher prices for European call options for the models (\ref{Model_GeneralJumpDiffusion}). To this end, we remark that our L\'evy model has a finite variation (i.e. $\int\vert{h}(x)\vert\nu_L(dx)<\infty$) and for any $\psi\in\mathbb{R}$, such that ${\cal{Z}}^{EE}(S,\psi)\not=\emptyset$, equation (\ref{Martingale_Equation}) becomes
\begin{equation}\label{Martingale_Equation_4GJD}
b+ \frac{\sigma^2}{2} + \theta \sigma^2 + \lambda\int(e^x - 1) \exp(\theta x +\psi x^2)F_J(dx) = r \quad \mathbb P \otimes dt \text{-a.e.} .\end{equation}
Furthermore, for any $\psi\in\mathbb{R}$ such that ${\cal{Z}}^{EE}(S,\psi)\not=\emptyset$, we introduce the following useful notations
\begin{equation}\label{Lambda(psi)b(psi)}
\begin{split}
&\theta(\psi):=\mbox{Root of}\ (\ref{Martingale_Equation_4GJD}),\quad b^{\psi}:=b+\theta(\psi)\sigma^2,\quad  \lambda^\psi:= \lambda \int\exp(\theta(\psi) x+\psi x^2) F_J(dx),\\
&  r^{\psi}:=\lambda\int(1-e^x)\exp(\theta(\psi)x+\psi{x}^2)F_J(dx),\quad F^\psi_J(dx) :={\frac{\lambda}{\lambda^{\psi}}}\exp(\theta(\psi) x+\psi x^2) F_J(dx).
\end{split}
\end{equation}
\begin{definition}\label{Esscher_Price_Definition}  Let $(x,\Sigma, R, T')\in (0,\infty)\times(0,\infty)\times\mathbb{R}\times(0,\infty)$.\\
{\rm{(a)}}  $C^{BS}(x,\Sigma,R,T')$ denotes the Black-Scholes price at time zero of the European call option maturing at time $T'$ when the stock's price follows the Black-Scholes model  with initial price $x$, volatility $\Sigma$ and interest rate $R$. I.e.
\begin{equation}\label{BS-Function}
\begin{split}
&C^{BS}(x,\Sigma,R,T')=x\Phi(d_+(x,\Sigma,R,T'))-Ke^{-RT'}\Phi(d_-(x,\Sigma,R,T')),\\
& d_{\pm}(x,\Sigma,R,T'):={\frac{\ln(x/K)+(R\pm\Sigma^2/2)T'}{\Sigma\sqrt{T'}}}.
\end{split}
\end{equation}
{\rm{(b)}}  Let $\psi\in\mathbb{R}$ such that $\mathcal Q^{EE} (S,\psi) \neq \emptyset$. Then $C^{EE}_t(\psi,T')$ (respectively $C^{EE}_0(\psi,x,T')$) denotes the price at time $t\in[0,T']$ (respectively at time zero when $S_0=x$) under $\mathbb Q^\psi \in \mathcal Q^{EE}(S,\psi)$ for the European call option with maturity $T'$ and strike price $K$.
\end{definition}
Below, we state our main results of this section.
\begin{theorem}\label{Pricing_General_Levy} Let $\psi \in \mathbb R$ such that $\mathcal Q^{EE} (S,\psi) \neq \emptyset$ and $K$ be a strike price ($K>0$).  
 Then  the following assertions hold.\\
    {\rm{(a)}} For any $T\in(0,\infty)$, we have 
    \begin{equation*}
    C^{EE}_t(\psi,T)=C^{EE}_0(\psi,S_t,T-t),\quad t\in[0,T].
    \end{equation*}
  {\rm{(b)}} For any $(x,T)\in(0,\infty)\times(0,\infty)$, we have 
     \begin{equation}\label{Price_Jump-diffusion}
            \begin{split}
               &C^{EE}_0(\psi,x,T)\\
               &= \exp\left\{r^{\psi}T\right\} \mathbb E^{\mathbb Q^\psi} \big[C^{BS}(x\exp(\widetilde{\Gamma}_{T}),\sigma, r^\psi+r,T) \big],\\
                &=\exp\left\{(r^\psi-\lambda^\psi)T\right\} \sum_{n\geq 0} \frac{\big(\lambda{T}\big)^n}{n!} \int_{\mathbb R^n} C^{BS}\big(x e^{\mathbf{1}_n\mathbf{y}^{tr} }, \sigma, r^\psi+r,T\big) e^{\theta \mathbf{1}_n\mathbf{y}^{tr}  + \psi |\mathbf y|^2} F^{\otimes n}_J(d\mathbf y).
            \end{split}
        \end{equation}
        Herein, $\mathbf{1}_n:=(1,\ldots,1)\in\mathbb{R}^n$ for $n\geq 1$, $\mathbf{1}_0:=0,\ \mathbb{R}^0:=\{0\}$, $F_J^{\otimes 0}(dx):=\delta_{0}(dx)$ and $\widetilde\Gamma$ is given by 
        \begin{equation*}
    \widetilde\Gamma_t :=  \sum_{i=1}^{N_{t}} J_i,\quad t\in[0,T].
\end{equation*}
        {\rm{(c)}}  For any $(x,T)\in(0,\infty)\times(0,\infty)$, we have
        \begin{equation}\label{Price_Comparaison}
            C^{BS}(x, \sigma, r,T)\leq C^{EE}_0(\psi,x,T) < x.
        \end{equation}
\end{theorem}
\begin{proof} The proof is achieved in three parts, where we prove assertions (a), (b) and (c) respectively.\\
{\bf Part 1.} Here we prove assertion (a). 
    \begin{equation}\label{Equality1}
    \begin{split}
            C^{EE}_t( \psi,T) &= e^{-r(T-t)} \mathbb E^{\mathbb Q^\psi}\big[(S_T - K)^+ \big| \mathcal F_t\big]= e^{-r(T-t)} \mathbb E^{\mathbb Q^\psi}\big[(S_t e^{X_T-X_t} - K)^+ \big| \mathcal F_t\big]\\
            &= e^{-r(T-t)} \mathbb E\big[{\frac{Z^{\psi}_T}{Z^{\psi}_t}}(S_t e^{X_T-X_t} - K)^+ \big| \mathcal F_t\big].  \end{split}
    \end{equation}
Due to the fact that $\ln(Z^{\psi}_T/Z^{\psi}_t)=\theta(X_T-X_t)+\psi\sum_{t<s\leq T}(\Delta{X}_s)^2-\kappa(\theta,\psi)(T-t)$ (see Theorem \ref{General_Levy_Existence}-(c)) and the independent increments of $X$, we get 
 \begin{equation}\label{Equality2}
    \mathbb E\left[{\frac{Z^{\psi}_T}{Z^{\psi}_t}}(S_t e^{X_T-X_t} - K)^+ \big| \mathcal F_t\right]=f^{EE}(S_t,\psi,t,T), \end{equation}
    where the function  $f^{EE}$ is given by 
    \begin{equation}\label{F(EE)}\begin{split} 
  f^{EE}(x,\psi,t,T)&:=  \mathbb E\left[\exp\left(\theta(X_T-X_t)+\psi\sum_{t<s\leq T}(\Delta{X}_s)^2-\kappa(\theta,\psi)(T-t)\right)(x e^{X_T-X_t} - K)^+\right]\\
  &= \mathbb E\left[\exp\left(\theta{X}_{T-t}+\psi\sum_{0<s\leq {T-t}}(\Delta{X}_s)^2-\kappa(\theta,\psi)(T-t)\right)(xe^{X_{T-t}} - K)^+\right]\\
   &= \mathbb E^{\mathbb Q^\psi}\left[(xe^{X_{T-t} }- K)^+\right]=e^{r(T-t)} C^{EE}_0(\psi,x,T-t).
\end{split} \end{equation}
 The second equality above is due to the stationarity property of the process $X$. Therefore, by combining (\ref{Equality1}), (\ref{Equality2}) and (\ref{F(EE)}), assertion (a) follows immediately. \\
 {\bf Part 2.} This part proves assertion (b). To this end,  we suppose $S_0=x$ and get $S_T=x\exp(X_T)$. Thanks to Theorem \ref{Levy_Psi_Constant_Existence}-(b), the process $X$ under $\mathbb{Q}^{\psi}$ has the same distribution as the process $X^{\psi}$ under $\mathbb{P}$ where         
    \begin{equation}\label{model: general jump-diffusion X^psi}
    X^\psi_t := b^\psi t +\sigma W_t +\sum_{i=1}^{N^\psi_t} J^\psi_i=:  b^\psi t +\sigma W_t +\Gamma^{\psi}_t.
\end{equation}
Here $W, N^\psi$ and $J^\psi=(J^{\psi}_i)_{i\geq 1}$ are independent with $N^\psi$ being the Poisson process with intensity $\lambda^\psi$, $J^{\psi}=(J^\psi_i)_{i\geq1}$ are independent and identically distributed random variables with distribution function $F^\psi_J(dx)$, where $(b^{\psi},\lambda^{\psi} F^\psi_J)$ is given in (\ref{Lambda(psi)b(psi)}). As a result, on the one hand, by combining the above fact and the independence between $W$ and $\Gamma^{\psi}$, we derive\\
\begin{equation}\label{C(EE1)}
 \begin{split}
               C^{EE}_0(\psi,x,T)&= e^{-rT} \mathbb E^{\mathbb Q^\psi}\left[(x\exp(X_T) - K)^+ \right]= e^{-rT} \mathbb E\left[(x\exp(X_T^{\psi}) - K)^+ \right]\\
               &= e^{-rT} \mathbb E\left[(x\exp( b^\psi{T} +\sigma W_T +\Gamma^{\psi}_T) - K)^+ \right]= e^{-rT} \mathbb E\left[g_{\psi}(x,\Gamma^{\psi}_T)\right]\\
               &\mbox{where}\\
               g_{\psi}(x,y)&:=\mathbb E\left[\left(x\exp( b^\psi{T} +\sigma W_T +y) - K\right)^+ \right],\quad\mbox{for any}\ y\in\mathbb{R}.
\end{split} \end{equation}
On the other hand, in virtue of (\ref{Martingale_Equation_4GJD}) and (\ref{Lambda(psi)b(psi)}) which lead to $b^{\psi}=r+r^{\psi}-{\frac{\sigma^2}{2}}$, we obtain
\begin{equation*}
 \begin{split}
  g_{\psi}(x,y)&=\mathbb E\left[\left(xe^{y}\exp\left((r^\psi+r-{\frac{\sigma^2}{2}})T +\sigma W_T\right) - K\right)^+ \right]=\exp((r^{\psi}+r)T)C^{BS}(xe^y,\sigma,r^{\psi}+r,T).
  \end{split} \end{equation*}
  Therefore, the first equality in (\ref{Price_Jump-diffusion}) follows immediately from combining the latter equality with (\ref{C(EE1)}) and the fact that the distribution of $\Gamma^{\psi}_T$ under $\mathbb{P}$  and that of $\widetilde\Gamma_T$ under $\mathbb{Q}^{\psi}$ are the same.\\  
  Thanks to the first equality in  (\ref{Price_Jump-diffusion}) and the independence of $N^{\psi}_T$ with $(J^{\psi}_i)_i$, we get 
  \begin{equation}\label{C(EE2)}
 \begin{split}
               C^{EE}_0(\psi,x,T)&= e^{r^{\psi}T}\mathbb E\left[C^{BS}(x\exp(\Gamma^{\psi}_T),\sigma,r^{\psi}+r,T)\right]=e^{r^{\psi}T}\mathbb E\left[\mathbb E\left[C^{BS}(x\exp(\Gamma^{\psi}_T),\sigma,r^{\psi}+r,T)\big| N^{\psi}_T\right]\right]\\
               &=\sum_{n=0}^\infty \exp((r^{\psi}-\lambda^{\psi})T){\frac{({\lambda^{\psi}T})^n}{n!}}\mathbb E\left[C^{BS}(x\exp(\Gamma^{\psi}_T),\sigma,r^{\psi}+r,T)\big| N^{\psi}_T=n\right]\\
                 &=\sum_{n=0}^\infty  \exp((r^{\psi}-\lambda^{\psi})T){\frac{({\lambda^{\psi}T})^n}{n!}}\mathbb E\left[C^{BS}\left(x\exp(\sum_{i=1}^n J^{\psi}_i),\sigma,r^{\psi}+r,T\right)\right]
 \end{split} \end{equation} 
In virtue of $F^{\psi}_J(dx)=(\lambda/\lambda^{\psi})\exp(\theta(\psi)x+\psi{x}^2)F_J(dx)$ (see (\ref{Lambda(psi)b(psi)})), it is easy to prove that for any  nonnegative function $l:\ \mathbb{R}\longrightarrow\mathbb{R}$, for $n\geq 1$ we have 
\begin{equation*}\begin{split}
\mathbb E\left[l\left(\sum_{i=1}^n J^{\psi}_i\right)\right]&=\int_{\mathbb{R}^n}l\left(\sum_{i=1}^n x_i\right) F^{\psi}_J(dx_1)\ldots{F}^{\psi}_J(dx_n)\\
&=\left({\frac{\lambda}{\lambda^{\psi}}}\right)^n \int_{\mathbb{R}^n}l\left(\sum_{i=1}^n x_i\right)\exp\left(\theta(\psi)\sum_{i=1}^nx_i+\psi\sum_{i=1}^nx_i^2\right) F_J(dx_1)\ldots{F}_J(dx_n),\\
&=\left({\frac{\lambda}{\lambda^{\psi}}}\right)^n \int_{\mathbb{R}^n}l\left(\mathbf{1}_n\mathbf{x}^{tr} \right)\exp\left(\theta(\psi)\mathbf{1}_n\mathbf{x}^{tr} +\psi\vert\mathbf{x}\vert^2\right) F_J^{\otimes{n}}(d\mathbf{x}).
\end{split} \end{equation*}
  Therefore, the second equality in (\ref{Price_Jump-diffusion})  follows from combining this equality with (\ref{C(EE2)}). This proves assertion (b) and part 2 is complete.\\
 {\bf{P}art 3.} Recall that the function $x \rightarrow C^{BS}(x, \Sigma, R,T)$ is strictly increasing and strictly convex. Then, on the one hand,  by combining Jensen's inequality and assertion (b), we deduce 
    \begin{equation}\label{C(EE)3}\begin{split}
   C^{EE}_0(\psi,S_0,T)&= \exp\left\{r^{\psi}T\right\} \mathbb E^{\mathbb Q^\psi} \big[C^{BS}(S_0\exp(\widetilde{\Gamma}_{T}),\sigma, r^\psi+r) \big]\\
&\geq   \exp\left\{r^{\psi}T\right\} C^{BS}\left (S_0 \mathbb E^{\mathbb Q^\psi} [\exp(\widetilde{\Gamma}_{T})], \sigma, r^\psi+r,T\right)
\end{split} \end{equation}
On the other hand, direct calculations give 
\begin{equation}\label{Expectation}\begin{split}
\mathbb E^{\mathbb Q^\psi} [\exp(\widetilde{\Gamma}_{T})]&=\mathbb E [\exp({\Gamma}^{\psi}_{T})]=\exp\left(-\lambda^{\psi}T+\lambda^{\psi}T \mathbb E [\exp(J^{\psi}_1)]\right)\\
&=\exp\left(\lambda^{\psi}T\int(e^x-1)F^{\psi}_J(dx)\right)=\exp(-r^{\psi}T).
\end{split} \end{equation}
Hence, on the one hand, a combination of this latter equality with (\ref{C(EE)3}) and the easy fact that $$ \exp\left\{r^{\psi}T\right\} C^{BS}\left (S_0\exp(-r^{\psi}T), \sigma, r^\psi+r,T\right)=C^{BS}\left (S_0, \sigma, r,T\right)$$ yields the left-hand-side inequality in (\ref{Price_Comparaison}). On the other hand, the right-hand-side inequality in (\ref{Price_Comparaison}) is a direct consequence of the fact that $C^{BS}(S_0\exp(\widetilde{\Gamma}_{T}),\sigma, r^\psi+r,T) <S_0\exp(\widetilde{\Gamma}_{T})$ and (\ref{Expectation}). This ends the proof of theorem.\end{proof}
\begin{remark} {\rm{(a)}} By combining assertions (a) and (b) of Theorem \ref{Pricing_General_Levy}, for $t\in[0,T]$ we get
\begin{equation*}
            \begin{split}
&C^{EE}_t(\psi,T)=\exp\left\{(r^\psi-\lambda^\psi)(T-t)\right\}C^{BS}\big(S_t, \sigma, r^\psi+r,T-t\big)\\
& \exp\left\{(r^\psi-\lambda^\psi)(T-t)\right\}\sum_{n\geq 1} \frac{\big(\lambda{(T-t)}\big)^n}{n!} \int_{\mathbb R^n} C^{BS}\big(S_t{e}^{ \mathbf{1}_n\mathbf{y}^{tr}}, \sigma, r^\psi+r,T-t\big) e^{\theta \mathbf{1}_n\mathbf{y}^{tr} + \psi |\mathbf y|^2} F^{\otimes n}_J(d\mathbf y).
 \end{split}
\end{equation*}
 {\rm{(b)}}  As $X$ under $\mathbb{Q}^{\psi}$ has the same distribution as $X^{\psi}$ under $\mathbb{P}$, we also write 
\begin{equation*}C^{EE}_t(\psi,T) = \exp\left\{r^{\psi}(T-t)\right\} \mathbb E \big[C^{BS}(S_t\exp({\Gamma}_{T-t}^{\psi}),\sigma, r^\psi+r, T-t) \big],\quad \end{equation*}
 {\rm{(c)}} If $F_n(dy)$ denotes the distribution function of $\Gamma^{(n)}:=\sum_{i=1}^nJ_i$ ($n\geq 1)$, then we also have
        \begin{equation}
            \begin{split}
                C^{EE}_0(\psi,x,r,T) &=\exp\left\{(r^\psi -\lambda^\psi)T\right\} C^{BS}(x,\sigma, r^\psi+r,T) \\
                &+\exp\left\{(r^\psi -\lambda^\psi)T\right\}  \sum_{n\geq 1} \frac{\big(\lambda{T}\big)^n}{n!} \int_{\mathbb R} C^{BS}\big(xe^y, \sigma, r^\psi+r,T \big)\exp(\theta{y}+\psi{y}^2)  F_n(dy). \end{split}\end{equation}
\end{remark}
The remaining part of this section illustrates the main theorem on particular cases such as constant jump-diffusion models or Merton's model (log-normal jump-diffusions).
\begin{corollary}\label{corollary: CJD}
Let $\gamma \in(-1,0)\cup(0,\infty)$, and suppose that $F_J(dx)=\delta_{\gamma}(dx)$ (i.e.  $X$ follows a constant jump-diffusion model with constant jump size equal to $\gamma$). Then the following assertions hold.\\
   {\rm{(a)}} For any $\psi\in\mathbb{R}$, there exists a second-order Esscher density with parameter $\psi$ (i.e. ${\cal{Z}}(S,\psi)\not=\emptyset$). Furthermore, the assumption (\ref{Integrability_Condition2}) is fulfilled and hence ${\cal{Z}}(S,\psi)\subset{\cal{M}}^p$ for any $p\in(1,\infty)$\\
{ \rm{(b)}}  The option's price $C^{EE}_t(\psi,T)$ is given by 
    \begin{equation}\label{Esscher_Price_LevyConstant}
    \begin{split}
        C^{EE}_t(\psi,T)& = \exp\left\{(r^{\psi}-\lambda^\psi )(T-t)\right\} \sum_{n\geq 0} \frac{(\lambda^\psi (T-t))^n}{n!} C^{BS}(S_t e^{n \gamma}, \sigma, r^\psi+r,T-t),\\
        & =  \sum_{n\geq 0} e^{-\lambda^\psi (T-t)} \frac{(\lambda^\psi (T-t))^n}{n!} C^{BS}(S_t e^{n \gamma - \lambda^\psi(e^\gamma-1)(T-t)}, \sigma, r,T-t).\end{split}\end{equation}
        where $\lambda^\psi := \lambda e^{\theta \gamma + \psi \gamma^2}$ and $r^{\psi}:=\lambda(1-e^{\gamma})\exp(\theta\gamma+\psi\gamma^2)$.\\
{\rm{(c)}} For any $(x,T)\in (0,\infty)\times (0,\infty)$, the function $\psi\longrightarrow{C}^{EE}_0(\psi,x, T)$  is strictly increasing, and $\displaystyle\lim_{\psi\downarrow-\infty}C^{EE}_0(\psi,x, T)=C^{BS}(x,\sigma,r,T)$.\end{corollary}
\begin{proof} On the one hand, remark  that assertion (a) is obvious due to $F_J(dx)=\delta_{\gamma}(dx)$ for this case of L\'evy. On the other hand, for $n\geq 1$, we obtain $ \int_{\mathbb{R}^n}f(x)F^{\otimes n}_J(dx)=f(\gamma,...,\gamma)$ for any function $f:\mathbb{R}^n\longrightarrow \mathbb{R}$. Thus, the proof of the first equality in (\ref{Esscher_Price_LevyConstant}) follows immediately from combining this latter fact with Theorem \ref{Pricing_General_Levy}. The second equality in  (\ref{Esscher_Price_LevyConstant})  follows from the first equality and direct calculations after using the definition of the Black-Scholes pricing function $C^{BS}$. This proves assertion (b).  Hence the rest of this proof focuses on proving assertion (c). To this end, on the one hand, thanks to \cite[lemma  4.2]{choulli2025second} and its proof, we have 
\begin{equation}\label{Psi_limits}
\Lambda:= \lambda^{\psi}\ \mbox{increases strictly with}\  \psi,\ \mbox{and}\ \lim_{\psi\longrightarrow-\infty}\lambda^{\psi}=0.
\end{equation}
 On the other hand, for any $(x,T)\in (0,\infty)\times (0,\infty)$, we consider the following real-valued functions defined on $\mathbb{R}^+$ and given by 
 \begin{equation}\label{eq: Fn(Lambda)}
 \begin{split}
        f_k(\Lambda)&:= \frac{e^{-\Lambda T}[\Lambda T]^k}{k!} C^{BS} (x_k(\Lambda),\sigma,r,T),\quad\mbox{where}\quad x_k(\Lambda):=xe^{k\gamma}\exp(-\Lambda(e^{\gamma}-1)T),\quad k\geq 0,\\
       \Sigma_n(\Lambda) &:= \sum_{k=0}^nf_k(\Lambda),\quad \Sigma(\Lambda):= \sum_{k=0}^\infty f_k(\Lambda);\ \Lambda\in[0,\infty[,\quad n\geq 0.\end{split}
    \end{equation}
Then, we denote by $\varphi$ is the standard normal probability density function, we derive  for any $n\ge0$ 
\begin{align*}
f'_n(\Lambda)& =T\frac{e^{-\Lambda{T}}[\Lambda{T}]^n}{n!}\left[
	C^{BS} (x_{n+1}(\Lambda),\sigma,r,T)-C^{BS} (x_n(\Lambda),\sigma,r,T) -x_n(\Lambda)\kappa\Phi(d_+(x_n(\Lambda))
	\right],\\
& = T\frac{e^{-\Lambda{T}}(\Lambda{T})^n}{n!}\frac 12 (x_n(\Lambda)\kappa)^2 \frac{d^2 C^{BS} }{d(x_n(\Lambda))^2}(\xi_n, \sigma, r, T),\\
& = \frac{Te^{-\Lambda{T}}(\Lambda{T})^n}{2n!}\frac{(x_n(\Lambda)\kappa)^2}{\xi_n\sigma\sqrt{\tau}}\varphi(d_+{(\xi_n)}) >0 \quad \mbox{with\ } d_+(y)=\frac{1}{\sigma\sqrt{T}}\bigg[\ln(y/K)+(r+\frac{\sigma^2}{2})T \bigg], 
\end{align*}
where $\kappa:=e^{\gamma}-1$, $\xi_n$ is between $x_n(\Lambda)$ and $x_{n+1}(\Lambda)$. By combining the above equality with direct calculations, we conclude that 
\begin{equation}
f'_n(\Lambda)>0,\ \mbox{for any}\ \Lambda\in(0,\infty),\  \ \sup_{0\leq\Lambda\leq C}f'_n(\Lambda)\leq  \frac{\kappa^2}{\sigma \sqrt{T}\min(1,1+\kappa)} \frac{C^n}{n!}.
\end{equation}
Therefore, in virtue of \cite[Theorem 7.10]{Rudin1976}, we deduce that the sequence $\Sigma_n(\Lambda)':={\frac{d}{d\Lambda}}\Sigma_n(\Lambda)$ converges uniformly on any compact subset of $[0,\infty)$. Hence, by combining this with \cite[Theorem 7.17]{Rudin1976},  we get 
\begin{equation}\label{Limits}
{\frac{d}{d\Lambda}}\sum_{n\geq0}f_n(\Lambda)=\sum f'_n(\Lambda)>0,\quad\mbox{and}\quad  \lim_{\Lambda\longrightarrow0}\sum_{n\geq0}f_n(\Lambda)=\sum_{n\geq 0}\lim_{\Lambda\longrightarrow0}f_n(\Lambda)=C^{BS}(x,\sigma,r,T).   
 \end{equation}
 Therefore, the proof of assertion (c) follows immediately from combining (\ref{Limits}) and (\ref{Psi_limits}). This completes the proof the corollary. \end{proof}
The next example is the Merton Model, where the jumps' sizes are normally distributed. 
\begin{corollary} Let $\mu_J\in\mathbb{R}$ and $\sigma_J\in(0,\infty)$ and suppose that $J_1$ is normally distributed with mean $\mu_J$ and variance $\sigma^2_J$ (i.e. $X$ follows  the Merton model). Then the following assertions hold.\\
 \rm{(a)} For any $\psi<1/(2\sigma_J^2)$, we have ${\cal{Z}}(S,\psi)\not=\emptyset$. Furthermore, assumption (\ref{Integrability_Condition2}) is fulfilled and hence ${\cal{Z}}(S,\psi)\subset{\cal{M}}^p$ for any $p\in(1,\infty)$.\\
  \rm{(b)}  Let  $\psi<1/(2\sigma_J^2)$. Then we have  \begin{equation}\label{option price LJD}
        C^{EE}_t(\psi,T) =  \sum_{n\geq 0} e^{(r^\psi-\lambda^\psi) (T-t)} \frac{\big(\lambda^\psi (T-t)\big)^n}{n!} C^{BS}(S_t^{(n)}, \sigma^{(n)}, r^\psi+r, T-t)
    \end{equation}
    where
    \begin{equation}
        \begin{split}
            &S_t^{(n)} = S_t \exp \bigg(n[\mu_\psi + \frac{\sigma_\psi^2}{2}] \bigg)\quad d_\pm^{(n)} = \frac{\ln \big(\frac{S_t^{(n)}}{K}\big) + (r^\psi+r \pm \frac{(\sigma^{(n)})^2}{2})(T-t)}{\sigma^{(n)} \sqrt{T-t}} \\ 
            &(\sigma^{(n)})^2 = \sigma^2 + \frac{n\sigma_\psi^2}{T-t}
        \end{split}
    \end{equation}
\end{corollary}
\begin{proof}
    Since $J^\psi_1$ follows a normal distribution
    \begin{equation*}
   \mu_\psi:= \mathbb{E}[J^{\psi}_1]=\frac{\mu_J + \theta \sigma^2_J}{1-2\psi \sigma^2_J},\quad \mbox{and}\quad \sigma_\psi^2:=\mbox{Var}(J^{\psi}_1)=\frac{\sigma^2_J}{1-2\psi \sigma^2_J}.\end{equation*} 
   As a result, $\Gamma_n^{\psi}=\sum_{i=1}^n J^{\psi}_i$ follows a normal distribution with mean $n\mu_{\psi}$ and variance $n\sigma_{\psi}^2$,  and hence
    \begin{equation}
        \begin{split}
            \mathbb E[C^{BS}(x\exp(\Gamma^\psi_n), \sigma, r^\psi+r,T-t)] = \int C^{BS}(x e^y, \sigma, r^\psi+r,T-t) f_{\psi}(y)dy= x C_1(x) - Ke^{-(r^\psi+r)(T-t)} C_0(x),
        \end{split}
    \end{equation}
    where, for any $\xi\in\{0,1\}$, 
    \begin{equation}
        \begin{split}
            C_\xi(u)&:= \int \exp(y\xi) \Phi \big(d(ue^y, \xi) \big) f_{\psi}(y)dy,\quad f_{\psi}(y) := \frac{1}{\sqrt{2\pi n} \sigma_\psi} \exp\bigg(-\frac{(y - n\mu_\psi)^2}{2n \sigma_{\psi}^2} \bigg),\\
            d(z, \xi)&:= \frac{\ln\big(\frac{z}{K}\big) + (r^\psi+r - \frac{\sigma^2}{2} + \sigma^2 \xi)(T-t)}{\sigma \sqrt{T-t}},\quad z>0.
        \end{split}
    \end{equation}
    For the proof, we define the following notations
    \begin{equation}
        \begin{split}
            g(u) &:= \frac{\ln(u)}{\sigma \sqrt{T-t}},\quad\quad  B:= \frac{\ln(K) - (r^\psi +r- \frac{\sigma^2}{2} +\sigma^2 \xi)(T-t)}{\sigma \sqrt{T-t}},\\
            \widetilde \sigma^2 &:= \frac{\sigma^2 (T-t) n \sigma_\psi^2}{\sigma^2 (T-t) + n \sigma_\psi^2},\quad \widetilde \mu = \widetilde \mu(u,\xi):= \widetilde \sigma^2 \bigg(\xi - \frac{g(u) - B}{\sigma \sqrt{T-t} + \frac{\mu_\psi}{\sigma_\psi^2}} \bigg)\\
            \widehat \sigma^2 &:= \frac{\sigma^2 (T-t) + n\sigma_\psi^2}{\sigma^2 (T-t)},\quad\widehat \mu = \widehat \mu(\xi) := -\frac{\widehat \sigma^2 \widetilde \sigma^2}{\sigma \sqrt{T-t}} \bigg( \frac{\mu_\psi}{\sigma_\psi^2} + \xi \bigg)
        \end{split}
    \end{equation}
    Now, to derive the option price equation, we calculate the derivative of $C_\xi$ with respect to $u$,
    \begin{equation}
        C'_\xi(u) = \int \frac{\exp(I_\xi(u,y))}{\sigma \sqrt{T-t} \ u\sqrt{2\pi} \sqrt{2\pi n} \sigma_\psi} dy
    \end{equation}
    where with simple algebra, we have
    \begin{equation}
        I_\xi (u,y) = -\frac{1}{2} \bigg(\frac{y-\widetilde \mu}{\widetilde \sigma} \bigg)^2 -\frac{1}{2} \bigg(\frac{(g(u) - B) - \widehat \mu}{\widehat \sigma} \bigg)^2 + n\bigg[\xi \mu_\psi +\frac{\xi^2 \sigma_\psi^2}{2}\bigg]
    \end{equation}
    Hence,
    \begin{equation}
        C'_\xi(u) = \frac{\widetilde \sigma \exp\bigg({n\bigg[\xi \mu_\psi +\frac{\xi^2 \sigma_\psi^2}{2}\bigg]}\bigg)}{\sqrt{2\pi n} \sigma_\psi} g'(u) \exp \bigg(-\frac{1}{2} \bigg[\frac{g(u) - B - \widehat \mu}{\widehat \sigma} \bigg]^2 \bigg)
    \end{equation}
    where $g'(u) = 1/\big(\sigma \sqrt{T-t}\ u \big)$. Now, by integrating the last expression over the interval $[0,x]$ and performing a change of variable we get
    \begin{equation}
        C_\xi(x) = \frac{\widetilde \sigma \widehat \sigma }{\sqrt{n} \sigma_\psi}\exp\bigg({n\bigg[\xi \mu_\psi +\frac{\xi^2 \sigma_\psi^2}{2}\bigg]}\bigg)\Phi \bigg(\frac{g(x) - B - \widehat \mu}{\widehat \sigma} \bigg)
    \end{equation}
    By simple calculation we find
    \begin{equation}
        \frac{\widetilde \sigma \widehat \sigma }{\sqrt{n} \sigma_\psi} = 1
    \end{equation}
    and
    \begin{equation}
        \frac{g(x) - B - \widehat \mu}{\widehat \sigma} = \frac{\ln \big(\frac{x\exp\big(n\mu_\psi +n\sigma_\psi^2/2\big)}{K}\big) + [r^\psi+r - \frac{(\sigma^{(n)})^2}{2} + (\sigma^{(n)})^2 \xi](T-t)}{\sigma^{(n)} \sqrt{T-t}} =: d_\xi^{(n)}
    \end{equation}
    where 
    \begin{equation}
        (\sigma^{(n)})^2 = \sigma^2 + \frac{n\sigma_\psi^2}{T-t}
    \end{equation}
    Plugging these back into $C_\xi(x)$ and then substituting $C_\xi(x)$ in the price equation, we get the option price (\ref{option price LJD}).
\end{proof}    
    
\section{Risk management application}\label{section: risk management}

In this section, we investigate how the Esscher martingale measure and jump-diffusion modeling can be used to support risk management in illiquid markets. As mentioned in \cite{bondi2020comparing}, if there are many liquid options around, like plain vanilla calls and puts on liquid stocks, we would expect the calibration method to perform best. However, in new financial markets, like insurance derivatives, cryptocurrencies, energy, or electricity markets, there are often only a few derivatives or
	any at all available, or the underlying (like electricity) is difficult to trade in since it is not storable. In these cases the calibration method is not implementable. In such settings, the Esscher measure becomes a practical and theoretically grounded way to construct a risk-neutral pricing measure as has been done, e.g., in the book \cite{benth2008stochastic} 
	about electricity and related markets. 
	Besides that, there are few, if at all, studies about practical implementation of hedging strategies in incomplete markets. In \cite{bondi2020comparing} the authors main result is that while the Esscher martingale measure based pricing method in a liquid market does underperform the calibration method, as is to be expected, it does so only by less than 5\%. So it might be a feasible choice of method in new financial markets.\\

	A second key modeling choice in illiquid commodity markets is the specification of the underlying price dynamics. Jump-diffusion models are particularly suitable for commodities because they capture abrupt price movements arising from supply and demand shocks, geopolitical events, and other market disruptions. As argued in \cite{hilliard1998valuation} and \cite{feng2008pricing}, incorporting jumps allows the model to reflect skewness, heavy tails, and sudden spikes - features that are prominent in empirical commodity returns and essential for realistic risk assessment.\\
	
\begin{figure}[H]
    \centering
    \includegraphics[width=.65\textwidth]{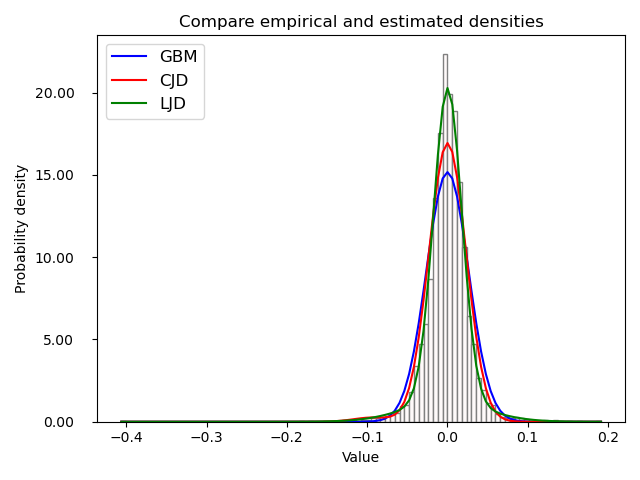}
    \caption{empirical vs.\ model densities of WTI log-returns.}
    \medskip
    \small
    Blue is the geometric Brownian motion model (GBM), red is for the constant jump-diffusion model (CJD) and green is for the log-normal jump-diffusion model (LJD)
    \label{fig: densities}
\end{figure}

	Motivated by these considerations, we analyze daily spot prices of the WTI crude oil from 1986-01-02 to 2010-08-31. We then fit three models to the log-return prices and estimate model parameters using maximum log-likelihood estimation (MLE), see Table \ref{table: mle params}, compare their empirical fit, see Figure \ref{fig: densities}, and then use these models to explore two questions. First, how does the second-order Esscher parameter determine a range of option prices, the so-called Esscher pricing interval? Second, which role does this Esscher parameter play in stress testing?

\begin{table}[H]
    \begin{center}
        \begin{filecontents*}{data/mle_params.csv}
model,$\mu$,$\sigma$,$\lambda$,$\mu_J$,$\sigma_J$,$\gamma$
GBM,5.12E-04,2.63E-02,,,,
CJD,4.99E-04,2.32E-02,1.51E-02,,,-9.34E-02
LJD,5.81E-04,1.77E-02,1.69E-01,-3.65E-03,4.63E-02,
\end{filecontents*}
\csvautotabular{data/mle_params.csv}
    \end{center}
        \caption{
            MLE estimated parameters
        }
        \label{table: mle params}
\end{table}

\subsection{The Esscher pricing interval}\label{subsection: pricing interval}

With the theoretical models fitted to the empirical log-returns, we calculate European option prices under the second-order Esscher measure. To determine the Esscher pricing interval we herein vary the second-order parameter denoted by $\psi$.
	We focus on the constant jump-diffusion (CJD) model for which the price is given by the infinite series in   equation \eqref{Esscher_Price_LevyConstant}. 
	For the numerical evaluation a finite series approximation with $n$ terms is employed. 
		Hereto note that according to \eqref{eq: Fn(Lambda)} 
	\[ 0\leq f_k(\Lambda)\leq x\exp( -\Lambda e^{\gamma}\tau) \frac{(\Lambda e^{\gamma}  \tau)^k}{k!} \rightarrow 0 \quad \mbox{for} \;k\rightarrow +\infty.
	\]
	
We analyse the number of terms $n$ to be fixed while truncating the remaining terms 
	when varying the second-order Esscher parameter $\psi$.
	
	As an example, we consider the model and option parameters given in Table \ref{table: Esscher interval params} where some of the values were taken from Table \ref{table: mle params} above.

\begin{table}[H]
	\begin{center}
		\begin{filecontents*}{data/Esscher_pricing_interval_params.csv}
model parameters,$\mu$,$\sigma$,$\lambda$,$\gamma$,$\mu_J$,$\sigma_J$
,4.99E-04,2.32E-02,1.51E-02,-9.34E-02,-9.34E-02,4.63E-02
,,,,,,
option parameters,$S_0$,$K$,$r$,$T$ (days),,
,\$60,\$60,0.000136986,63,,
\end{filecontents*}
\csvautotabular{data/Esscher_pricing_interval_params.csv}
	\end{center}
	\caption{
		Parameters for the Esscher pricing interval
	}
	\label{table: Esscher interval params}
\end{table}

\begin{figure}[H]
    \centering
    \includegraphics[width=.65\textwidth]{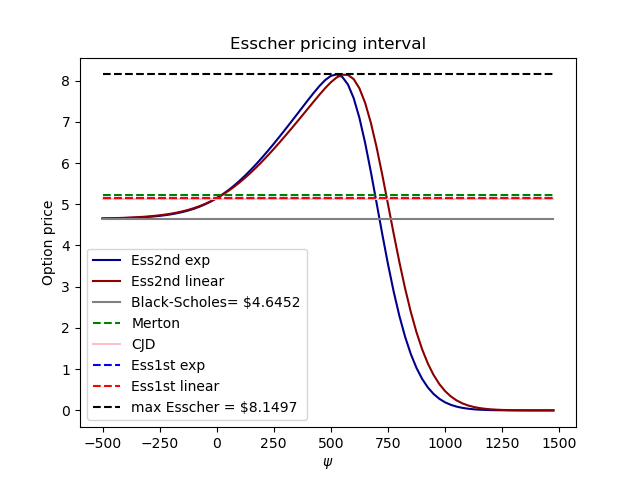}
    \caption{Esscher pricing interval}
    \label{fig: interval}
\end{figure}

We start by considering the number of jumps to be $n=10$. This number is motivated by \cite{kou2002jump} where the author calculates the price of an option with time to expiry of six months and suggests that numerically only the first 10 to 15 terms in the series are needed for most applications. While this is true for models such as Kou's model and Merton's model, it doesn't hold true for our case with a free parameter. Indeed, large values of $\psi$ yield inaccurate prices  contradicting the monotonicity with respect of the  parameter $\psi$  
in  Corollary  \ref{corollary: CJD}. In Figure \ref{fig: interval} for example, we observe first an increase of the option price with  increasing parameter values for  $\psi$ followed by a decrease to the value zero.

The behavior of the option price for this fixed number $n=10$  when $\psi$ and hence, according to \eqref{Psi_limits}, $\Lambda$ tends to infinity can be explained  as follows. 
	Consider the finite series $\Sigma_n(\Lambda)$ in \eqref{eq: Fn(Lambda)}, then for any $k\in \{0,\ldots,n\}$ we find
	\[
	0\leq f_k(\Lambda)\leq x\exp( -\Lambda e^{\gamma}\tau) \frac{(\Lambda e^{\gamma}  \tau)^k}{k!} \rightarrow 0 \quad \mbox{for} \;\Lambda\rightarrow +\infty
	\]
	and thus also  $\Sigma_n(\Lambda) \rightarrow 0$.

\begin{figure}[!htbp]
    \centering
    \includegraphics[width=.75\textwidth]{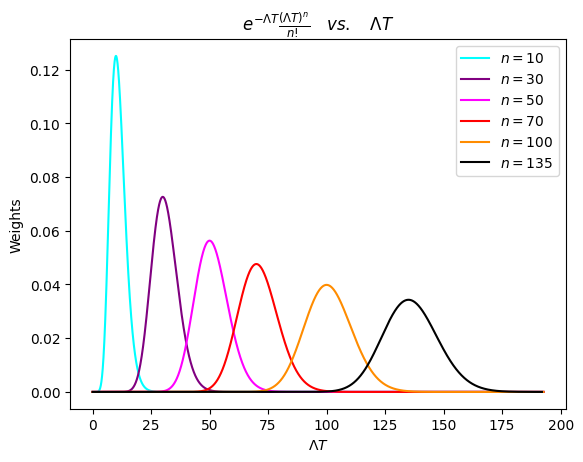}
    \caption{Weights vs. $\Lambda T$}
    \label{fig: weights}
\end{figure}

Thus, the remaining question is, \textit{how many terms to consider while increasing $\Lambda$ or equivalently $\psi$} and related to it \textit{how large is the Esscher pricing interval for a range of $\psi$}?\\ 
The option price $C^{EE}$ can be seen as an infinite weighted average of Black-Scholes prices with weights $ \frac{e^{-\Lambda \tau} (\Lambda \tau)^k}{k!}$ for $k>0$ depending on $\Lambda$ and hence on $\psi$. We analyse these weights as functions of $\Lambda$. To simplify the notations we denote $w(x):= \frac{e^{-x} x^k}{k!}$ and consider $x>0$, then it easily follows that 
\begin{align*}
	w'(x)& =w(x)(\frac k x-1) \implies w'(x)=0 \iff x=k\\
	w''(x) &= w'(x)(\frac k x-1)-w(x) \frac{k}{x^2}  \implies w''(k)=-\frac{w(k)}{k} <0.
\end{align*}
Thus, $w$ is a concave function of $x$ which reaches its maximum at $x=k$ and which tends to zero for $x$ tending to zero respectively to $+\infty$. This implies that we should choose $n$ in relation to the value of $\psi$ such that  the terms with the highest weights are included. This implies that when increasing $\psi$, and hence $\Lambda$, more terms in the expansion should be used. This can also be financially motivated as $\psi$ is related to measuring the jump risk. Illustration of the behavior of the weights is given in Figure \ref{fig: weights}.\\

\begin{figure}[!htbp]
    \centering
    \includegraphics[width=.75\textwidth]{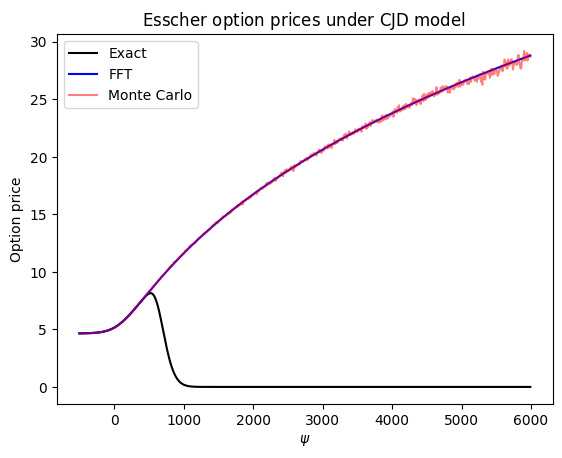}
    \caption{Exponential Esscher prices before adding extra terms}
    \label{fig: compare pricing methods n=10}
\end{figure}

\begin{figure}[H]
    \centering
    \includegraphics[width=.75\textwidth]{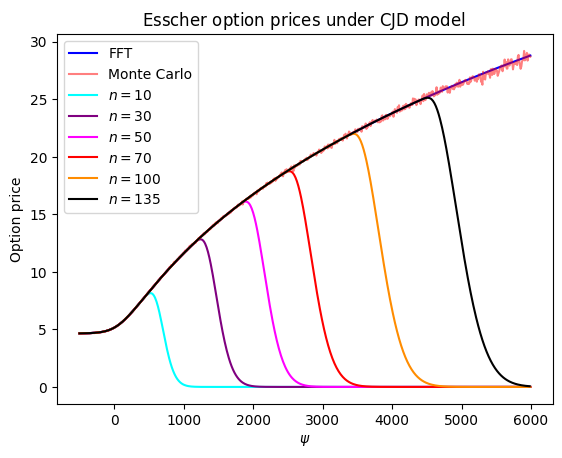}
    \caption{Exponential Esscher prices after adding extra terms}
    \label{fig: compare pricing methods n different}
\end{figure}

For extra validity of the Esscher pricing interval, we compare the pricing interval obtained by equation (\ref{Esscher_Price_LevyConstant}) with prices obtained by alternative methods such as Fast Fourier transform and Monte Carlo. The result of this comparison is illustrated in Figure \ref{fig: compare pricing methods n=10} when we fix $n$ to $10$ and in Figure \ref{fig: compare pricing methods n different} when we add extra terms to the series in accordance to  the previous argument.

We can conclude that when increasing $\psi$ more terms have to be taken into account in the finite series approximation of the option price \eqref{Esscher_Price_LevyConstant} and that the pricing interval with the  Black-Scholes price as a lower bound becomes larger.\\

Next, we study how the Esscher pricing interval for the CJD model compares to option prices under other known models in the literature.
%
%
\begin{figure}[!htbp]
    \centering
    \includegraphics[width=.8\textwidth]{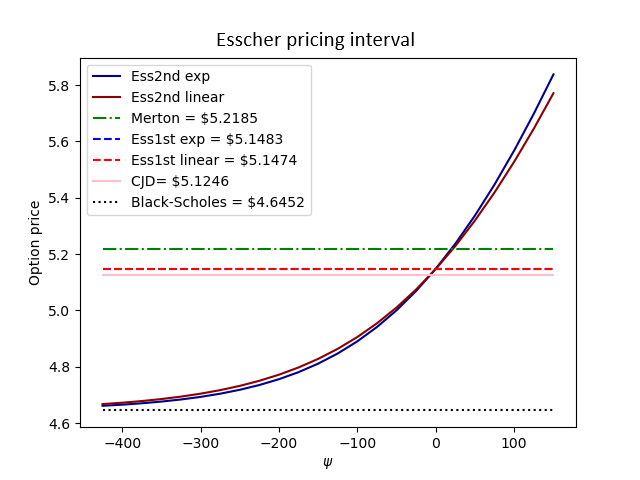}
    \caption{Esscher pricing interval vs.\ other models}
    \label{fig: interval zoom}
\end{figure}
%
%
In Figure \ref{fig: interval zoom}, we consider a pricing interval corresponding to a choice for $\psi$ in $[-425, 150]$. 
This interval appears sufficiently wide, as it encompasses the prices of the models of interest when compared with the second‑order Esscher price.
We observe that, as $\psi$ decreases, the option price approaches the Black–Scholes price, in agreement with Corollary \ref{corollary: CJD}. 
The model named CJD in Figure \ref{fig: interval zoom}, is a jump-diffusion model with constant jump and where the jump risk is not priced (meaning that the market price of risk is the same as Black-Scholes' one, i.e., $\eta = (\mu - r)/\sigma^2$). Thus, due to the presence of jumps (that adds more uncertainty to the model), it leads to a higher price compared to the Black-Scholes price. However, it has a lower value compared to a jump-diffusion model with constant jump when the jump risk is priced using the classical first-order Esscher. Finally, we can see that even though the Merton's jump-diffusion model does not price the jump risk, it has a higher price compared to the constant jump-diffusion model 
with the jump risk priced by the classical Esscher.
This is due to the extra uncertainty coming from the random jump sizes.

\subsection{Stress Testing: CJD \texorpdfstring{$\psi$}-Variation vs. Benchmarks}
In this subsection, we use the second-order Esscher framework to perform a stress-testing exercise for delta-hedging performance. In standard risk-management practice, stress testing refers to evaluating portfolio losses under extreme but plausible market conditions, complementing traditional risk measures such as Value-at-Risk (VaR) and Expected Shortfall (ES). Within our model, such stressed environments arise when the second-order Esscher parameter $\psi$ takes sufficiently large values, as $\psi$ directly influences the effect of the jump intensity $\lambda^\psi$. 
Empirical evidence suggests that values of $\psi$ sufficiently above zero correspond to regimes characterized by markedly  elevated jump risk, thereby representing stressed market conditions in our setting (see Section \ref{section: fear}). Consequently, for the stress‑testing experiment, we consider $\psi\in [-400, 400]$, for which $n=10$ remains an admissible choice in light of the discussion in the preceding subsection.\\

\begin{figure}[!htbp]
	\centering
	\includegraphics[scale=0.95]{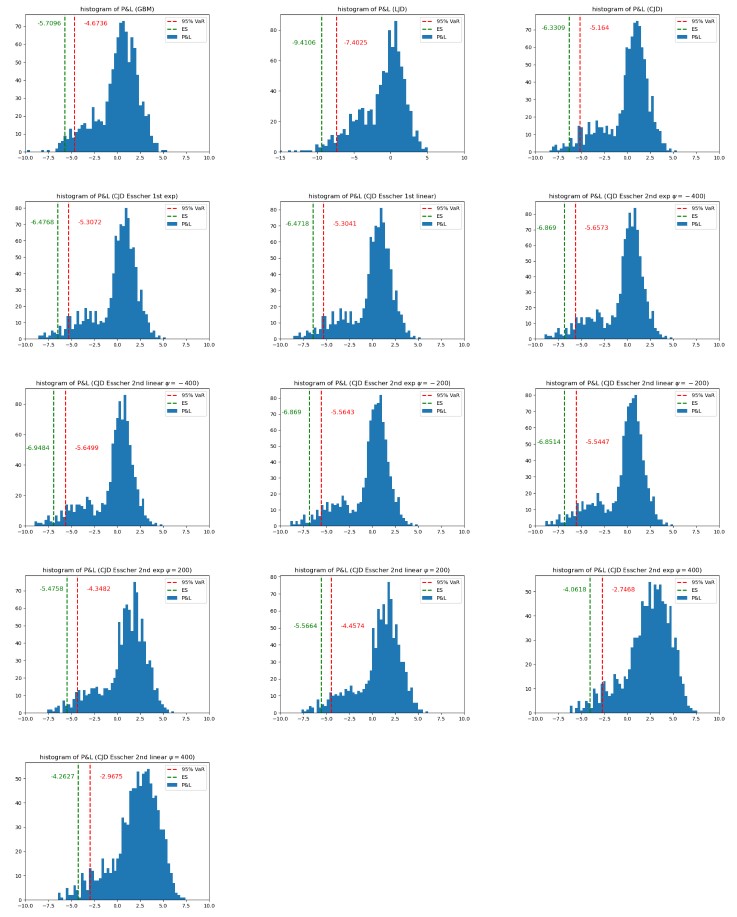}
	\caption{Hedging errors}
	\label{fig: pnl} 
\end{figure}

Using the MLE estimated parameters given in Table \ref{table: mle params}, we follow a similar idea to that in \cite{kuen2014hedging}, where we simulate P\&L of a delta-hedging strategy under each different model and compare 5\% VaR and ES. 
The model classes that we consider are geometric Brownian motion (GBM), constant jump-diffusion (CJD) and log-normal jump-diffusion (LJD). 
In addition, 
we  evaluate how VaR and ES vary across different choices of $\psi$. This allows us to illustrate how the second-order Esscher method naturally generates a range of risk outcomes, including those associated with stressed jump-risk scenarios. The distributions of the hedging errors for the considered models are illustrated in Figure \ref{fig: pnl}.
%
%
\begin{table}[!htbp]
    \begin{center}
        \begin{filecontents*}{data/Var_ES.csv}
model,Value-at-risk,Expected shortfall
GBM,-4.6736,-5.7096
LJD,-7.4025,-9.4106
CJD (no jump risk),-5.164,-6.3309
CJD $\psi = -400$ (exp),-5.6573,-6.869
CJD $\psi = -400$ (linear),-5.6499,-6.9484
CJD $\psi = -200$ (exp),-5.5643,-6.869
CJD $\psi = -200$ (linear),-5.5447,-6.8514
CJD $\psi = 0$ (exp),-5.3072,-6.4768
CJD $\psi = 0$ (linear),-5.3041,-6.4718
CJD $\psi = 200$ (exp),-4.3482,-5.4758
CJD $\psi = 200$ (linear),-4.4574,-5.5664
CJD $\psi = 400$ (exp),-2.7468,-4.0618
CJD $\psi = 400$ (linear),-2.9675,-4.2627
\end{filecontents*}
\csvautotabular{data/Var_ES.csv}
    \end{center}
        \caption{
            5\% VaR and ES
        }
        \label{table: VaR}
\end{table}
Table \ref{table: VaR}, lists the VaR and ES of the several models (GBM, CJD, LJD).
There are several cases under the CJD model all of which depend on how we choose the pricing measure. For example, the case of no jump risk means that the pricing measure for this case doesn't consider jump risk (similar to Merton's assumption) and the market price of risk is the same as the Black-Scholes' one, i.e., $\eta = (\mu - r)/\sigma^2$. Other cases of the CJD model are under the second-order Esscher measure except for the case when $\psi =0$ which retrieves the classical first-order Esscher. \\

To interpret the values in Table \ref{table: VaR}, we take for example the case of the GBM. The VaR for the GBM model, means that there is a 5\% probability that the portfolio will fall in value by more than \$4.67 over the life of the option. The expected shortfall measures the average loss that exceeds the 5\% VaR threshold. Thus, for the GBM case, it means that on average, losses exceeding the 5\% VaR value is \$5.71.

We observe that varying $\psi$  induces a range of VaR and ES values within the same model specification.
This flexibility allows risk managers to evaluate and manage risk exposure across various pricing scenarios of the option. Additionally, based on other internal risk considerations, such as liquidity constraints, operational risks, credit risk, etc., risk managers can select a $\psi$ that yields a risk measurement aligning with their risk appetite and these internal considerations. This approach enables a more tailored and comprehensive risk assessment strategy, beyond what is specified by the model dynamics alone. 

Further, we  notice that the VaR and ES of the LJD model are higher compared with the GBM and CJD models. This is due to the dynamics of the return process where the jumps are normally distributed with mean $\mu_J$ and variance $\sigma_J^2$. This allows this LJD model to incorporate extreme values of the underlying  arising from jump dynamics that are not captured by the constant jump component of the constant jump‑diffusion model.

\section{Fear quantification}\label{section: fear}

Classical option pricing assumes risk‐neutral agents, yet observed market prices often diverge from theory especially during periods of stress.  These deviations may reflect investor sentiment—fear, optimism, or uncertainty—modulating prices beyond purely model dynamics.  We investigate whether the second‐order Esscher parameter $\psi$ captures this behavioral component.  By calibrating $\psi$ to real‐world S\&P\,500 option prices and comparing it against sentiment proxies, we aim to unveil a sentiment‐adjusted mechanism within risk‐neutral valuation.
	
	Recall that the free parameter $\psi$  plays a critical role in shaping the risk-neutral measure used for derivative pricing. By modifying the exponential tilting of the likelihood ratio, $\psi$  directly influences the form of the risk-neutral density, thereby enabling the pricing kernel to respond more flexibly to prevailing market conditions. Importantly, $\psi$  also affects the market price of risk $\theta$, which is no longer determined solely by the structural parameters of the model. Instead, $\theta$  becomes a function of both the model dynamics and the value of $\psi$, making it more responsive to pricing discrepancies. Given this dual influence—on both the measure transformation and the risk premium—it is essential to investigate how $\psi$  interacts with market sentiment and whether it captures latent behavioral or regime-dependent effects that are not observable through traditional model parameters alone.
	
	In particular, we explore whether $\psi$ encodes market fear by calibrating the constant jump-diffusion (CJD), the log-normal jump-diffusion (LJD) and the variance gamma (VG) models to S\&P\,500 option prices over January 2019–December 2020, extracting daily $\psi$ values and comparing them to both news‐tone sentiment indices and the VIX.  Our analysis reveals regime‐dependent volatility clustering in $\psi$ that align with shifts in investor mood. These findings suggest $\psi$ serves as a behavioral amplifier for jump risk under stress, bridging quantitative pricing and sentiment dynamics.

\subsection{Data Description}

\begin{figure}[H]
  \centering
  \includegraphics[width=1\linewidth]{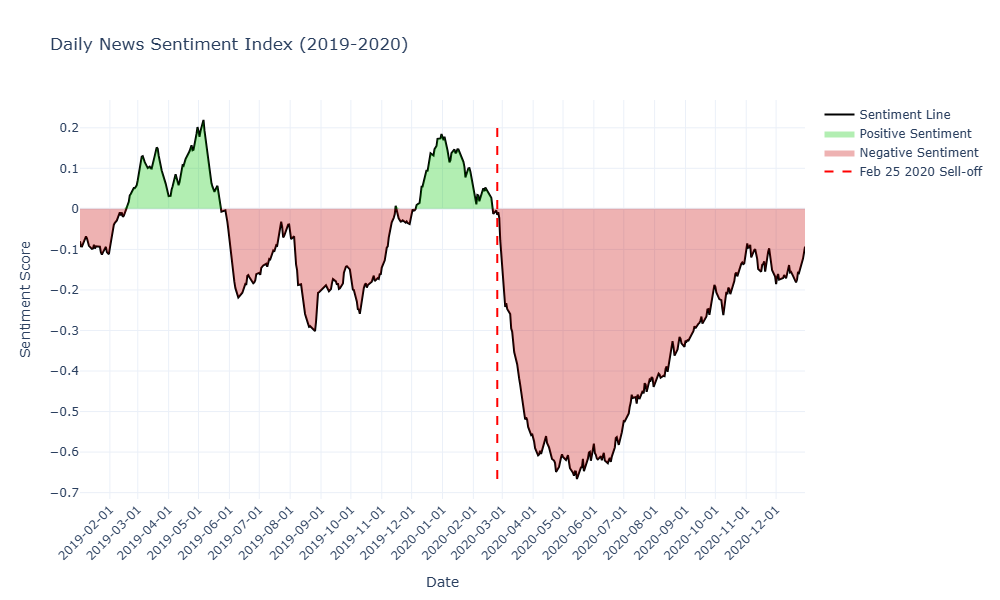}
  \caption{Daily News Sentiment Index.}
  \label{fig: sentiment_score_timeseries}
\end{figure}

To capture the dynamics of distinct market regimes, we focus on the period spanning early 2019 through the end of 2020. This interval encompasses several notable phases: a phase of stable growth and low volatility in early 2019; heightened uncertainty in late 2019 associated with the U.S.–China trade conflict; the pronounced market dislocation and volatility spikes during the first and second quarters of 2020 amid the COVID-19 pandemic; and, finally, the gradual recovery in the latter half of 2020 driven by vaccine-related optimism. For the empirical analysis, we employ data from the S\&P 500 Index, S\&P 500 Index option prices (SPX), the VIX Index, and the Daily News Sentiment Index compiled by the Federal Reserve Bank of San Francisco. Although multiple providers offer sentiment measures, the San Francisco Fed’s index was selected primarily for its accessibility and methodological transparency. This index delivers daily sentiment scores derived from lexical analysis of 24 major U.S. newspapers, spans the full 2019–2020 horizon, and incorporates macro-oriented filtering and smoothing techniques to mitigate noise. Figure \ref{fig: sentiment_score_timeseries} presents the sentiment scores over this period, with positive values reflecting optimism and negative values reflecting pessimism.

The vertical dashed line in Figure \ref{fig: sentiment_score_timeseries} marks the point at which localized uncertainty surrounding COVID-19 escalated into widespread panic. While early signs of stress appeared around February 20, 2020, the most pronounced inflection occurred on February 25. The sell-off was compounded by a collapse in oil prices to year-long lows and unprecedented troughs in U.S. Treasury yields—1.31\% for the 10-year and 1.80\% for the 30-year. Governments responded with emergency fiscal measures, such as Indonesia’s \$742 million stimulus package, yet investor sentiment continued to deteriorate as COVID-19 cases surged outside China and global supply chains faced severe disruption \href{https://en.wikipedia.org/wiki/2020_stock_market_crash}{Wikipedia}. Reports of rapid viral spread in South Korea, Italy, and Iran further intensified anxiety, triggering mass sell-offs across Asia-Pacific and European markets. The severity of the downturn was underscored by cumulative two-day losses ending February 25: the Dow declined 6.59\%, the S\&P 500 fell 6.28\%, and the Nasdaq dropped 6.38\%—the steepest such declines since 2015–2018 \href{https://www.cnbc.com/2020/02/25/stock-market-today-live.html}{CNBC}.

Taken together, these developments mark a decisive shift in market sentiment from localized concern to systemic risk. Accordingly, we anchor our temporal split at February 25, 2020, to distinguish the initial reaction phase from the second-wave COVID-19 stress period. This partition provides a clearer framework for analyzing investor behavior and asset dynamics before and after the onset of global panic.

The option dataset complements the sentiment and index measures by providing the necessary derivative market inputs for calibration. Option data is sourced from OptionMetrics, with maturities fixed in the 28–32 day window to ensure a consistent pricing environment across trading days. This choice reflects standard exchange conventions, where monthly contracts typically expire on the third Friday of the month, resulting in slight variations in calendar duration for “one‑month” options. To mitigate illiquidity, contracts with zero trading volume are excluded, and a minimum open interest threshold of 100 is imposed, a benchmark widely used in the literature to capture meaningful market participation and tighter bid–ask spreads. Strike prices are restricted to the near‑the‑money range, defined by moneyness $\frac{K}{S}\in [0.9,1.1]$, since deep in‑the‑money or far out‑of‑the‑money contracts often suffer from poor liquidity and unreliable pricing. This focus on the ATM region also aligns with the requirements of the \cite{breeden1978} framework and related risk‑neutral density extraction methods, which rely on accurate curvature around the strike. Market option prices are computed as the midpoint between bid and ask quotes, thereby reducing the influence of spread noise and ensuring cleaner input data for calibration and analysis.

\subsection{Methodology}
We decouple structural and behavioral parameters in a two‐step procedure:
Unlike the structural model parameters $\Theta := (\mu, \sigma, \lambda, \mu_J, \sigma_J)$, which are statistically grounded in historical asset behavior, the second Esscher transform parameter $\psi$ arises exclusively from the construction of an equivalent martingale measure. Consequently, $\psi$ lacks a historical footprint and manifests solely within the option market, reflecting latent factors such as supply-demand imbalances, trader sentiment, and macroeconomic uncertainty.

To explore this phenomenon, we deviate from the conventional approach in the literature that jointly calibrates all model parameters to option prices. Instead, we seek to decouple $\psi$ from $\Theta$ and interpret it as a "sentiment-shadow" variable—an abstract quantity that encapsulates market mood. This is implemented via a two-step procedure:

\paragraph{1. Historical MLE of Structural Parameters}  
The first step is to estimate the model parameters $\Theta$. This is done by performing the maximum likelihood estimation based on historical asset returns. Let $D = \{S(0), S(1), S(2), \cdots , S(M)\}$ denote the the realizations of stock price at equally-spaced times $k=0, 1,2, \cdots , M$. The one period rate of return $x_i = \ln S(i) - \ln S(i-1)$ is i.i.d. Then the MLE estimator is defined as follows:
$$\hat{\Theta} = \arg\max_{\Theta} \, \mathcal{L}(\Theta; \{x_i\}_{i=1}^{M}) = \arg\max_{\Theta} \, \prod_{i=1}^{M} f(x_i \mid \Theta)$$
where $\mathcal{L}$ is the likelihood function and $f(x_i \mid \Theta)$ denotes the probability density of the return $x_i$ given parameters $\Theta$.

Model parameters are estimated on a daily basis using a two-year rolling window spanning the period from January 2, 2019, to December 30, 2020. Specifically, each day’s estimation utilizes data from the preceding two years; for instance, parameters on January 2, 2019, are derived from observations between January 2, 2017, and January 1, 2019. The choice of a two-year window is because it have enough data points for a stable MLE, it captures short to mid-term regimes and it avoids including outdated structural patterns.

\paragraph{2. Calibration of \(\psi\) via Penalized RMSE}

Once $\hat{\Theta}$ are obtained, these values are substituted into the theoretical pricing formula $C^{\text{model}}(\psi; \hat{\Theta})$. Then, to calibrate the second Esscher parameter $\psi$ using observed option prices, the optimization objective is typically framed as a minimization of the root mean square error (RMSE) between market prices and model-implied values as follows:
$$\hat{\psi} = \arg\min_{\psi} \, \sqrt{ \frac{1}{N} \sum_{i=1}^{N} \left( C_i^{\text{model}}(\psi; \hat{\Theta}) - C_i^{\text{market}} \right)^2 }$$
where $C_i^{\text{market}}$ and $C_i^{\text{model}}$ represent the observed and theoretical option prices, respectively, for contract $i$. However, calibration may suffer from the drawback associated with the ill-posedness of of some inverse pricing problems, hence, instability may occur. To mitigate this issue, \textit{Ridge penalization} (also known as $L_2$ regularization) is introduced into the calibration objective. This regularization technique imposes a quadratic penalty on the magnitude of $\psi$, effectively discouraging extreme parameter values and stabilizing the estimation landscape. 
The chosen penalization coefficient, $\alpha = 10^{-8}$, is sufficiently small to preserve the fidelity of the RMSE fit while promoting smoother optimization convergence. It acts as a soft constraint rather than a rigid prior, ensuring that $\psi$ remains in a statistically and economically plausible range. Therefore, the calibration problem becomes.
$$
\hat{\psi} = \arg\min_{\psi} \left[ \sqrt{ \frac{1}{N} \sum_{i=1}^{N} \left(  C_i^{\text{model}}(\psi; \hat{\Theta}) -C_i^{\text{market}} \right)^2 } + \alpha \psi^2 \right]
$$

\subsection{Empirical Results}
In this section, we present the outcomes of the calibration procedure and examine the empirical behavior of the second-order Esscher parameter $\psi$ . For each of the CJD, LJD, and VG models, calibrated values of $\psi$  are obtained and analyzed through scatter plots against both the market sentiment index and the VIX. Time-series plots are then constructed to track the evolution of $\psi$  across dates, allowing direct comparison with sentiment dynamics and volatility levels. Since the market price of risk is a function of $\psi$ , we extend this analysis by plotting its trajectory under each model to investigate regime-dependent patterns. Finally, option prices computed using the calibrated $\psi$  are compared with observed market option prices and with valuations derived from the first-order Esscher framework, highlighting the improvements in pricing accuracy and interpretability afforded by the second-order approach.

\subsubsection{Market price of variance risk \texorpdfstring{$\psi$} vs. VIX and market sentiment}

\begin{figure}[H]
  \centering
  \includegraphics[width=1\linewidth]{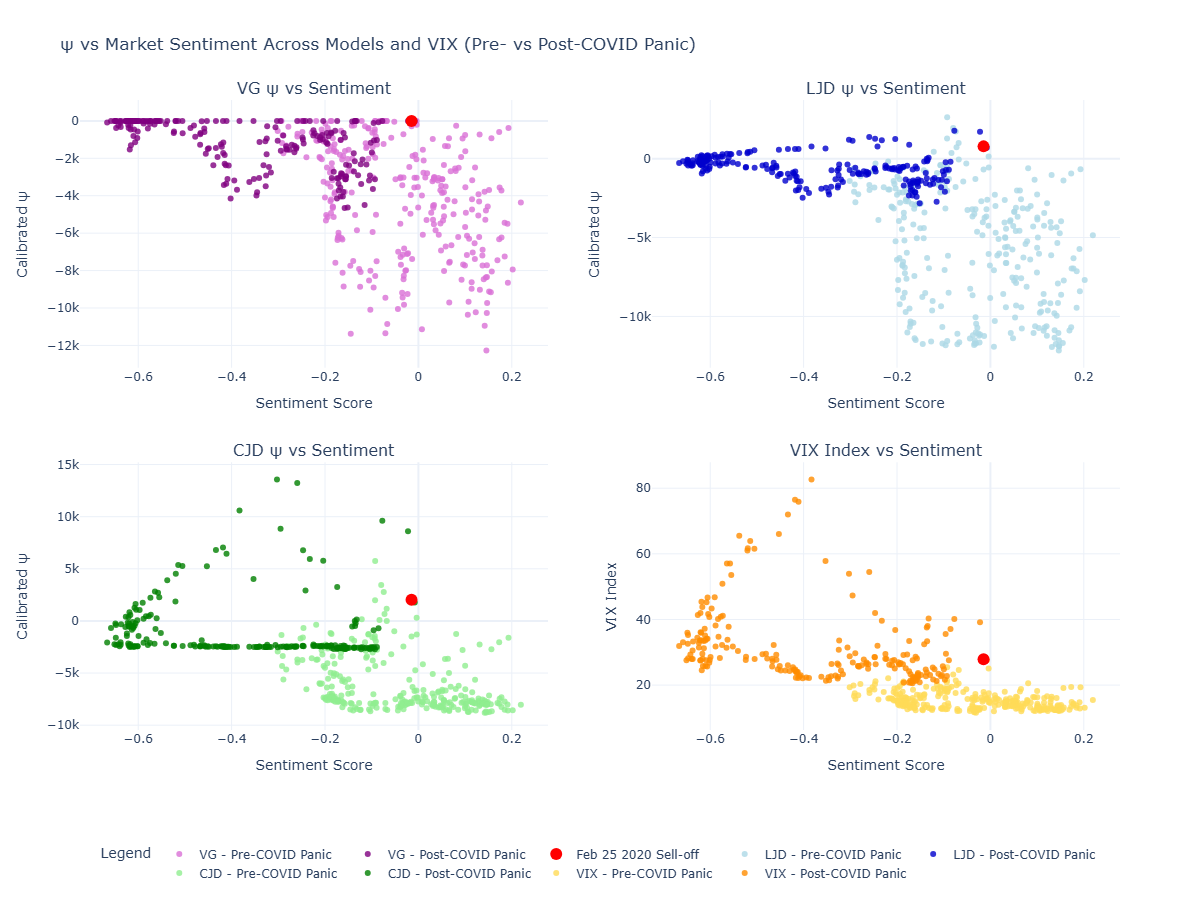}
  \caption{Calibrated $\psi$ vs.\ Daily News Sentiment Index.}
  \label{fig: all models vs sentiment scatter}
\end{figure}

\begin{figure}[H]
  \centering
  \includegraphics[width=1\linewidth]{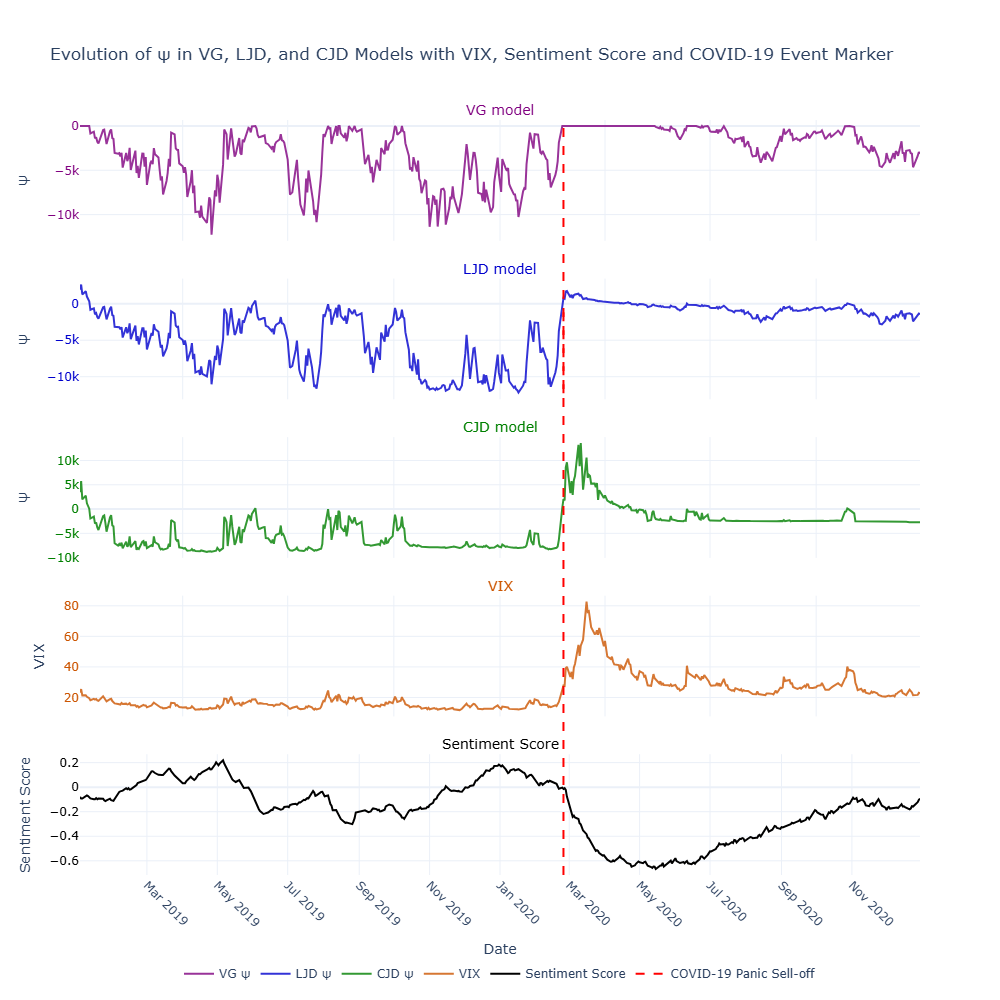}
  \caption{Calibrated $\psi$ across time for all models.}
  \label{fig: all models vs sentiment timeseries}
\end{figure}

\begin{figure}[H]
  \centering
  \includegraphics[width=1\linewidth]{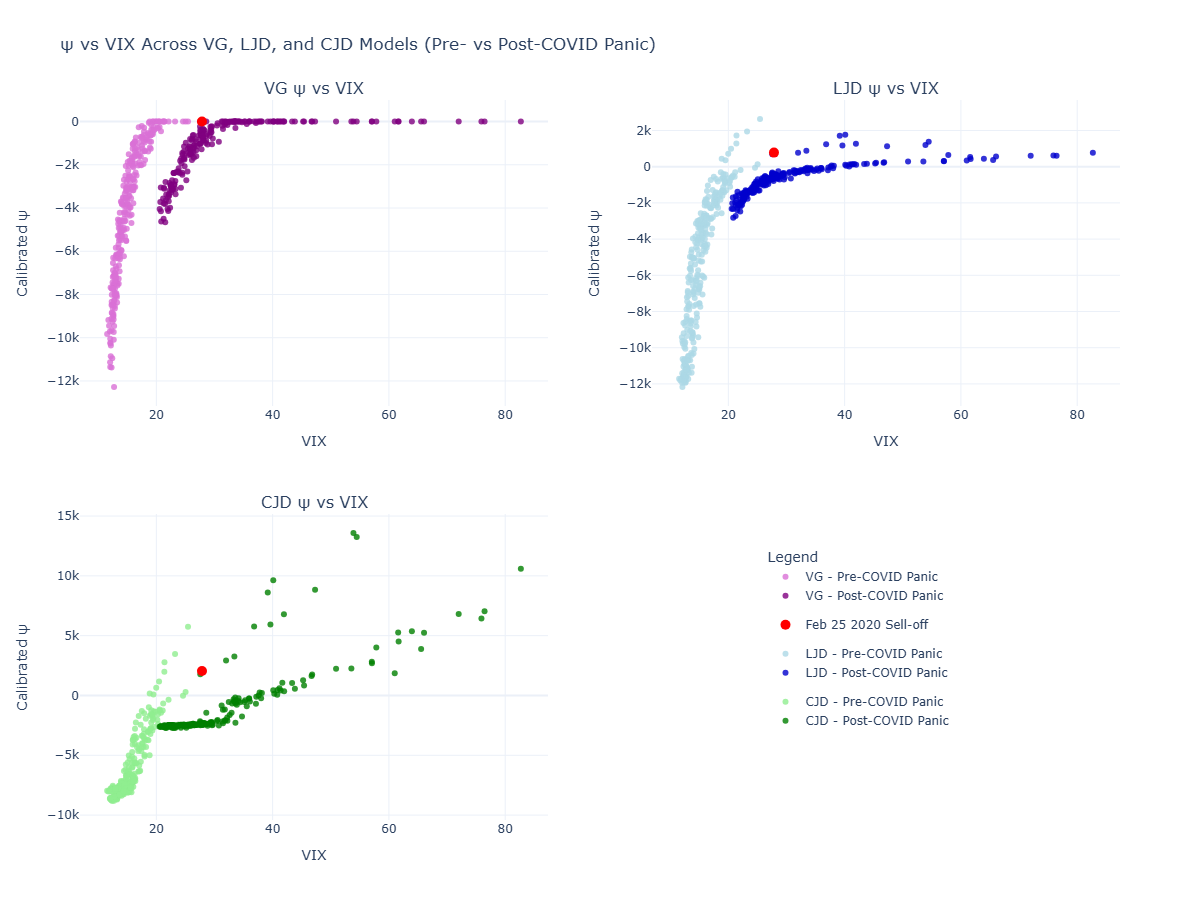}
  \caption{Calibrated $\psi$ vs. VIX Index.}
  \label{fig: all models vs vix}
\end{figure}

Figures \ref{fig: all models vs sentiment scatter}, \ref{fig: all models vs sentiment timeseries}, and \ref{fig: all models vs vix} jointly illustrate the empirical relationship between the calibrated second-order Esscher parameter $\psi$ , the Daily News Sentiment Index, and the VIX index across three models—VG, LJD, and CJD—using complementary visualization techniques. The scatter plots in Figure \ref{fig: all models vs sentiment scatter} reveal a regime-dependent inverse association: lower sentiment scores, indicative of market pessimism and stress, tend to correspond with higher values of $\psi$ . This pattern suggests that $\psi$  effectively captures latent behavioral risk, intensifying during periods of negative sentiment. The clustering of data points into pre- and post-COVID regimes further highlights the structural shift in investor behavior, with tighter groupings during stressed conditions and more dispersed values during optimistic periods.

Figure \ref{fig: all models vs sentiment timeseries} complements this analysis by presenting the evolution of $\psi$  for each model alongside both the sentiment score and the VIX index. The time-series plots reinforce the regime separation observed in the scatter plots, with sharp increases in $\psi$  coinciding with drops in sentiment—particularly around the COVID-19 panic sell-off marked on February 25, 2020. Notably, $\psi$  and VIX exhibit similar patterns, both spiking during periods of heightened market stress and receding as sentiment improves. However, while VIX reflects implied volatility extracted directly from option prices, $\psi$ modifies the pricing kernel to incorporate sentiment-sensitive distortions, thereby providing a structurally distinct lens on market stress within the second-order Esscher framework. This reinforces the interpretation of $\psi$  as a behavioral stress proxy that complements traditional volatility measures.

Figure \ref{fig: all models vs vix} further deepens this interpretation by directly comparing $\psi$  to the VIX index across models. Unlike the sentiment-based scatter plots, these plots exhibit clearer and more nonlinear patterns, with $\psi$  rising sharply with VIX before plateauing—particularly in the VG and LJD models. The clustering of pre- and post-COVID data is more pronounced, suggesting that $\psi$  responds more systematically to volatility than to sentiment alone. This nonlinear structure implies that $\psi$  may encode a threshold-like sensitivity to market volatility, where behavioral distortions intensify rapidly once VIX exceeds certain levels. Unlike sentiment scores, which capture qualitative shifts in investor mood, VIX quantifies expected future volatility derived from option prices. The stronger relationship between $\psi$  and VIX indicates that $\psi$  internalizes volatility dynamics such as clustering and regime shifts, adjusting the pricing kernel to reflect elevated market price of risks. In this way, $\psi$  complements traditional risk indicators like VIX by embedding behavioral distortions directly into model pricing, offering a structurally grounded measure of market stress that bridges both statistical volatility and sentiment-driven pricing asymmetries.

Together, these plots demonstrate that $\psi$  not only reflects regime-dependent market stress but also tracks both sentiment-driven and volatility-driven distortions in option pricing. The VIX index, shown across all figures, exhibits a consistently negative correlation with sentiment and a nonlinear association with $\psi$, reinforcing its role as a fear gauge and validating the behavioral interpretation of $\psi$  within the second-order Esscher framework.

\begin{figure}[H]
  \centering
  \includegraphics[width=1\linewidth]{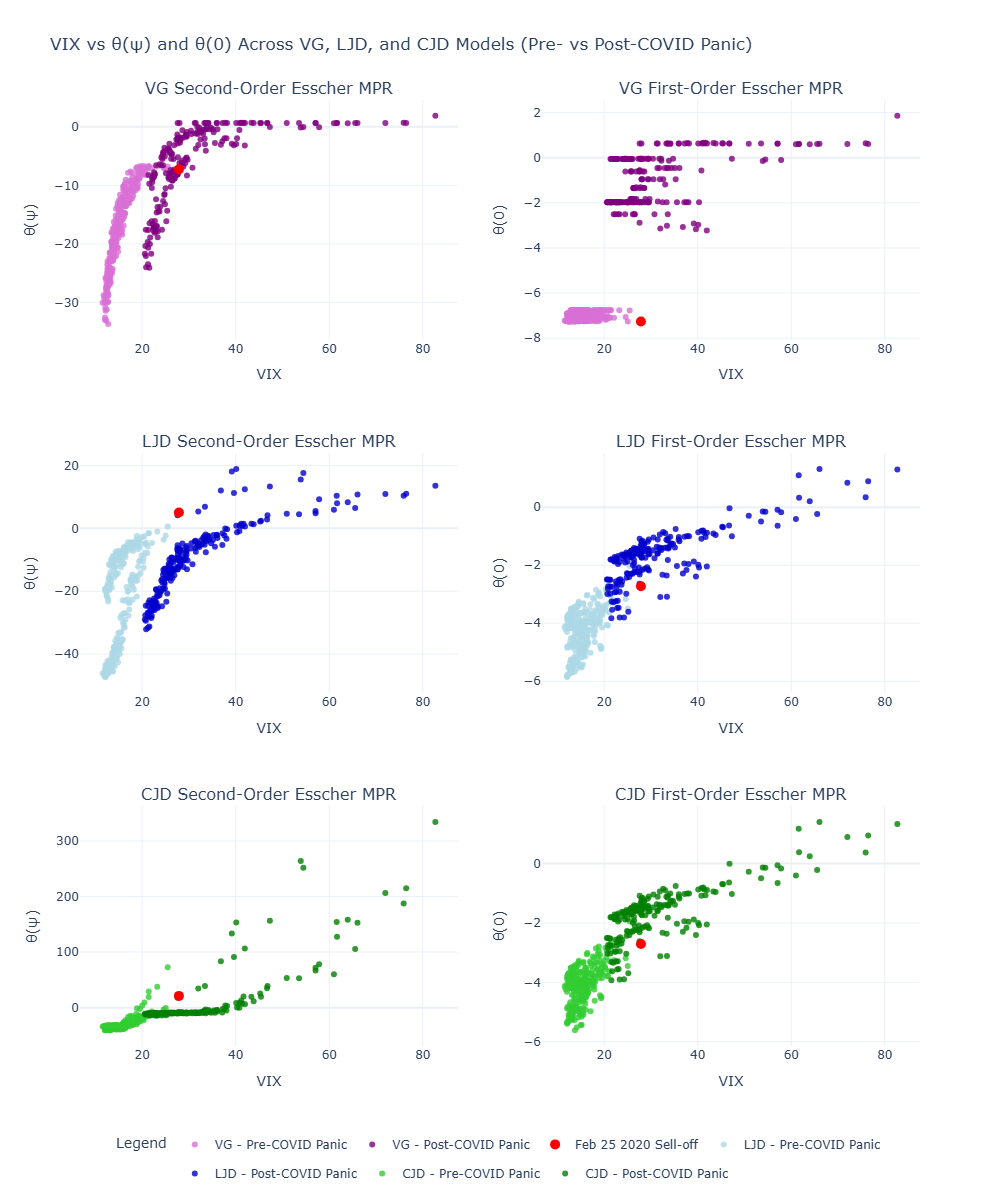}
  \caption{Market Price of Risk for First and Second Order Esscher vs. VIX.}
  \label{fig: all models mpr scatter}
\end{figure}

\paragraph{Modeling VIX in terms $\psi$}

This subsection explores the relationship between the calibrated distortion parameter $\psi$ and the VIX index by expressing VIX in terms of model parameters and $\psi$. Following the formulation in \cite{ballotta2024vix},  the VIX under a risk-neutral measure $\mathbb{Q}$ for a L\'evy-driven asset process is given by

\begin{equation}\label{eq: model for vix}
    \mathcal{V}(0,T) = \sqrt{\sigma^2 + 2\int \left(e^x - 1 -x \right)\nu^{\mathbb{Q}}(dx)}
\end{equation}

where the asset price evolves as

\begin{equation}
    S_t = S_0 \exp \left( (r-\varphi^\mathbb{Q}(-i))t + L_t \right)
\end{equation}
with $L_t$ denoting a L\'evy process and $\varphi^\mathbb{Q}$ the characteristic exponent under the risk-neutral measure
\begin{equation}
    \varphi^\mathbb{Q} (u) = iu\mathbb{E}^\mathbb{Q}[L_1] - \frac{u^2\sigma^2}{2} + \int \left(e^{iux} - 1 -iux \right)\nu^\mathbb{Q}(dx)
\end{equation}

When the risk-neutral measure is the second-order Esscher measure, the L\'evy measure $\nu^\mathbb{Q}(dx)$ becomes $e^{\theta x + \psi x^2}\nu(dx)$, allowing the VIX to be expressed directly as a function of $\psi$. For the three considered momdels -CJD, LJD, and VG- the resulting expressions are:

\begin{equation*}
    \begin{split}
        \mathcal{V}_{cjd}(\psi) &= \sqrt{\sigma^2_{cjd} + 2\lambda\left(e^\gamma - 1 -\gamma \right)e^{\theta \gamma + \psi \gamma^2}}\\
        \mathcal{V}_{ljd}(\psi) &=\sqrt{\sigma^2_{ljd} + 2\lambda\int \left(e^x - 1 -x \right)e^{\theta x + \psi x^2}F_{ljd}(dx)}\\
        \mathcal{V}_{vg}(\psi) &= \sqrt{ 2\int \left(e^x - 1 -x \right)e^{\theta x + \psi x^2}\nu_{vg}(dx)}
    \end{split}
\end{equation*}

where $F_{ljd}(dx)$  and $\nu_{vg}(dx)$ denote the jump size distributions specific to the LJD and VG models, respectively.

\begin{figure}[H]
  \centering
  \includegraphics[width=1\linewidth]{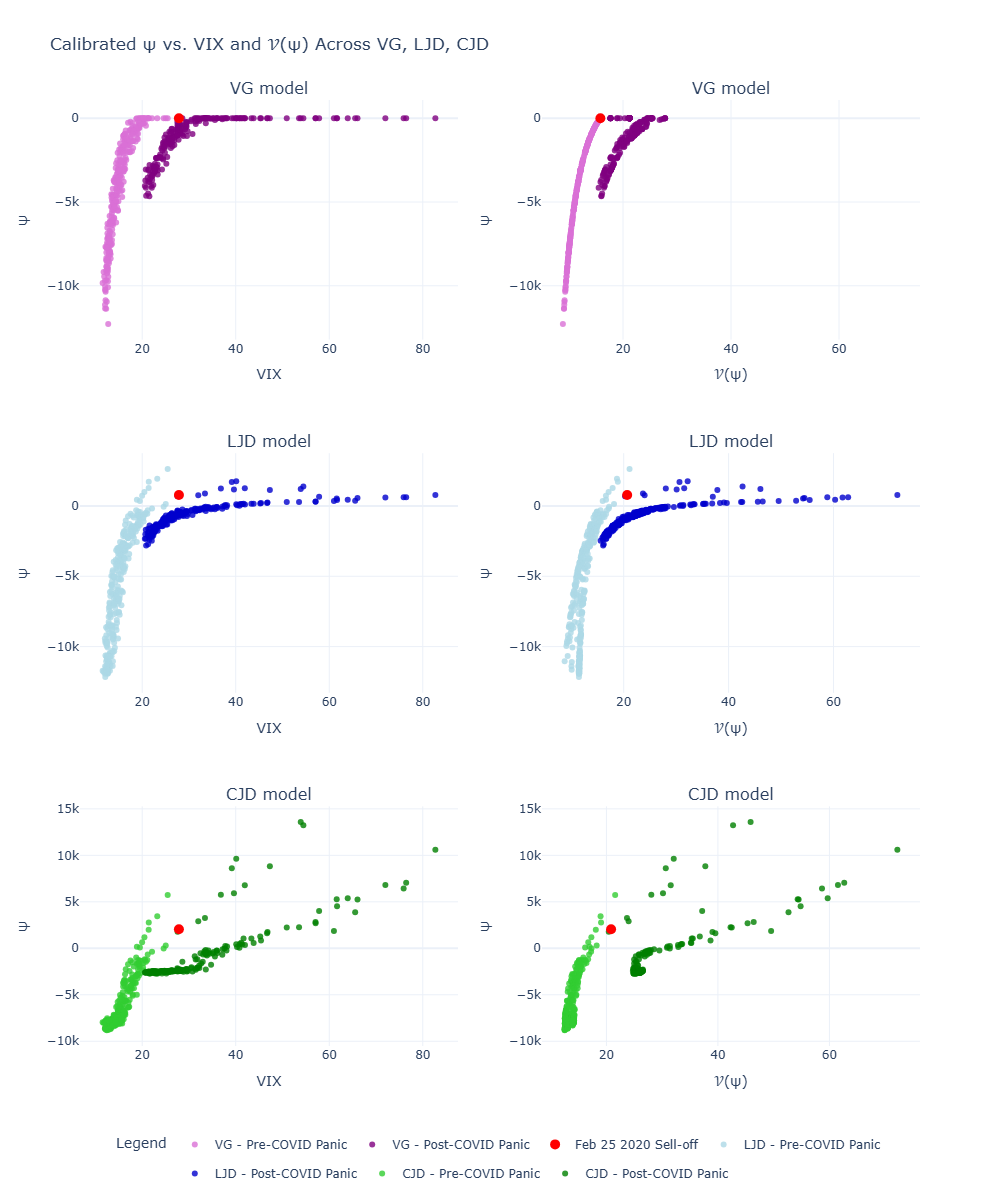}
  \caption{$\psi$ vs. VIX and $\mathcal{V}(\psi)$.}
  \label{fig: all models vix vs modeled vix}
\end{figure}

Figure \ref{fig: all models vix vs modeled vix} compares the scatter plots of calibrated $\psi$  values against both the observed VIX (left column) and the model-implied $\mathcal{V}(\psi )$ (right column). For the all models, the modeled VIX values closely replicate the patterns observed in the empirical VIX scatter plots, suggesting that $\mathcal{V}(\psi )$ effectively captures the volatility dynamics embedded in $\psi$, especially for the LJD and CJD models.

\subsubsection{Market price of risk \texorpdfstring{$\theta (\psi)$} vs. VIX and market sentiment}

\begin{figure}[H]
  \centering
  \includegraphics[width=1\linewidth]{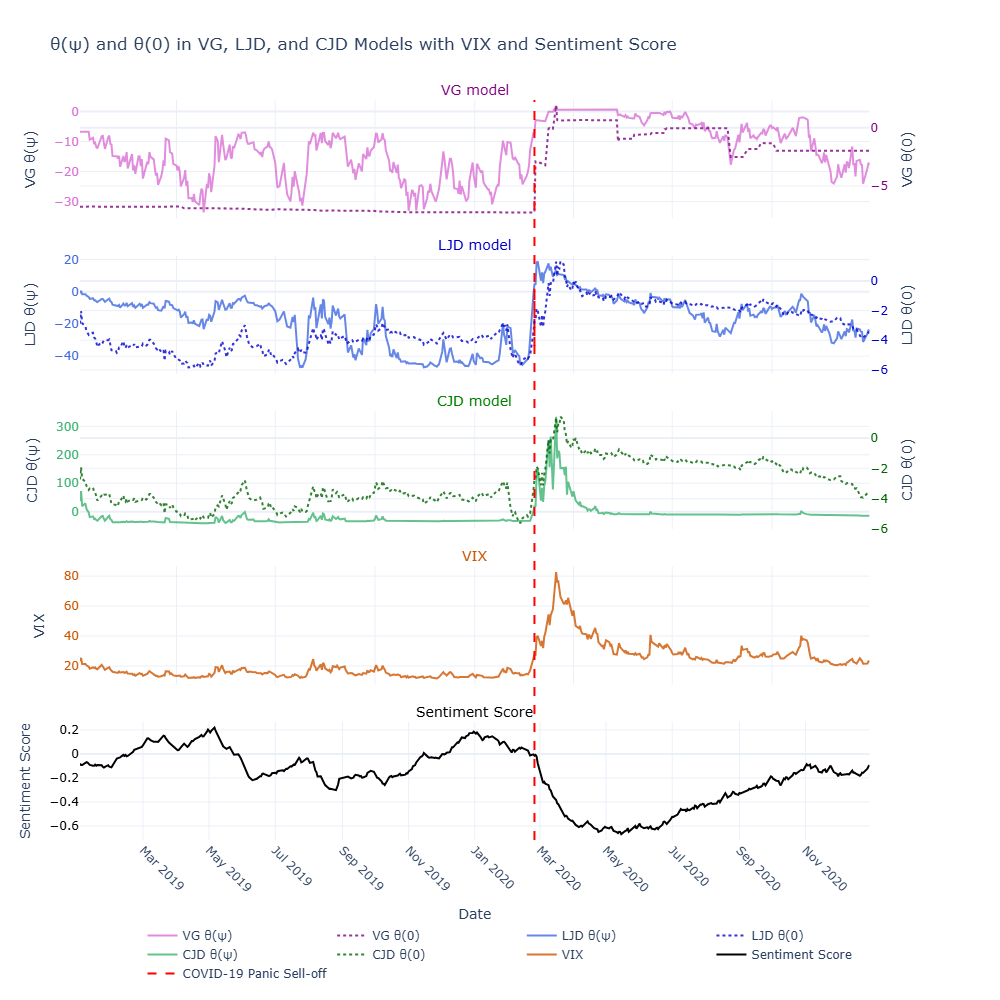}
  \caption{Market Price of Risk for First and Second Order Esscher.}
  \label{fig: all models mpr timeseries}
\end{figure}

The behavior of the market price of risk (MPR), denoted $\theta (\psi )$, is examined across models using both time-series plots and scatter plots—Figures \ref{fig: all models mpr scatter} and \ref{fig: all models mpr timeseries}, respectively. In both visualizations, $\theta (\psi )$ represents the second-order Esscher MPR, which incorporates the calibrated $\psi$  through a martingale condition, while $\theta (0)$ corresponds to the first-order Esscher MPR, derived solely from model parameters.

Figure \ref{fig: all models mpr scatter} offers a clear visual comparison between the MPR $\theta$  derived from first-order Esscher transforms (right column) and second-order Esscher transforms (left column), across three models—CJD, LJD, and VG—and segmented by pre- and post-COVID-19 market panic regimes. The key insight lies in the structural differences between the two columns: while the first-order Esscher $\theta$ values show weak or inconsistent correlation with VIX, the second-order Esscher $\theta$  values exhibit a strong, monotonic, and regime-sensitive relationship with VIX across all models.

In the first-order Esscher plots, $\theta$  appears relatively flat or erratic with respect to VIX, especially in the post-panic regime. This reflects a fundamental limitation: the first-order Esscher transform ties $\theta$  rigidly to model parameters estimated from asset returns, making it insensitive to real-time shifts in market sentiment or pricing stress. As a result, it fails to capture the behavioral and volatility-driven distortions that dominate option markets during crises. By contrast, the second-order Esscher plots show that $\theta$  increases smoothly with VIX, and the pre- and post-panic regimes form distinct, interpretable clusters. This confirms that the second-order framework—by introducing $\psi$  as a free parameter—allows $\theta$  to adapt dynamically to changing market conditions. It bridges the gap between historical asset dynamics and forward-looking investor expectations, especially under stress. In essence, the plot demonstrates that second-order Esscher pricing not only improves empirical fit but also restores economic interpretability to $\theta$. It transforms the market price of risk from a static byproduct into a volatility-aware, sentiment-sensitive diagnostic, making it a more reliable tool for pricing, risk management, and regime detection.

Viewed in conjunction with the time-series plots in Figure \ref{fig: all models mpr timeseries}, the advantage of the second-order Esscher framework becomes even more apparent. For all models, the second-order MPR (solid lines) fluctuates more visibly than the first-order MPR (dotted lines), which remains relatively flat. This contrast underscores the interpretive strength of $\theta (\psi )$: by incorporating $\psi$, it reflects evolving market conditions and behavioral stress. In the VG model, the first-order MPR shows minimal variation, consistent with the observation that its estimated parameters change little across dates. In contrast, the second-order MPR responds more dynamically, revealing latent pricing distortions that the first-order measure overlooks. Moreover, the range of values attained by $\theta (\psi )$ is substantially wider than that of $\theta (0)$, indicating that the second-order Esscher framework allows for a broader spectrum of risk compensation. This expanded range reflects the model’s ability to accommodate extreme market conditions and nonlinear investor responses—features that are suppressed under the first-order approach. For the CJD model, although the second-order MPR appears less volatile in the plot, this is an artifact of scaling—its values are significantly larger, which compresses visual fluctuations at lower levels. When properly rescaled, the CJD second-order MPR exhibits variation comparable to those seen in VG and LJD models, further supporting the claim that $\theta (\psi )$ offers a more sensitive and regime-aware measure of market risk.

Taken together, these plots demonstrate that the second-order Esscher MPR $\theta (\psi )$ provides a richer and more responsive characterization of market risk than the first-order Esscher MPR $\theta (0)$. By incorporating $\psi$, $\theta (\psi )$ captures both volatility-driven and sentiment-driven distortions, adapting to structural shifts in market conditions. Crucially, it transforms the market price of risk from a static artifact into a dynamic, economically interpretable signal—one that reflects the evolving compensation investors require for bearing risk under uncertainty. This flexibility makes it a more informative tool for pricing, risk management, and regime detection.

\subsubsection{Esscher-based option pricing accuracy}

Having calibrated the second-order Esscher parameter $\psi$, we now assess its impact on option pricing accuracy. This section compares model-implied option prices under the second-order Esscher transformation with both observed market prices and those derived from the first-order Esscher framework. By evaluating pricing errors across time and strike levels, we highlight the empirical advantages of incorporating $\psi$  into the pricing kernel. The results reveal that second-order Esscher pricing consistently improves fit, particularly during periods of market stress, offering a more adaptive and behavior-sensitive approach to derivative valuation.

\begin{figure}[H]
  \centering
  \includegraphics[width=1\linewidth]{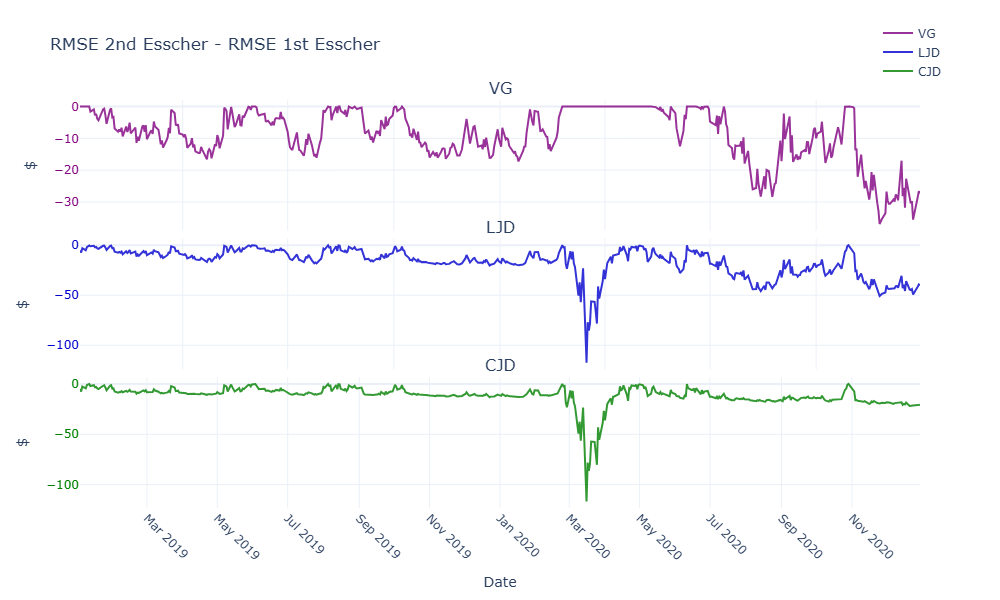}
  \caption{Difference between 2nd Esscher and 1st Esscher pricing errors}
  \label{fig: pricing errors}
\end{figure}

The comparative performance of first- and second-order Esscher pricing is visualized in Figure \ref{fig: pricing errors}, which plots the difference in RMSE between the two methods across time and models. A negative value indicates that the second-order Esscher yields lower pricing error than the first-order counterpart. Across all three models—VG, LJD, and CJD—the second-order Esscher consistently delivers more accurate option prices, especially during periods of elevated market stress. This improvement is most pronounced in the LJD and CJD models, where the second-order RMSE drops sharply below the first-order RMSE around the COVID-19 panic. In contrast, the VG model shows no divergence between the two methods during stress periods. This is a direct consequence of the structural constraint that $\psi$  in the VG model cannot exceed zero, limiting its ability to adjust the pricing kernel upward during crises. As a result, the VG second-order Esscher measure coincides with the first-order measure in stressed regimes, offering no additional flexibility.

\begin{figure}[H]
  \centering
  \includegraphics[width=1\linewidth]{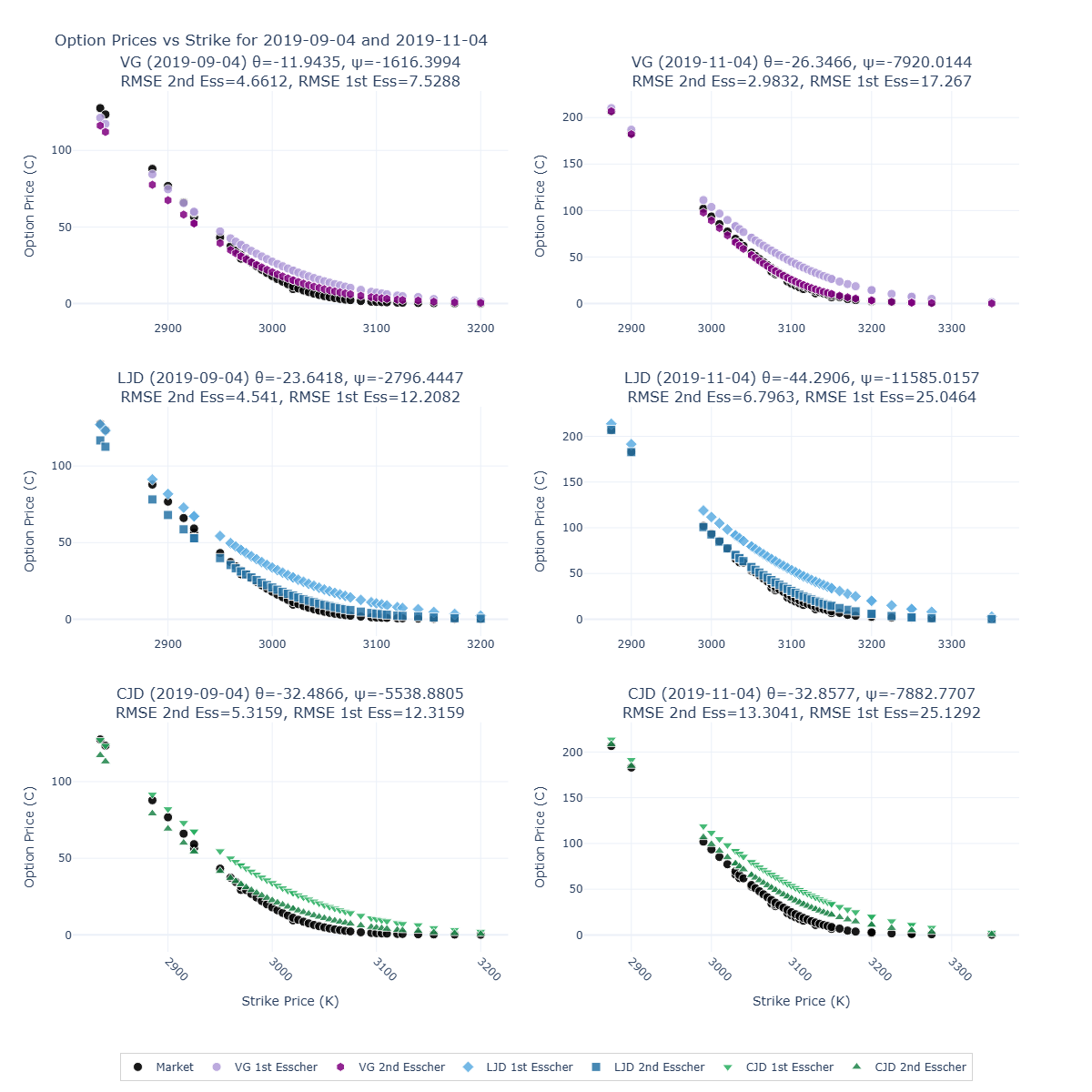}
  \caption{Option prices for dates 2019-09-04 and 2019-11-04}
  \label{fig: option prices dates 1}
\end{figure}

\begin{figure}[H]
  \centering
  \includegraphics[width=1\linewidth]{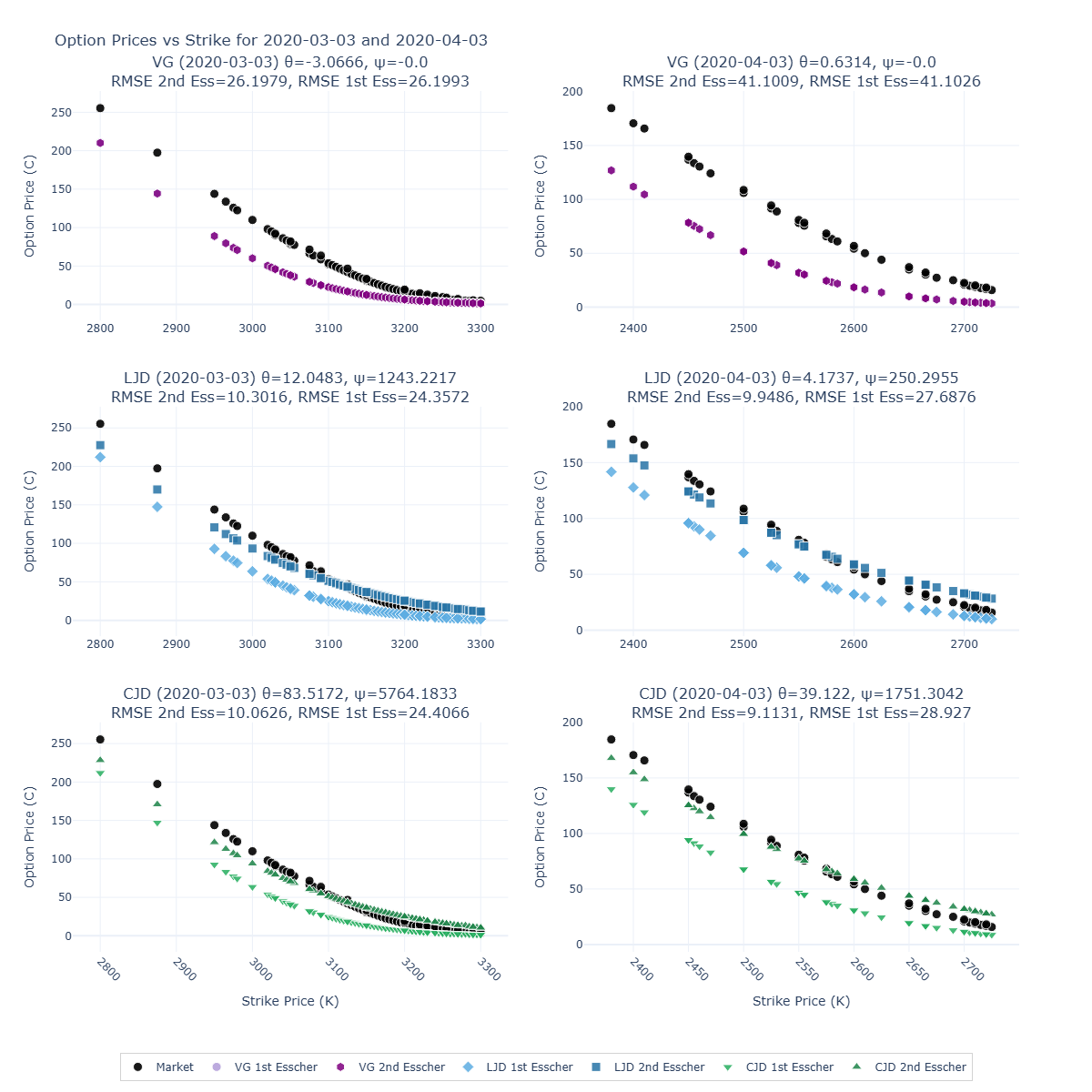}
  \caption{Option prices for dates 2020-03-03 and 2020-04-03}
  \label{fig: option prices dates 2}
\end{figure}

\begin{figure}[H]
  \centering
  \includegraphics[width=1\linewidth]{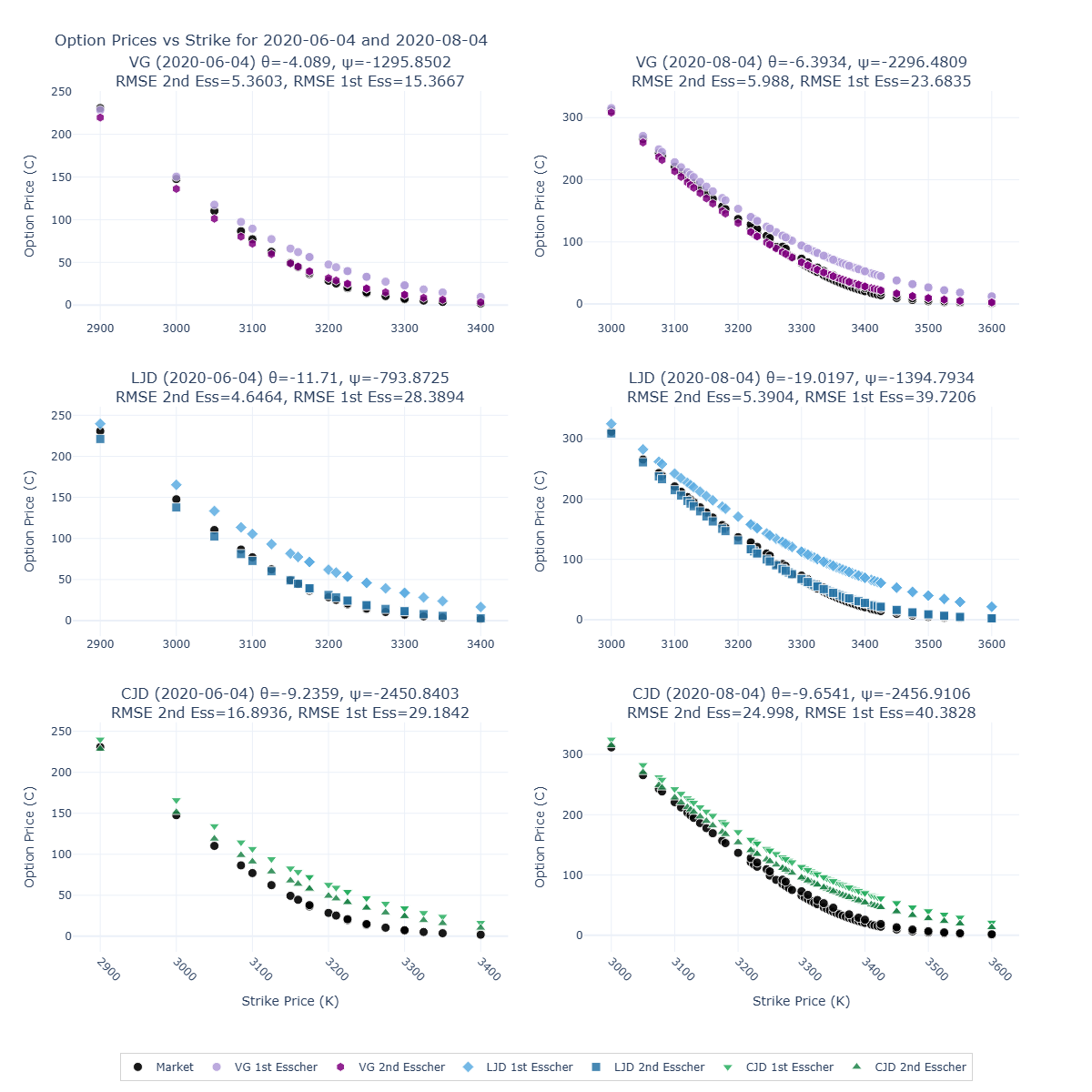}
  \caption{Option prices for dates 2020-06-04 and 2020-08-04}
  \label{fig: option prices dates 3}
\end{figure}

Figures \ref{fig: option prices dates 1}, \ref{fig: option prices dates 2}, and \ref{fig: option prices dates 3} present option prices against strike prices for multiple dates, comparing market prices with model-implied prices under both Esscher transformations. These plots reinforce the insights from the RMSE difference plot. In cases where the models overprice options—typically in low-volatility regimes—the second-order Esscher pricing consistently yields closer alignment with market prices across all models. During high-volatility periods, when models tend to underprice options, the second-order Esscher transformation continues to outperform the first-order for LJD and CJD. For VG, however, the pricing curves under both transformations coincide, again reflecting the upper bound constraint on $\psi$  that limits its responsiveness to market dislocations.

These results collectively underscore the practical value of second-order Esscher pricing as a flexible and empirically grounded enhancement to traditional risk-neutral valuation. By introducing the free parameter $\psi$, the second-order framework enables option prices to respond more accurately to prevailing market conditions—especially during periods of heightened volatility or sentiment-driven dislocation. Across diverse regimes, it consistently yields lower pricing errors and better alignment with observed market prices. This adaptability makes second-order Esscher pricing a robust tool for capturing behavioral stress and regime sensitivity in derivative valuation, offering a meaningful improvement over the static structure of first-order Esscher transforms.

\section{Conclusion}

This paper demonstrates the practical power and conceptual flexibility of second-order Esscher pricing in both option valuation and risk management. By extending the classical Esscher framework to include a free parameter \(\psi\), we derive tractable pricing formulas for a wide range of L\'evy models—including CJD, LJD, VG, and NIG—and show how \(\psi\) enables a richer calibration to market data. In particular, our detailed analysis of the constant jump-diffusion model reveals how the market price of risk \(\theta\) evolves as a function of \(\psi\), offering a transparent mechanism for embedding sentiment and behavioural distortions into the pricing kernel. Visual sensitivity studies further illustrate the responsiveness of \(\theta\) to changes in \(\psi\), especially under stressed market conditions.

In our risk management application, we show that \(\psi\) serves as a robust stress-testing variable, allowing practitioners to simulate P\&L, value-at-risk (VaR), and expected shortfall (ES) across a spectrum of market regimes. This approach enhances model adaptability without requiring more complex dynamics, enabling a more strategic and comprehensive assessment of risk exposures. By embedding external financial information into \(\psi\), risk managers can tailor pricing and hedging strategies to reflect latent market conditions that are not captured by structural parameters alone.

Our empirical findings demonstrate that incorporating the second-order Esscher parameter \(\psi\) markedly enhances option pricing accuracy. Models calibrated solely to historical spot returns via maximum likelihood estimation tend to produce substantial pricing errors when compared to observed market option prices. In contrast, direct calibration to option prices—facilitated by the inclusion of \(\psi\)—results in significantly tighter fits and more credible valuations. Although we calibrated \(\psi\) using market option data, this procedure primarily serves to illustrate that adjusting \(\psi\) in response to evolving market conditions enables more accurate pricing, particularly in illiquid environments. While machine learning techniques can offer further refinement in data-rich settings, they often lack interpretability and do not guarantee arbitrage-free outcomes. In contrast, the second-order Esscher framework retains theoretical rigour and martingale consistency, while allowing \(\psi\) to flexibly encode behavioural and regime-sensitive distortions. As such, \(\psi\) functions not only as a calibration parameter but also as a dynamic proxy for market sentiment—comparable to the VIX—making it a powerful and versatile tool for both academic modelling and practical risk management.

\paragraph{Acknowledgement.} Part of this research was designed when the first author visited the Department of Applied Mathematics, Computer Science and Statistics at Ghent University. The visit was fully funded by the FWO mobility grant V500324N. \\The third author (Mich\`ele Vanmaele) gratefully acknowledges Professor Tahir Choulli, Ella Elazkany and the Mathematical and Statistical Sciences Department at the University of Alberta for their warm welcome and hospitality during a research visit to Edmonton to collaborate on this paper. This research visit was funded by the Research Foundation – Flanders (FWO) under grant FWO‑SAB K803124, which is gratefully acknowledged for its financial support. Ella Elazkay is fully funded by NSERC RGPIN-2025-05129-Choulli.

\begin{thebibliography}{10}

\bibitem[Aguilar(2021)]{aguilar2021explicit}
J.-P. Aguilar.
\newblock Explicit option valuation in the exponential {NIG} model.
\newblock \emph{Quantitative Finance}, 21\penalty0 (8):\penalty0 1281--1299,
  2021.

\bibitem[Ballotta(2024)]{ballotta2024vix}
L.~Ballotta.
\newblock Is the {VIX} just volatility? {T}he {D}evil is in the (de)tails.
\newblock \emph{Wilmott}, 2024\penalty0 (130):\penalty0 18--20, 2024.

\bibitem[Barndorff-Nielsen(1994)]{barndo1995normal}
O.~E. Barndorff-Nielsen.
\newblock Normal$\backslash$$\backslash$inverse {G}aussian processes and the
  modelling of stock returns.
\newblock Research Reports 300, Department of Theoretical Statistics, Institute
  of Mathematics. University of Aarhus, Denmark, 1994.

\bibitem[Barndorff-Nielsen(1997)]{barndorff1997normal}
O.~E. Barndorff-Nielsen.
\newblock Normal inverse {G}aussian distributions and stochastic volatility
  modelling.
\newblock \emph{Scandinavian Journal of Statistics}, 24\penalty0 (1):\penalty0
  1--13, 1997.

\bibitem[Barndorff-Nielsen(1997)process]{Barndorff1997NIGtype}
O.E. Barndorff-Nielsen. 
\newblock Processes of normal inverse Gaussian type.
\newblock \emph{Finance and Stochastics,} 2(1): 41-68, 1997.

\bibitem[Barndorff-Nielsen and Shiryaev(2015)]{barndorff2015change}
O.~E. Barndorff-Nielsen and A.~N. Shiryaev.
\newblock \emph{Change of {T}ime and {C}hange of {M}easure}, volume~21.
\newblock World Scientific Publishing Company, 2015.

\bibitem[Benth and Sgarra(2012)]{benth2012risk}
F.~E. Benth and C.~Sgarra.
\newblock The market price of risk and the {E}sscher transform in power markets.
\newblock \emph{Stochastic Analysis and Applications}, 30\penalty0
  (1):\penalty0 20--43, 2012.

\bibitem[Benth et~al.(2008)Benth, Benth, and Koekebakker]{benth2008stochastic}
F.~E. Benth, J.~S. Benth, and S.~Koekebakker.
\newblock \emph{Stochastic Modelling of Electricity and Related Markets},
  volume~11.
\newblock World Scientific, 2008.

\bibitem[Bondi et~al.(2020)Bondi, Radoji{\v{c}}i{\'c}, and
  Rheinl{\"a}nder]{bondi2020comparing}
A.~Bondi, D.~Radoji{\v{c}}i{\'c}, and T.~Rheinl{\"a}nder.
\newblock Comparing two different option pricing methods.
\newblock \emph{Risks}, 8\penalty0 (4):\penalty0 108, 2020.

\bibitem[Boughamoura and Trabelsi(2014)]{boughamoura2014two}
W.~Boughamoura and F.~Trabelsi.
\newblock On two-parametric {E}sscher transform for geometric {CGMY} {L}{\'e}vy
  processes.
\newblock \emph{International Journal of Operational Research}, 19\penalty0
  (3):\penalty0 280--301, 2014.

\bibitem[Breeden and Litzenberger(1978)]{breeden1978}
D.~R. Breeden and R.~H. Litzenberger.
\newblock Prices of state-contingent claims implicit in option prices.
\newblock \emph{The Journal of Business}, 51\penalty0 (4):\penalty0 621--651,
  1978.

\bibitem[Carr and Madan(1999)]{carr1999option}
P.~Carr and D.~Madan.
\newblock Option valuation using the fast {F}ourier transform.
\newblock \emph{Journal of Computational Finance}, 2\penalty0 (4):\penalty0
  61--73, 1999.

\bibitem[Choulli et~al.(2025)Choulli, Elazkany, and
  Vanmaele]{choulli2025second}
T.~Choulli, E.~Elazkany, and M.~Vanmaele.
\newblock The second-order Esscher martingale densities for continuous-time
  market models.
\newblock \emph{Frontiers of Mathematical Finance}, 6:\penalty0
  16--65, 2025.

\bibitem[Eberlein and Kallsen(2019)]{eberlein2019mathematical}
E.~Eberlein and J.~Kallsen.
\newblock \emph{Mathematical Finance}.
\newblock Springer, 2019.

\bibitem[Feng and Linetsky(2008)]{feng2008pricing}
L.~Feng and V.~Linetsky.
\newblock Pricing options in jump-diffusion models: {A}n extrapolation
  approach.
\newblock \emph{Operations Research}, 56\penalty0 (2):\penalty0 304--325, 2008.

\bibitem[He(2019)]{HeSemimartingale}
S. W. He, J. G. Wang, \& J. -A. Yan. \newblock\emph{Semimartingale Theory and Stochastic Calculus. Routledge, 2019}

\bibitem[Hilliard and Reis(1998)]{hilliard1998valuation}
J.~E. Hilliard and J.~Reis.
\newblock Valuation of commodity futures and options under stochastic
  convenience yields, interest rates, and jump diffusions in the spot.
\newblock \emph{Journal of Financial and Quantitative Analysis}, 33\penalty0
  (1):\penalty0 61--86, 1998.

\bibitem[Honore(1998)]{Honore1998Pitfalls}
P. Honor\'e. 
\newblock Pitfalls in estimating jump-diffusion models. \newblock \emph{Available at SSRN 61998,} 1998.

\bibitem[Jacod and Shiryaev(2013)]{jacod2013limit}
J.~Jacod and A.~Shiryaev.
\newblock \emph{Limit Theorems for Stochastic Processes}, volume 288.
\newblock Springer Science \& Business Media, 2013.

\bibitem[Jeanblanc et~al.(2009)Jeanblanc, Yor, and
  Chesney]{jeanblanc2009mathematical}
M.~Jeanblanc, M.~Yor, and M.~Chesney.
\newblock \emph{Mathematical Methods for Financial Markets}.
\newblock Springer Science \& Business Media, 2009.

\bibitem[Ken-Iti(1999)]{ken1999levy}
S.~Ken-Iti.
\newblock \emph{L{\'e}vy Processes and Infinitely Divisible Distributions},
  volume 68.
\newblock Cambridge University Press, 1999.

\bibitem[Kou(2002)]{kou2002jump}
S.~G. Kou.
\newblock A jump-diffusion model for option pricing.
\newblock \emph{Management Science}, 48\penalty0 (8):\penalty0 1086--1101,
  2002.

\bibitem[Kuen~Siu et~al.(2014)Kuen~Siu, Nawar, and Ewald]{kuen2014hedging}
T.~Kuen~Siu, R.~Nawar, and C.~O. Ewald.
\newblock Hedging crude oil derivatives in {GARCH}-type models.
\newblock \emph{Journal of Energy Markets}, 7\penalty0 (1): 3-26, 2014.

\bibitem[Madan and Milne(1991)]{madan1991option}
D.~B. Madan and F.~Milne.
\newblock Option pricing with {VG} martingale components.
\newblock \emph{Mathematical Finance}, 1\penalty0 (4):\penalty0 39--55, 1991.

\bibitem[Madan(1990)]{madanVG1990}
D.B. Madan, and E. Seneta.
\newblock The variance gamma (VG) model for share market returns. 
\newblock \emph{Journal of Business,} 63\penalty0(4): 511-524, 1990.

\bibitem[Madan et~al.(1998)Madan, Carr, and Chang]{madan1998variance}
D.~B. Madan, P.~P. Carr, and E.~C. Chang.
\newblock The {V}ariance {G}amma process and option pricing.
\newblock \emph{Review of Finance}, 2\penalty0 (1):\penalty0 79--105, 1998.

\bibitem[Matsuda(2004)]{matsuda2004introduction}
K.~Matsuda.
\newblock Introduction to {M}erton jump diffusion model.
\newblock Research report, Department of Economics, The Graduate Center, The
  City University of New York, New York, 2004.

\bibitem[Merton(1976)]{merton1976option}
R.~C. Merton.
\newblock Option pricing when underlying stock returns are discontinuous.
\newblock \emph{Journal of Financial Economics}, 3\penalty0 (1-2):\penalty0
  125--144, 1976.
  
  \bibitem[Rudin(1976)]{Rudin1976}
W. Rudin. 
\newblock Principles of Mathematical Analysis. (3d. ed.). \newblock \emph{McGraw-Hill,} 1976.

\bibitem[Salhi(2017)]{salhi2017pricing}
K.~Salhi.
\newblock Pricing {E}uropean options and risk measurement under exponential
  {L}\'evy models: {A} practical guide.
\newblock \emph{International Journal of Financial Engineering}, 4\penalty0
  (02n03):\penalty0 1750016, 2017.

\bibitem[Schmelzle(2010)]{schmelzle2010option}
M.~Schmelzle.
\newblock Option pricing formulae using {F}ourier transform: {T}heory and
  application.
\newblock Preprint, http://pfadintegral.com, 2010.

\bibitem[Schoutens(2003)]{schoutens2003levy}
W.~Schoutens.
\newblock \emph{L{\'e}vy Processes in Finance: Pricing Financial Derivatives}.
\newblock John Wiley $\&$ sons LTD, 2003.

\end{thebibliography}

\begin{thebibliography}{100}

    \bibitem{aksamit2017no}
{\sc Aksamit, A., Choulli, T., Deng, J., and Jeanblanc, M.}
\newblock No-arbitrage up to random horizon for quasi-left-continuous models.
\newblock {\em Finance and Stochastics 21\/} (2017), 1103--1139.


\bibitem{aksamit2015optional}
{\sc Aksamit, A., Choulli, T., and Jeanblanc, M.}
\newblock On an optional semimartingale decomposition and the existence of a deflator in an enlarged filtration.
\newblock In {\em In Memoriam Marc Yor-S{\'e}minaire de Probabilit{\'e}s XLVII}. Springer, 2015, pp.~187--218.

\bibitem{alharbi2022log}
{\sc Alharbi, F., and Choulli, T.}
\newblock Log-optimal portfolio after a random time: Existence, description and sensitivity analysis.
\newblock {\em arXiv preprint arXiv:2204.03798\/} (2022).




\bibitem{avanzi2007optimal}
{\sc Avanzi, B., Gerber, H.~U., and Shiu, E.~S.}
\newblock Optimal dividends in the dual model.
\newblock {\em Insurance: Mathematics and Economics 41}, 1 (2007), 111--123.

\bibitem{azema1993theoreme}
{\sc Az{\'e}ma, J., Yor, M., Meyer, P.~A., Az{\'e}ma, J., Jeulin, T., Knight, F., and Yor, M.}
\newblock Le th{\'e}or{\`e}me d'arr{\^e}t en une fin d'ensemble pr{\'e}visible.
\newblock In {\em S{\'e}minaire de Probabilit{\'e}s XXVII\/} (1993), Springer, pp.~133--158.

\bibitem{badescu2009esscher}
{\sc Badescu, A., Elliott, R.~J., and Siu, T.~K.}
\newblock Esscher transforms and consumption-based models.
\newblock {\em Insurance: Mathematics and Economics 45}, 3 (2009), 337--347.

\bibitem{ball1985jumps}
{\sc Ball, C.~A., and Torous, W.~N.}
\newblock On jumps in common stock prices and their impact on call option pricing.
\newblock {\em The Journal of Finance 40}, 1 (1985), 155--173.

\bibitem{baptiste2018pricing}
{\sc Baptiste, J., Carassus, L., and L{\'e}pinette, E.}
\newblock Pricing without martingale measure.
\newblock {\em Available at SSRN 3190878\/} (2018).

\bibitem{barles1997backward}
{\sc Barles, G., Buckdahn, R., and Pardoux, E.}
\newblock Backward stochastic differential equations and integral-partial differential equations.
\newblock {\em Stochastics: An International Journal of Probability and Stochastic Processes 60}, 1-2 (1997), 57--83.

\bibitem{barlow1978study}
{\sc Barlow, M.~T.}
\newblock Study of a filtration expanded to include an honest time.
\newblock {\em Zeitschrift f{\"u}r Wahrscheinlichkeitstheorie und verwandte Gebiete 44}, 4 (1978), 307--323.

\bibitem{barndorff1997normal}
{\sc Barndorff-Nielsen, O.~E.}
\newblock Normal inverse gaussian distributions and stochastic volatility modelling.
\newblock {\em Scandinavian Journal of statistics 24}, 1 (1997), 1--13.

\bibitem{barron2003conditional}
{\sc Barron, E.~N., Cardaliaguet, P., and Jensen, R.}
\newblock Conditional essential suprema with applications.
\newblock {\em Applied Mathematics and optimization 48\/} (2003), 229--253.

\bibitem{bauer2008universal}
{\sc Bauer, D., Kling, A., and Russ, J.}
\newblock A universal pricing framework for guaranteed minimum benefits in variable annuities1.
\newblock {\em ASTIN Bulletin: The Journal of the IAA 38}, 2 (2008), 621--651.

\bibitem{bayraktar2010no}
{\sc Bayraktar, E., and Sayit, H.}
\newblock No arbitrage conditions for simple trading strategies.
\newblock {\em Annals of Finance 6}, 1 (2010), 147--156.

\bibitem{belanger2004general}
{\sc B{\'e}langer, A., Shreve, S.~E., and Wong, D.}
\newblock A general framework for pricing credit risk.
\newblock {\em Mathematical Finance: An International Journal of Mathematics, Statistics and Financial Economics 14}, 3 (2004), 317--350.

\bibitem{bender2008pricing}
{\sc Bender, C., Sottinen, T., and Valkeila, E.}
\newblock Pricing by hedging and no-arbitrage beyond semimartingales.
\newblock {\em Finance and Stochastics 12\/} (2008), 441--468.

\bibitem{bensaid1992derivative}
{\sc Bensaid, B., Lesne, J.-P., Pages, H., and Scheinkman, J.}
\newblock Derivative asset pricing with transaction costs 1.
\newblock {\em Mathematical Finance 2}, 2 (1992), 63--86.

\bibitem{benth2008stochastic}
{\sc Benth, F.~E., Benth, J.~S., and Koekebakker, S.}
\newblock {\em Stochastic modelling of electricity and related markets}, vol.~11.
\newblock World Scientific, 2008.

\bibitem{benth2012risk}
{\sc Benth, F.~E., and Sgarra, C.}
\newblock The market price of risk and the esscher transform in power markets.
\newblock {\em Stochastic Analysis and Applications 30}, 1 (2012), 20--43.

\bibitem{bernhart2012swing}
{\sc Bernhart, M., Pham, H., Tankov, P., and Warin, X.}
\newblock Swing options valuation: A bsde with constrained jumps approach.
\newblock In {\em Numerical Methods in Finance: Bordeaux, June 2010}. Springer, 2012, pp.~379--400.

\bibitem{bielecki2013credit}
{\sc Bielecki, T.~R., and Rutkowski, M.}
\newblock {\em Credit risk: modeling, valuation and hedging}.
\newblock Springer Science \& Business Media, 2013.

\bibitem{black1973pricing}
{\sc Black, F., and Scholes, M.}
\newblock The pricing of options and corporate liabilities.
\newblock {\em Journal of political economy 81}, 3 (1973), 637--654.

\bibitem{blanchet2004hazard}
{\sc Blanchet-Scalliet, C., and Jeanblanc, M.}
\newblock Hazard rate for credit risk and hedging defaultable contingent claims.
\newblock {\em Finance and Stochastics 8\/} (2004), 145--159.

\bibitem{bo2010markov}
{\sc Bo, L., Wang, Y., and Yang, X.}
\newblock Markov-modulated jump--diffusions for currency option pricing.
\newblock {\em Insurance: Mathematics and Economics 46}, 3 (2010), 461--469.

\bibitem{bondi2020comparing}
{\sc Bondi, A., Radoji{\v{c}}i{\'c}, D., and Rheinl{\"a}nder, T.}
\newblock Comparing two different option pricing methods.
\newblock {\em Risks 8}, 4 (2020), 108.

\bibitem{boughamoura2014two}
{\sc Boughamoura, W., and Trabelsi, F.}
\newblock On two-parametric esscher transform for geometric cgmy l{\'e}vy processes.
\newblock {\em International Journal of Operational Research 19}, 3 (2014), 280--301.

\bibitem{briand2003lp}
{\sc Briand, P., Delyon, B., Hu, Y., Pardoux, E., and Stoica, L.}
\newblock Lp solutions of backward stochastic differential equations.
\newblock {\em Stochastic Processes and their Applications 108}, 1 (2003), 109--129.

\bibitem{buckdahn1998hedging}
{\sc Buckdahn, R., and Hu, Y.}
\newblock Hedging contingent claims for a large investor in an incomplete market.
\newblock {\em Advances in Applied Probability 30}, 1 (1998), 239--255.

\bibitem{buckdahn1998pricing}
{\sc Buckdahn, R., and Hu, Y.}
\newblock Pricing of american contingent claims with jump stock price and constrained portfolios.
\newblock {\em Mathematics of operations research 23}, 1 (1998), 177--203.

\bibitem{buckdahn1994bsde}
{\sc Buckdahn, R., and Pardoux, E.}
\newblock Bsde’s with jumps and associated integro-partial differential equations.
\newblock {\em preprint 79\/} (1994).

\bibitem{buhlmann1980economic}
{\sc B{\"u}hlmann, H.}
\newblock An economic premium principle.
\newblock {\em ASTIN Bulletin: The Journal of the IAA 11}, 1 (1980), 52--60.

\bibitem{buhlmann1996no}
{\sc B{\"u}hlmann, H., Delbaen, F., Embrechts, P., and Shiryaev, A.~N.}
\newblock No-arbitrage, change of measure and conditional esscher transforms.
\newblock {\em CWI quarterly 9}, 4 (1996), 291--317.

\bibitem{buhlmann1998esscher}
{\sc B{\"u}hlmann, H., Delbaen, F., Embrechts, P., and Shiryaev, A.~N.}
\newblock On esscher transforms in discrete finance models.
\newblock {\em ASTIN Bulletin: The Journal of the IAA 28}, 2 (1998), 171--186.

\bibitem{carassus2022pricing}
{\sc Carassus, L., and L{\'e}pinette, E.}
\newblock Pricing without no-arbitrage condition in discrete time.
\newblock {\em Journal of Mathematical Analysis and Applications 505}, 1 (2022), 125441.

\bibitem{carr2002fine}
{\sc Carr, P., Geman, H., Madan, D.~B., and Yor, M.}
\newblock The fine structure of asset returns: An empirical investigation.
\newblock {\em The Journal of Business 75}, 2 (2002), 305--332.

\bibitem{carr2000valuation}
{\sc Carr, P., and Linetsky, V.}
\newblock The valuation of executive stock options in an intensity-based framework.
\newblock {\em Review of Finance 4}, 3 (2000), 211--230.

\bibitem{carr1999option}
{\sc Carr, P., and Madan, D.}
\newblock Option valuation using the fast fourier transform.
\newblock {\em Journal of computational finance 2}, 4 (1999), 61--73.

\bibitem{chan1990ruin}
{\sc Chan, B.}
\newblock Ruin probability for translated combination of exponential claims.
\newblock {\em ASTIN Bulletin: The Journal of the IAA 20}, 1 (1990), 113--114.

\bibitem{chan2006classical}
{\sc Chan, B., Gerber, H.~U., and Shiu, E.~S.}
\newblock “on a classical risk model with a constant dividend barrier”, xiaowen zhou, october 2005.
\newblock {\em North American Actuarial Journal 10}, 2 (2006), 133--139.

\bibitem{chan1999pricing}
{\sc Chan, T.}
\newblock Pricing contingent claims on stocks driven by l{\'e}vy processes.
\newblock {\em Annals of Applied Probability\/} (1999), 504--528.

\bibitem{cheridito2003arbitrage}
{\sc Cheridito, P.}
\newblock Arbitrage in fractional brownian motion models.
\newblock {\em Finance and stochastics 7}, 4 (2003), 533--553.

\bibitem{chiang2019valuation}
{\sc Chiang, S.~L., and Tsai, M.~S.}
\newblock Valuation of an option using non-parametric methods.
\newblock {\em Review of Derivatives Research 22\/} (2019), 419--447.

\bibitem{choulli2017no}
{\sc Choulli, T., and Deng, J.}
\newblock No-arbitrage for informational discrete time market models.
\newblock {\em Stochastics 89}, 3-4 (2017), 628--653.

\bibitem{choulli2020structure}
{\sc Choulli, T., and Deng, J.}
\newblock Structure conditions under progressively added information.
\newblock {\em Theory of Probability \& Its Applications 65}, 3 (2020), 418--453.

\bibitem{choulli2007minimal}
{\sc Choulli, T., Stricker, C., and Li, J.}
\newblock Minimal hellinger martingale measures of order q.
\newblock {\em Finance and Stochastics 11\/} (2007), 399--427.

\bibitem{colantoni2023inverse}
{\sc Colantoni, F., and D’Ovidio, M.}
\newblock On the inverse gamma subordinator.
\newblock {\em Stochastic Analysis and Applications 41}, 5 (2023), 999--1024.

\bibitem{cui2017general}
{\sc Cui, Z., Kirkby, J.~L., and Nguyen, D.}
\newblock A general framework for discretely sampled realized variance derivatives in stochastic volatility models with jumps.
\newblock {\em European Journal of Operational Research 262}, 1 (2017), 381--400.

\bibitem{cvitanic1998backward}
{\sc Cvitanic, J., Karatzas, I., and Soner, H.~M.}
\newblock Backward stochastic differential equations with constraints on the gains-process.
\newblock {\em Annals of Probability\/} (1998), 1522--1551.

\bibitem{dalang1990equivalent}
{\sc Dalang, R.~C., Morton, A., and Willinger, W.}
\newblock Equivalent martingale measures and no-arbitrage in stochastic securities market models.
\newblock {\em Stochastics: An International Journal of Probability and Stochastic Processes 29}, 2 (1990), 185--201.

\bibitem{dassios2003pricing}
{\sc Dassios, A., and Jang, J.-W.}
\newblock Pricing of catastrophe reinsurance and derivatives using the cox process with shot noise intensity.
\newblock {\em Finance and Stochastics 7\/} (2003), 73--95.

\bibitem{delbaen1994general}
{\sc Delbaen, F., and Schachermayer, W.}
\newblock A general version of the fundamental theorem of asset pricing.
\newblock {\em Mathematische annalen 300}, 1 (1994), 463--520.

\bibitem{dellacheriemeyer92} {\sc Dellacherie, M., Maisonneuve, B. and  Meyer} \newblock P-A. Probabilit\'es et Potentiel, chapitres XVII-XXIV: Processus de Markov (fin). Compl\'ements de calcul stochastique, Hermann, Paris, 1992.

\bibitem{dellacheriemeyer80}{\sc Dellacherie, C. and Meyer} \newblock P-A. Probabilit\'e et Potentiel, Th\'eorie des martingales. Chapter V-VIII. Hermann, 1980.

\bibitem{dufresne2007fitting}
{\sc Dufresne, D.}
\newblock Fitting combinations of exponentials to probability distributions.
\newblock {\em Applied Stochastic Models in Business and Industry 23}, 1 (2007), 23--48.

\bibitem{dufresne2007stochastic}
{\sc Dufresne, D.}
\newblock Stochastic life annuities.
\newblock {\em North American Actuarial Journal 11}, 1 (2007), 136--157.

\bibitem{dufresne1988probability}
{\sc Dufresne, F., and Gerber, H.~U.}
\newblock The probability and severity of ruin for combinations of exponential claim amount distributions and their translations.
\newblock {\em Insurance: Mathematics and Economics 7}, 2 (1988), 75--80.

\bibitem{dufresne1989three}
{\sc Dufresne, F., and Gerber, H.~U.}
\newblock Three methods to calculate the probability of ruin.
\newblock {\em ASTIN Bulletin: The Journal of the IAA 19}, 1 (1989), 71--90.

\bibitem{dufresne1991rational}
{\sc Dufresne, F., and Gerber, H.~U.}
\newblock Rational ruin problems—a note for the teacher.
\newblock {\em Insurance: Mathematics and Economics 10}, 1 (1991), 21--29.

\bibitem{dufresne1991risk}
{\sc Dufresne, F., and Gerber, H.~U.}
\newblock Risk theory for the compound poisson process that is perturbed by diffusion.
\newblock {\em Insurance: mathematics and economics 10}, 1 (1991), 51--59.

\bibitem{eberlein2009jump}
{\sc Eberlein, E.}
\newblock Jump--type l{\'e}vy processes.
\newblock In {\em Handbook of financial time series}. Springer, 2009, pp.~439--455.

\bibitem{el2008backward}
{\sc El~Karoui, N., Hamad{\`e}ne, S., and Matoussi, A.}
\newblock Backward stochastic differential equations and applications, 2008.

\bibitem{el1997reflected}
{\sc El~Karoui, N., Kapoudjian, C., Pardoux, E., Peng, S., and Quenez, M.-C.}
\newblock Reflected solutions of backward sde's, and related obstacle problems for pde's.
\newblock {\em the Annals of Probability 25}, 2 (1997), 702--737.

\bibitem{elkaroui1997reflected}
{\sc El~Karoui, N., Pardoux, {\'E}., and Quenez, M.~C.}
\newblock Reflected backward sdes and american options.
\newblock {\em Numerical methods in finance 13\/} (1997), 215--231.

\bibitem{el1997backward}
{\sc El~Karoui, N., Peng, S., and Quenez, M.~C.}
\newblock Backward stochastic differential equations in finance.
\newblock {\em Mathematical finance 7}, 1 (1997), 1--71.

\bibitem{el2020conditional}
{\sc El~Mansour, M., and L{\'e}pinette, E.}
\newblock Conditional interior and conditional closure of random sets.
\newblock {\em Journal of Optimization Theory and Applications 187}, 2 (2020), 356--369.

\bibitem{elie2014adding}
{\sc Elie, R., and Kharroubi, I.}
\newblock Adding constraints to bsdes with jumps: an alternative to multidimensional reflections.
\newblock {\em ESAIM: Probability and Statistics 18\/} (2014), 233--250.

\bibitem{elliott2022generalized}
{\sc Elliott, R., and Siu, T.}
\newblock A generalized esscher transform for option valuation with regime switching risk.
\newblock {\em Quantitative Finance 22}, 4 (2022), 691--705.

\bibitem{elliott2005option}
{\sc Elliott, R.~J., Chan, L., and Siu, T.~K.}
\newblock Option pricing and esscher transform under regime switching.
\newblock {\em Annals of Finance 1}, 4 (2005), 423.

\bibitem{elliott2006option}
{\sc Elliott, R.~J., Siu, T.~K., and Chan, L.}
\newblock Option pricing for garch models with markov switching.
\newblock {\em International Journal of Theoretical and Applied Finance 9}, 06 (2006), 825--841.

\bibitem{embrechts2000actuarial}
{\sc Embrechts, P.}
\newblock Actuarial versus financial pricing of insurance.
\newblock {\em The Journal of Risk Finance 1}, 4 (2000), 17--26.

\bibitem{esche2005minimal}
{\sc Esche, F., and Schweizer, M.}
\newblock Minimal entropy preserves the l{\'e}vy property: how and why.
\newblock {\em Stochastic processes and their applications 115}, 2 (2005), 299--327.

\bibitem{escher1932probability}
{\sc Escher, F.}
\newblock On the probability function in the collective theory of risk.
\newblock {\em Skand. Aktuarie Tidskr. 15\/} (1932), 175--195.

\bibitem{essaky2008reflected}
{\sc Essaky, E.}
\newblock Reflected backward stochastic differential equation with jumps and rcll obstacle.
\newblock {\em Bulletin des sciences mathematiques 132}, 8 (2008), 690--710.

\bibitem{essaky2011general}
{\sc Essaky, E., and Hassani, M.}
\newblock General existence results for reflected bsde and bsde.
\newblock {\em Bulletin des Sciences Math{\'e}matiques 135}, 5 (2011), 442--466.

\bibitem{fan2016bounded}
{\sc Fan, S.}
\newblock Bounded solutions, lp (p> 1) solutions and l1 solutions for one dimensional bsdes under general assumptions.
\newblock {\em Stochastic Processes and their Applications 126}, 5 (2016), 1511--1552.

\bibitem{fard2015analytical}
{\sc Fard, F.~A.}
\newblock Analytical pricing of vulnerable options under a generalized jump--diffusion model.
\newblock {\em Insurance: Mathematics and Economics 60\/} (2015), 19--28.

\bibitem{feng2008pricing}
{\sc Feng, L., and Linetsky, V.}
\newblock Pricing options in jump-diffusion models: an extrapolation approach.
\newblock {\em Operations Research 56}, 2 (2008), 304--325.

\bibitem{fernholz2009stochastic}
{\sc Fernholz, R., and Karatzas, I.}
\newblock Stochastic portfolio theory: an overview.
\newblock {\em Handbook of numerical analysis 15}, 89-167 (2009), 1180--91267.

\bibitem{filipovic2009separation}
{\sc Filipovi{\'c}, D., Kupper, M., and Vogelpoth, N.}
\newblock Separation and duality in locally l0-convex modules.
\newblock {\em Journal of Functional Analysis 256}, 12 (2009), 3996--4029.

\bibitem{follmer1997optional}
{\sc F{\"o}llmer, H., and Kramkov, D.}
\newblock Optional decompositions under constraints.
\newblock {\em Probability Theory and Related Fields 109\/} (1997), 1--25.

\bibitem{follmer2011stochastic}
{\sc F{\"o}llmer, H., and Schied, A.}
\newblock {\em Stochastic finance: an introduction in discrete time}.
\newblock Walter de Gruyter, 2011.

\bibitem{follmer1991hedging}
{\sc F{\"o}llmer, H., and Schweizer, M.}
\newblock Hedging of contingent claims under incomplete information.
\newblock {\em Applied stochastic analysis 5}, 389-414 (1991), 19--31.

\bibitem{fontana2014arbitrages}
{\sc Fontana, C., Jeanblanc, M., and Song, S.}
\newblock On arbitrages arising with honest times.
\newblock {\em Finance and Stochastics 18}, 3 (2014), 515--543.

\bibitem{fujiwara2003minimal}
{\sc Fujiwara, T., and Miyahara, Y.}
\newblock The minimal entropy martingale measures for geometric l{\'e}vy processes.
\newblock {\em Finance and Stochastics 7\/} (2003), 509--531.

\bibitem{gerber1970extension}
{\sc Gerber, H.~U.}
\newblock An extension of the renewal equation and its application in the collective theory of risk.
\newblock {\em Scandinavian Actuarial Journal 1970}, 3-4 (1970), 205--210.

\bibitem{gerber1972games}
{\sc Gerber, H.~U.}
\newblock Games of economic survival with discrete-and continuous-income processes.
\newblock {\em Operations research 20}, 1 (1972), 37--45.

\bibitem{gerber1996actuarial}
{\sc Gerber, H.~U., and Shiu, E.~S.}
\newblock Actuarial bridges to dynamic hedging and option pricing.
\newblock {\em Insurance: Mathematics and Economics 18}, 3 (1996), 183--218.

\bibitem{gerber1998time}
{\sc Gerber, H.~U., and Shiu, E.~S.}
\newblock On the time value of ruin.
\newblock {\em North American Actuarial Journal 2}, 1 (1998), 48--72.

\bibitem{gerber2005time}
{\sc Gerber, H.~U., and Shiu, E.~S.}
\newblock The time value of ruin in a sparre andersen model.
\newblock {\em North American Actuarial Journal 9}, 2 (2005), 49--69.

\bibitem{gerber1993option}
{\sc Gerber, H.~U., Shiu, E.~S., et~al.}
\newblock {\em Option pricing by Esscher transforms}.
\newblock HEC Ecole des hautes {\'e}tudes commerciales, 1993.

\bibitem{gerber2006maximizing}
{\sc Gerber, H.~U., Shiu, E.~S., and Smith, N.}
\newblock Maximizing dividends without bankruptcy.
\newblock {\em ASTIN Bulletin: The Journal of the IAA 36}, 1 (2006), 5--23.

\bibitem{gerber2012valuing}
{\sc Gerber, H.~U., Shiu, E.~S., and Yang, H.}
\newblock Valuing equity-linked death benefits and other contingent options: a discounted density approach.
\newblock {\em Insurance: Mathematics and Economics 51}, 1 (2012), 73--92.

\bibitem{gerber2013valuing}
{\sc Gerber, H.~U., Shiu, E.~S., and Yang, H.}
\newblock Valuing equity-linked death benefits in jump diffusion models.
\newblock {\em Insurance: Mathematics and Economics 53}, 3 (2013), 615--623.

\bibitem{gerber2015geometric}
{\sc Gerber, H.~U., Shiu, E.~S., and Yang, H.}
\newblock Geometric stopping of a random walk and its applications to valuing equity-linked death benefits.
\newblock {\em Insurance: Mathematics and Economics 64\/} (2015), 313--325.

\bibitem{goovaerts2008actuarial}
{\sc Goovaerts, M.~J., and Laeven, R.~J.}
\newblock Actuarial risk measures for financial derivative pricing.
\newblock {\em Insurance: Mathematics and Economics 42}, 2 (2008), 540--547.

\bibitem{grandits1999p}
{\sc Grandits, P.}
\newblock The p-optimal martingale measure and its asymptotic relation with the minimal-entropy martingale measure.
\newblock {\em Bernoulli\/} (1999), 225--247.

\bibitem{guasoni2006no}
{\sc Guasoni, P.}
\newblock No arbitrage under transaction costs, with fractional brownian motion and beyond.
\newblock {\em Mathematical Finance 16}, 3 (2006), 569--582.

\bibitem{guo2001information}
{\sc Guo, X.}
\newblock Information and option pricings.
\newblock {\em Quantitative Finance 1}, 1 (2001), 38.

\bibitem{hamadene2003reflected}
{\sc Hamad{\`e}ne, S., and Ouknine, Y.}
\newblock Reflected backward stochastic differential equation with jumps and random obstacle.

\bibitem{hamadene2008reflected}
{\sc Hamadene, S., and Ouknine, Y.}
\newblock Reflected backward sdes with general jumps.
\newblock {\em arXiv preprint arXiv:0812.3965\/} (2008).

\bibitem{hamadene2012lp}
{\sc Hamad{\`e}ne, S., and Popier, A.}
\newblock Lp-solutions for reflected backward stochastic differential equations.
\newblock {\em Stochastics and Dynamics 12}, 02 (2012), 1150016.

\bibitem{haug2007complete}
{\sc Haug, E.~G.}
\newblock The complete guide to option pricing formulas.
\newblock {\em (No Title)\/} (2007).

\bibitem{he2019semimartingale}
{\sc He, S.-w., Wang, J.-g., and Yan, J.-a.}
\newblock {\em Semimartingale theory and stochastic calculus}.
\newblock Routledge, 2019.

\bibitem{hilliard1998valuation}
{\sc Hilliard, J.~E., and Reis, J.}
\newblock Valuation of commodity futures and options under stochastic convenience yields, interest rates, and jump diffusions in the spot.
\newblock {\em Journal of financial and quantitative analysis 33}, 1 (1998), 61--86.

\bibitem{hubalek2006esscher}
{\sc Hubalek, F., and Sgarra~{\S}, C.}
\newblock Esscher transforms and the minimal entropy martingale measure for exponential l{\'e}vy models.
\newblock {\em Quantitative finance 6}, 02 (2006), 125--145.

\bibitem{hui2003pricing}
{\sc Hui, C.~H., Lo, C.-F., and Lee, H.}
\newblock Pricing vulnerable black-scholes options with dynamic default barriers.
\newblock {\em The Journal of Derivatives 10}, 4 (2003), 62--69.

\bibitem{jacod2006calcul}
{\sc Jacod, J.}
\newblock {\em Calcul stochastique et problemes de martingales}, vol.~714.
\newblock Springer, 2006.

\bibitem{jacod2013limit}
{\sc Jacod, J., and Shiryaev, A.}
\newblock {\em Limit theorems for stochastic processes}, vol.~288.
\newblock Springer Science \& Business Media, 2013.

\bibitem{jarrow2009no}
{\sc Jarrow, R.~A., Protter, P., and Sayit, H.}
\newblock No arbitrage without semimartingales.

\bibitem{jeanblanc2009mathematical}
{\sc Jeanblanc, M., Yor, M., and Chesney, M.}
\newblock {\em Mathematical methods for financial markets}.
\newblock Springer Science \& Business Media, 2009.

\bibitem{jeulin2006semi}
{\sc Jeulin, T.}
\newblock {\em Semi-martingales et grossissement d'une filtration}, vol.~833.
\newblock Springer, 2006.

\bibitem{jiao2018modeling}
{\sc Jiao, Y., and Li, S.}
\newblock Modeling sovereign risks: From a hybrid model to the generalized density approach.
\newblock {\em Mathematical Finance 28}, 1 (2018), 240--267.

\bibitem{johnson1987pricing}
{\sc Johnson, H., and Stulz, R.}
\newblock The pricing of options with default risk.
\newblock {\em The Journal of Finance 42}, 2 (1987), 267--280.

\bibitem{kabanov2009markets}
{\sc Kabanov, Y.}
\newblock {\em Markets with Transaction Costs Mathematical Theory}.
\newblock Springer, 2009.

\bibitem{kabanov2013essential}
{\sc Kabanov, Y., and L{\'e}pinette, E.}
\newblock Essential supremum with respect to a random partial order.
\newblock {\em Journal of Mathematical Economics 49}, 6 (2013), 478--487.

\bibitem{kabanov2001teacher}
{\sc Kabanov, Y., and Stricker, C.}
\newblock A teacher's note on no-arbitrage criteria.
\newblock {\em S{\'e}minaire de probabilit{\'e}s de Strasbourg 35\/} (2001), 149--152.

\bibitem{kallsen2002time}
{\sc Kallsen, J., and Shiraev, A.}
\newblock Time change representation of stochastic integrals.
\newblock {\em Theory of Probability \& Its Applications 46}, 3 (2002), 522--528.

\bibitem{kallsen2002cumulant}
{\sc Kallsen, J., and Shiryaev, A.~N.}
\newblock The cumulant process and esscher's change of measure.
\newblock {\em Finance and stochastics 6}, 4 (2002), 397--428.

\bibitem{karatzas1998methods}
{\sc Karatzas, I., Shreve, S.~E., Karatzas, I., and Shreve, S.~E.}
\newblock {\em Methods of mathematical finance}, vol.~39.
\newblock Springer, 1998.

\bibitem{kharroubi2010backward}
{\sc Kharroubi, I., Ma, J., Pham, H., and Zhang, J.}
\newblock Backward sdes with constrained jumps and quasi-variational inequalities.

\bibitem{kijima2008extension}
{\sc Kijima, M., and Muromachi, Y.}
\newblock An extension of the wang transform derived from b{\"u}hlmann’s economic premium principle for insurance risk.
\newblock {\em Insurance: Mathematics and Economics 42}, 3 (2008), 887--896.

\bibitem{kirkby2015efficient}
{\sc Kirkby, J.~L.}
\newblock Efficient option pricing by frame duality with the fast fourier transform.
\newblock {\em SIAM Journal on Financial Mathematics 6}, 1 (2015), 713--747.

\bibitem{kirkby2016efficient}
{\sc Kirkby, J.~L.}
\newblock An efficient transform method for asian option pricing.
\newblock {\em SIAM Journal on Financial Mathematics 7}, 1 (2016), 845--892.

\bibitem{kirkby2017robust}
{\sc Kirkby, J.~L.}
\newblock Robust barrier option pricing by frame projection under exponential l{\'e}vy dynamics.
\newblock {\em Applied Mathematical Finance 24}, 4 (2017), 337--386.

\bibitem{kirkby2017unified}
{\sc Kirkby, J.~L., Nguyen, D., and Cui, Z.}
\newblock A unified approach to bermudan and barrier options under stochastic volatility models with jumps.
\newblock {\em Journal of Economic Dynamics and Control 80\/} (2017), 75--100.

\bibitem{kladivko2023mean}
{\sc Klad{\'\i}vko, K., and Zervos, M.}
\newblock Mean--variance hedging of contingent claims with random maturity.
\newblock {\em Mathematical Finance 33}, 4 (2023), 1213--1247.

\bibitem{klein1996pricing}
{\sc Klein, P.}
\newblock Pricing black-scholes options with correlated credit risk.
\newblock {\em Journal of Banking \& Finance 20}, 7 (1996), 1211--1229.

\bibitem{klein1999valuation}
{\sc Klein, P., and Inglis, M.}
\newblock Valuation of european options subject to financial distress and interest rate risk.
\newblock {\em The Journal of Derivatives 6}, 3 (1999), 44--56.

\bibitem{klein2001pricing}
{\sc Klein, P., and Inglis, M.}
\newblock Pricing vulnerable european options when the option’s payoff can increase the risk of financial distress.
\newblock {\em Journal of banking \& finance 25}, 5 (2001), 993--1012.

\bibitem{kou2002jump}
{\sc Kou, S.~G.}
\newblock A jump-diffusion model for option pricing.
\newblock {\em Management science 48}, 8 (2002), 1086--1101.

\bibitem{kou2007jump}
{\sc Kou, S.~G.}
\newblock Jump-diffusion models for asset pricing in financial engineering.
\newblock {\em Handbooks in operations research and management science 15\/} (2007), 73--116.

\bibitem{kreps1981arbitrage}
{\sc Kreps, D.~M.}
\newblock Arbitrage and equilibrium in economies with infinitely many commodities.
\newblock {\em Journal of Mathematical Economics 8}, 1 (1981), 15--35.

\bibitem{kruse2017lp}
{\sc Kruse, T., and Popier, A.}
\newblock Lp-solution for bsdes with jumps in the case p< 2: corrections to the paper ‘bsdes with monotone generator driven by brownian and poisson noises in a general filtration’.
\newblock {\em Stochastics 89}, 8 (2017), 1201--1227.

\bibitem{kuen2014hedging}
{\sc Kuen~Siu, T., Nawar, R., and Ewald, C.~O.}
\newblock Hedging crude oil derivatives in garch-type models.
\newblock {\em Journal of Energy Markets 7}, 1 (2014).

\bibitem{lau2008option}
{\sc Lau, J.~W., and Siu, T.~K.}
\newblock On option pricing under a completely random measure via a generalized esscher transform.
\newblock {\em Insurance: Mathematics and Economics 43}, 1 (2008), 99--107.

\bibitem{lepeltier2005penalization}
{\sc Lepeltier, J.-P., and Xu, M.}
\newblock Penalization method for reflected backward stochastic differential equations with one rcll barrier.
\newblock {\em Statistics \& probability letters 75}, 1 (2005), 58--66.

\bibitem{lepinette2019conditional}
{\sc L{\'e}pinette, E., and Molchanov, I.}
\newblock Conditional cores and conditional convex hulls of random sets.
\newblock {\em Journal of Mathematical Analysis and Applications 478}, 2 (2019), 368--392.

\bibitem{leung2009accounting}
{\sc Leung, T., and Sircar, R.}
\newblock Accounting for risk aversion, vesting, job termination risk and multiple exercises in valuation of employee stock options.
\newblock {\em Mathematical Finance: An International Journal of Mathematics, Statistics and Financial Economics 19}, 1 (2009), 99--128.

\bibitem{li2018exchange}
{\sc Li, W., Liu, L., Lv, G., and Li, C.}
\newblock Exchange option pricing in jump-diffusion models based on esscher transform.
\newblock {\em Communications in Statistics-Theory and Methods 47}, 19 (2018), 4661--4672.

\bibitem{liang2016valuing}
{\sc Liang, X., Tsai, C. C.-L., and Lu, Y.}
\newblock Valuing guaranteed equity-linked contracts under piecewise constant forces of mortality.
\newblock {\em Insurance: Mathematics and Economics 70\/} (2016), 150--161.

\bibitem{madan1998variance}
{\sc Madan, D.~B., Carr, P.~P., and Chang, E.~C.}
\newblock The Variance Gamma process and option pricing.
\newblock {\em Review of Finance 2}, 1 (1998), 79--105.

\bibitem{madan1991option}
{\sc Madan, D.~B., and Milne, F.}
\newblock Option pricing with vg martingale components 1.
\newblock {\em Mathematical finance 1}, 4 (1991), 39--55.

\bibitem{mao1995adapted}
{\sc Mao, X.}
\newblock Adapted solutions of backward stochastic differential equations with non-lipschitz coefficients.
\newblock {\em Stochastic Processes and their Applications 58}, 2 (1995), 281--292.

\bibitem{matsuda2004introduction}
{\sc Matsuda, K.}
\newblock Introduction to merton jump diffusion model.
\newblock {\em Department of Economics, The Graduate Center, The City University of New York, New York\/} (2004).

\bibitem{meister1995contributions}
{\sc Meister, S.}
\newblock Contributions to the mathematics of catastrophe insurance futures.
\newblock {\em Unpublished Diplomarbeit, ETH Z{\"u}rich\/} (1995).

\bibitem{merton1973theory}
{\sc Merton, R.~C.}
\newblock Theory of rational option pricing.
\newblock {\em The Bell Journal of economics and management science\/} (1973), 141--183.

\bibitem{merton1974pricing}
{\sc Merton, R.~C.}
\newblock On the pricing of corporate debt: The risk structure of interest rates.
\newblock {\em The Journal of finance 29}, 2 (1974), 449--470.

\bibitem{merton1976option}
{\sc Merton, R.~C.}
\newblock Option pricing when underlying stock returns are discontinuous.
\newblock {\em Journal of financial economics 3}, 1-2 (1976), 125--144.

\bibitem{milevsky2001titanic}
{\sc Milevsky, M.~A., and Posner, S.~E.}
\newblock The titanic option: valuation of the guaranteed minimum death benefit in variable annuities and mutual funds.
\newblock {\em Journal of Risk and Insurance\/} (2001), 93--128.

\bibitem{miyahara2001geometric}
{\sc Miyahara, Y.}
\newblock [geometric l{\'e}vy process \& memm] pricing model and related estimation problems.
\newblock {\em Asia-Pacific Financial Markets 8\/} (2001), 45--60.

\bibitem{monfort2012asset}
{\sc Monfort, A., and Pegoraro, F.}
\newblock Asset pricing with second-order esscher transforms.
\newblock {\em Journal of Banking \& Finance 36}, 6 (2012), 1678--1687.

\bibitem{monoyios2007minimal}
{\sc Monoyios, M.}
\newblock The minimal entropy measure and an esscher transform in an incomplete market model.
\newblock {\em Statistics \& probability letters 77}, 11 (2007), 1070--1076.

\bibitem{navas2003calculation}
{\sc Navas, J.~F.}
\newblock Calculation of volatility in a jump-diffusion model.
\newblock {\em Journal of Derivatives 11}, 2 (2003).

\bibitem{ng2011valuing}
{\sc Ng, A. C.-Y., and Li, J. S.-H.}
\newblock Valuing variable annuity guarantees with the multivariate esscher transform.
\newblock {\em Insurance: Mathematics and Economics 49}, 3 (2011), 393--400.

\bibitem{ouknine1998reflected}
{\sc Ouknine, Y.}
\newblock Reflected backward stochastic differential equations with jumps.
\newblock {\em Stochastics: An International Journal of Probability and Stochastic Processes 65}, 1-2 (1998), 111--125.

\bibitem{papapantoleon2018existence}
{\sc Papapantoleon, A., Possama{\"\i}, D., and Saplaouras, A.}
\newblock Existence and uniqueness results for bsde with jumps: the whole nine yards.

\bibitem{pardoux1990adapted}
{\sc Pardoux, E., and Peng, S.}
\newblock Adapted solution of a backward stochastic differential equation.
\newblock {\em Systems \& control letters 14}, 1 (1990), 55--61.

\bibitem{peng1999monotonic}
{\sc Peng, S.}
\newblock Monotonic limit theorem of bsde and nonlinear decomposition theorem of doob--meyers type.
\newblock {\em Probability theory and related fields 113\/} (1999), 473--499.

\bibitem{peng2010reflected}
{\sc Peng, S., and Xu, M.}
\newblock Reflected bsde with a constraint and its applications in an incomplete market.

\bibitem{ramezani2007maximum}
{\sc Ramezani, C.~A., and Zeng, Y.}
\newblock Maximum likelihood estimation of the double exponential jump-diffusion process.
\newblock {\em Annals of Finance 3\/} (2007), 487--507.

\bibitem{rong1997solutions}
{\sc Rong, S.}
\newblock On solutions of backward stochastic differential equations with jumps and applications.
\newblock {\em Stochastic Processes and their Applications 66}, 2 (1997), 209--236.

\bibitem{rouge2000pricing}
{\sc Rouge, R., and El~Karoui, N.}
\newblock Pricing via utility maximization and entropy.
\newblock {\em Mathematical Finance 10}, 2 (2000), 259--276.

\bibitem{royer2006backward}
{\sc Royer, M.}
\newblock Backward stochastic differential equations with jumps and related non-linear expectations.
\newblock {\em Stochastic processes and their applications 116}, 10 (2006), 1358--1376.

\bibitem{salhi2017pricing}
{\sc Salhi, K.}
\newblock Pricing european options and risk measurement under exponential levy models—a practical guide.
\newblock {\em International Journal of Financial Engineering 4}, 02n03 (2017), 1750016.

\bibitem{samuelson1965rational}
{\sc Samuelson, P.~A.}
\newblock Rational theory of warrant pricing. industr. managementrev. 6, 13-32. samuelson136industr.
\newblock {\em Management Rev\/} (1965).

\bibitem{schachermayer2008close}
{\sc Schachermayer, W., and Teichmann, J.}
\newblock How close are the option pricing formulas of bachelier and black--merton--scholes?
\newblock {\em Mathematical Finance: an international journal of mathematics, statistics and financial economics 18}, 1 (2008), 155--170.

\bibitem{schal1999martingale}
{\sc Sch{\"a}l, M.}
\newblock Martingale measures and hedging for discrete-time financial markets.
\newblock {\em Mathematics of operations research 24}, 2 (1999), 509--528.

\bibitem{schmelzle2010option}
{\sc Schmelzle, M.}
\newblock Option pricing formulae using fourier transform: Theory and application.
\newblock {\em Preprint, http://pfadintegral. com\/} (2010).

\bibitem{schoutens2003levy}
{\sc Schoutens, W.}
\newblock {\em L{\'e}vy processes in finance: pricing financial derivatives}.
\newblock Wiley Online Library, 2003.

\bibitem{schweizer1995minimal}
{\sc Schweizer, M.}
\newblock On the minimal martingale measure and the m{\"o}llmer-schweizer decomposition.
\newblock {\em Stochastic analysis and applications 13}, 5 (1995), 573--599.

\bibitem{sircar2007general}
{\sc Sircar, R., and Xiong, W.}
\newblock A general framework for evaluating executive stock options.
\newblock {\em Journal of Economic Dynamics and Control 31}, 7 (2007), 2317--2349.

\bibitem{siu2015valuing}
{\sc Siu, C.~C., Yam, S. C.~P., and Yang, H.}
\newblock Valuing equity-linked death benefits in a regime-switching framework.
\newblock {\em ASTIN Bulletin: The Journal of the IAA 45}, 2 (2015), 355--395.

\bibitem{siu2001bayesian}
{\sc Siu, T.~K., Tong, H., and Yang, H.}
\newblock Bayesian risk measures for derivatives via random esscher transform.
\newblock {\em North American Actuarial Journal 5}, 3 (2001), 78--91.

\bibitem{sottinen2001fractional}
{\sc Sottinen, T.}
\newblock Fractional brownian motion, random walks and binary market models.
\newblock {\em Finance and Stochastics 5}, 3 (2001), 343--355.

\bibitem{stockbridge2008discrete}
{\sc Stockbridge, R.}
\newblock The discrete binomial model for option pricing.
\newblock {\em Phdprogram in Applied mathematics, available online www. sematicscholar. org\/} (2008).

\bibitem{stricker1981quelques}
{\sc Stricker, C.}
\newblock Quelques remarques sur la topologie des semimartingales. applications aux int{\'e}grales stochastiques.
\newblock In {\em S{\'e}minaire de Probabilit{\'e}s XV 1979/80: Avec table g{\'e}n{\'e}rale des expos{\'e}s de 1966/67 {\`a} 1978/79\/} (1981), Springer, pp.~499--522.

\bibitem{szimayer2004reduced}
{\sc Szimayer, A.}
\newblock A reduced form model for eso valuation.
\newblock {\em Mathematical Methods of Operations Research 59\/} (2004), 111--128.

\bibitem{tacklind1942risque}
{\sc T{\"A}CKLIND, S.}
\newblock Sur le risque de ruine dans des jeux in{\'e}quitables. skand. aktuar-tidskr. 25, 1-42.
\newblock {\em T{\"a}cklind125Skand. Aktuartidskr\/} (1942).

\bibitem{tang1994necessary}
{\sc Tang, S., and Li, X.}
\newblock Necessary conditions for optimal control of stochastic systems with random jumps.
\newblock {\em SIAM Journal on control and optimization 32}, 5 (1994), 1447--1475.

\bibitem{tankov2003financial}
{\sc Tankov, P.}
\newblock {\em Financial modelling with jump processes}.
\newblock CRC press, 2003.

\bibitem{tankov2010financial}
{\sc Tankov, P.}
\newblock Financial modeling with l{\'e}vy processes.

\bibitem{tian2014pricing}
{\sc Tian, L., Wang, G., Wang, X., and Wang, Y.}
\newblock Pricing vulnerable options with correlated credit risk under jump-diffusion processes.
\newblock {\em Journal of Futures Markets 34}, 10 (2014), 957--979.

\bibitem{ulm2008analytic}
{\sc Ulm, E.~R.}
\newblock Analytic solution for return of premium and rollup guaranteed minimum death benefit options under some simple mortality laws.
\newblock {\em ASTIN Bulletin: The Journal of the IAA 38}, 2 (2008), 543--563.

\bibitem{ulm2014analytic}
{\sc Ulm, E.~R.}
\newblock Analytic solution for ratchet guaranteed minimum death benefit options under a variety of mortality laws.
\newblock {\em Insurance: Mathematics and Economics 58\/} (2014), 14--23.

\bibitem{wang2007normalized}
{\sc Wang, S.}
\newblock Normalized exponential tilting: pricing and measuring multivariate risks.
\newblock {\em North American Actuarial Journal 11}, 3 (2007), 89--99.

\bibitem{wang2000class}
{\sc Wang, S.~S.}
\newblock A class of distortion operators for pricing financial and insurance risks.
\newblock {\em Journal of risk and insurance\/} (2000), 15--36.

\bibitem{wang2003equilibrium}
{\sc Wang, S.~S.}
\newblock Equilibrium pricing transforms: new results using buhlmann’s 1980 economic model.
\newblock {\em ASTIN Bulletin: The Journal of the IAA 33}, 1 (2003), 57--73.

\bibitem{wang2019computing}
{\sc Wang, W., and Zhang, Z.}
\newblock Computing the gerber--shiu function by frame duality projection.
\newblock {\em Scandinavian Actuarial Journal 2019}, 4 (2019), 291--307.

\bibitem{wilmot2013jump}
{\sc Wilmot, N.~A., and Mason, C.~F.}
\newblock Jump processes in the market for crude oil.
\newblock {\em The Energy Journal 34}, 1 (2013).

\bibitem{zhang2016model}
{\sc Zhang, M., and Revie, M.}
\newblock Model selection with application to gamma process and inverse gaussian process.
\newblock In {\em Risk, Reliability and Safety: Innovating Theory and Practice: Proceedings of ESREL 2016 (Glasgow, Scotland, 25-29 September 2016)}. 2016.

\bibitem{zhang2019valuing}
{\sc Zhang, Z., and Yong, Y.}
\newblock Valuing guaranteed equity-linked contracts by laguerre series expansion.
\newblock {\em Journal of Computational and Applied Mathematics 357\/} (2019), 329--348.

\bibitem{zhang2020valuing}
{\sc Zhang, Z., Yong, Y., and Yu, W.}
\newblock Valuing equity-linked death benefits in general exponential l{\'e}vy models.
\newblock {\em Journal of Computational and Applied Mathematics 365\/} (2020), 112377.

\end{thebibliography}
\end{document}